\documentclass{aastex631}

\DeclareRobustCommand{\VAN}[3]{#2}
\let\VANthebibliography\thebibliography
\def\thebibliography{\DeclareRobustCommand{\VAN}[3]{##3}\VANthebibliography}

\hypersetup{linkcolor=blue,citecolor=blue,filecolor=cyan,urlcolor=magenta}
\usepackage{graphicx}   
\usepackage{amsmath}    
\usepackage{amssymb}    
\usepackage{bm}
\usepackage{newtxtext,newtxmath}
\usepackage{cancel}
\usepackage{color}
\usepackage{xcolor}
\usepackage{threeparttable}
\usepackage{hyperref}
\usepackage{booktabs}
\usepackage{multirow}
\usepackage[T1]{fontenc}
\usepackage{ae,aecompl}
\usepackage{newtxtext,newtxmath}

\def\d{\mathrm{d}}

\newcommand{\lta}{\lower 2pt \hbox{$\, \buildrel {\scriptstyle <}\over {\scriptstyle \sim}\,$}}
\newcommand{\gta}{\lower 2pt \hbox{$\, \buildrel {\scriptstyle >}\over {\scriptstyle \sim}\,$}}
\definecolor{blazeorange}{rgb}{1.0, 0.4, 0.0}
\definecolor{seagreen}{rgb}{0.18, 0.55, 0.34}
\definecolor{rufous}{rgb}{0.66, 0.11, 0.03}
\definecolor{royalfuchsia}{rgb}{0.79, 0.17, 0.57}
\definecolor{scarlet}{rgb}{1.0, 0.13, 0.0}
\definecolor{royalpurple}{rgb}{0.47, 0.32, 0.66}

\begin{document}

\title{Extragalactic Magnetar Giant Flares: Population Implications, Rates and Prospects for Gamma-Rays, Gravitational Waves and Neutrinos}

\correspondingauthor{Paz Beniamini}
\email{pazb@openu.ac.il}

\author[0000-0001-7833-1043]{Paz Beniamini}
\affiliation{Department of Natural Sciences, The Open University of Israel, P.O Box 808, Ra'anana 4353701, Israel}
\affiliation{Astrophysics Research Center of the Open university (ARCO), The Open University of Israel, P.O Box 808, Ra'anana 4353701, Israel}
\affiliation{Department of Physics, The George Washington University, 725 21st Street NW, Washington, DC 20052, USA}

\author[0000-0002-9249-0515]{Zorawar Wadiasingh}
\affiliation{Astrophysics Science Division, NASA/GSFC, Greenbelt, MD 20771, USA}
\affiliation{Department of Astronomy, University of Maryland, College Park, MD 20742, USA}
\affiliation{Center for Research and Exploration in Space Science and Technology, NASA/GSFC, Greenbelt, MD 20771, USA}

\author[0009-0006-8598-728X]{Aaron Trigg}
\affiliation{Department of Physics \& Astronomy, Louisiana State University, Baton Rouge, LA 70803, USA}

\author[0000-0003-2759-1368]{Cecilia Chirenti}
\affiliation{Department of Astronomy, University of Maryland, College Park, MD 20742, USA}
\affiliation{Astroparticle Physics Laboratory, NASA/GSFC, Greenbelt, 20771, MD, USA}
\affiliation{Center for Research and Exploration in Space Science and Technology, NASA/GSFC, Greenbelt, 20771, MD, USA}
\affiliation{Center for Mathematics, Computation and Cognition, UFABC, Santo Andre, 09210-170, SP, Brazil}

\author[0000-0002-2942-3379]{Eric Burns}
\affiliation{Department of Physics \& Astronomy, Louisiana State University, Baton Rouge, LA 70803, USA}

\author[0000-0002-7991-028X]{George Younes}
\affiliation{Astrophysics Science Division, NASA/GSFC, Greenbelt, MD 20771, USA}
\affiliation{CRESST, Center for Space Sciences and Technology, UMBC, Baltimore, MD 210250, USA}

\author[0000-0002-6548-5622]{Michela Negro}
\affiliation{Department of Physics \& Astronomy, Louisiana State University, Baton Rouge, LA 70803, USA}

\author[0000-0001-8530-8941]{Jonathan Granot}
\affiliation{Department of Natural Sciences, The Open University of Israel, P.O Box 808, Ra'anana 4353701, Israel}
\affiliation{Astrophysics Research Center of the Open university (ARCO), The Open University of Israel, P.O Box 808, Ra'anana 4353701, Israel}
\affiliation{Department of Physics, The George Washington University, 725 21st Street NW, Washington, DC 20052, USA}

\begin{abstract}
Magnetar Giant Flares (MGFs) are the most energetic non-catastrophic transients known to originate from stellar objects. The first discovered events were nearby. In recent years, several extragalactic events have been identified, implying an extremely high volumetric rate. We show that future instruments with a sensitivity $\lesssim 5\times 10^{-9}$ erg cm$^{-2}$ at $\sim 1$ MeV will be dominated by extragalactic MGFs over short gamma-ray bursts (sGRBs). Clear discrimination of MGFs requires intrinsic GRB localization capability to identify host galaxies. As MGFs involve a release of a sizable fraction of the neutron star's magnetic free energy reservoir in a single event, they provide us with invaluable tools for better understanding magnetar birth properties and the evolution of their magnetic fields. A major obstacle is to identify a (currently) small sub-population of MGFs in a larger sample of more energetic and distant sGRBs. We develop the tools to analyze the properties of detected events and their occurrence rate relative to sGRBs. Even with the current (limited) number of events, we can constrain the initial internal magnetic field of a typical magnetar at formation to be $B_0\approx 4\times 10^{14}-2\times 10^{15}$\,G. Larger samples will constrain the distribution of birth fields. We also estimate the contribution of MGFs to the gravitational wave (GW) stochastic background. Depending on the acceleration time of baryon-loaded ejecta involved in MGFs, their GW emission may reach beyond 10~kHz and, if so, will likely dominate over other conventional astrophysical sources in that frequency range. 
\end{abstract}

\keywords{stars: magnetars -- gamma ray burst: general -- gravitational waves}




\section{Introduction}
\label{sec:intro}
For many years, a key scientific mystery in the field of gamma-ray bursts (GRBs) regarded their typical energies \citep{1995PASP..107.1167P,1995PASP..107.1152L}. If most GRBs were less energetic, they would only be viewable from closer (Galactic) distances, while if they were more energetic, their typical distances should have been Cosmological. The conclusive evidence in favor of highly energetic, extragalactic origins came in the forms of the first detected electromagnetic (EM) counterpart known as the afterglow and redshift measurements \citep{1997Natur.387..783C,1997Natur.386..686V,1997Natur.387..878M}. That being said, already in 1979 a GRB was detected \citep{mazets1979observations}, originating from the Large Magellanic Cloud (LMC). This event, and subsequent (now known to be spurious) claims of cyclotron features in GRBs caused much confusion, with large portions of the community favoring local highly-magnetized neutron stars (``magnetars") for engines of classical GRBs \citep[e.g.,][]{1984Ap&SS.107..191U,1984Natur.310..121L,1992herm.book.....M}. The 1979 event was ultimately realized as the first discovery of a magnetar giant flare (MGF) - A relatively energetic but non-catastrophic burst associated with a highly magnetized neutron star. In the following years, two more MGFs were detected and localized to the Milky Way \citep{hurley1999giant,Palmer+05}. Clearly, while most detected GRBs are cosmological, a fraction of events tagged as GRBs, are much lower energy MGFs, hiding within the GRB population. This fact indicates that future more sensitive detectors would be detect mostly MGFs, and that the present situation where cosmological GRBs which numerically dominate GRB catalogs is a technological and historical happenstance. Such a counterfactual reveals MGFs in the Universe are (at least as refers to commonality) the ``true GRBs", and cosmological GRBs associated with collapsars or compact binary coalescences are rare, albeit interesting, sub-populations.

How easy is it to separate GRBs and MGFs? The ratio between the isotropic equivalent energies in GRBs and MGFs is huge, of order $10^4-10^7$. However, the situation is somewhat reminiscent of the early days of GRB science - it is not trivial to separate between nearby and weak vs. far-away and energetic events. Moreover, the sub-second durations of the main pulse of MGFs and their hard spectrum make them, at face value\footnote{A potential discriminant is the short (few ms) rise time of MGFs \citep{Hakkila2018,2021ApJ...907L..28B} or tens of microsecond timescale variability \citep{Roberts2021Natur.589..207R}. However, most detected events lack sufficient signal to noise to resolve the lightcurve on such a short timescale.}, somewhat comparable in their broad characteristics to the class of short-hard gamma-ray bursts (sGRB), and as such the two types of transients may be potentially confused. 
A related issue deals with correlations between burst properties. MGFs appear to exhibit a distinct scaling relation between their spectral peak energy ($E_{\rm{p}}$) and isotropic-equivalent energy $E_{\rm{iso}}$, approximately $E_{\rm{iso}} \propto E_{\rm{p}}^{4}$ \citep{Zhang2020ApJ...903L..32Z}, which contrasts with the $E_{\rm{iso}} \propto E_{\rm{p}}^{2}$ Amati relation for sGRBs \citep{Amati2002A&A...390...81A}. This places MGFs in a unique region of the $E_{\rm{iso}}-E_{\rm{p}}$ plane, likely reflecting differences in their underlying physical mechanisms. Notably, this relation holds for the time-integrated analyses of the full population of extragalactic MGFs. A similar $L_{\rm{iso}} \propto E_{\rm{p}}^{4}$ trend was observed in the time-resolved spectral analysis of the MGF candidate GRB\,200415A \citep{Chand2021RAA....21..236C} (an equivalent relation, again with a different scaling, $L_{\rm iso}\propto E_{\rm p}^2$ was reported for GRBs, \citealt{Yonetoku2004}). However, as pointed out in \cite{Trigg2024AA...687A.173T}, the result is sensitive to sampling effects in the choice of temporal intervals, limiting its utility as a discriminant between sGRBs and MGFs.

The pulsating tail is a tell-tale sign of MGFs \citep{1979A&A....79L..24B,hurley1999giant}. This is a longer lived but weaker emission episode than that coming from the initial spike. The tail's flux is modulated by the NS's rotational period. For Galactic events, these tails are readily observed. However, due to their dimness, they fall below the level of the noise for MGFs occurring at distances beyond a few Mpc (see \S \ref{sec:EMGF cands}). 
Similarly, longer lived EM counterparts associated with sGRBs, such as their multi-wavelength afterglow signal and the optical kilonova signal are missed (due to their faintness or lack of telescope coverage at the right time / band) in tens of percents of sGRBs (see, e.g., review by \citealt{2007PhR...442..166N}), and as such the lack of such signals is, on its own, not conclusive.
Gravitational wave (GW) emission may also be a potential discriminant. This too, is not so practical at this stage, considering that a direct measurement of GWs from an MGF is likely to only be detectable for a Galactic (or magellanic) event (as detailed in \S \ref{sec:GW}), while GWs from sGRBs are currently detectable up to $\sim200$ Mpc, much less than the typical distance of a gamma-ray detected sGRB. 
A final (and currently the most effective) tool for identifying MGFs, is a statistically significant spatial association with a nearby galaxy. This technique has so far led to the identification of 6 extragalactic MGF candidates \citep{ofek2006short,frederiks2007AstL...33...19F,mazets2008giant,Ofek2008ApJ...681.1464O,svinkin2021bright,Roberts2021Natur.589..207R,mereghetti2023magnetar,2024arXiv240906056T} and will improve with dedicated software \citep{2020ApJ...900...35T,2022ApJ...941..169D,2024arXiv241005720T} and instrumentation.

Magnetars stand apart from regular pulsars by virtue of their bursting and persistent X-ray activity, which is too luminous to be powered by the loss of rotational energy. Indeed it is thought that the decay of their large magnetic energy reservoirs is the main source of their power (see \citealt{2017ARA&A..55..261K} for a review). X-ray bursts from magnetars exhibit a wide range of energetics. MGFs, involving an energy release comparable to the dipolar magnetic energy reservoir of their underlying magnetars, represent the high end of the magnetar burst distribution. This, along with the fact that energy distribution of magnetar bursts appears to be top-heavy (i.e. the most energetic bursts dominate the overall energy release by bursts, see \citealt{1996Natur.382..518C,Gogus1999,Gogus2000}), means that quantifying the energies and rates of MGFs holds invaluable clues for understanding the underlying properties of these objects and answering such questions as what is the typical birth field of magnetars and what fraction of their magnetic energy is channeled to bursts.

We present here a phenomenological and largely model-independent study of MGFs and consider how the growing population of identified extragalactic MGFs and improved limits on their fraction within the sGRB population can be used to constrain key properties of magnetars. We begin, in \S \ref{sec:MGFsGRB} with a summary of previous works constraining the fraction of MGFs within the sGRB population and identifying extragalactic MGF candidates. We also present conservative and robust limits on the MGF/sGRB ratio and show how such constraints are sensitive to the limiting fluence of the underlying sample and the maximum distance to within which MGFs are searched for. Then, in \S \ref{sec:analysis} we present our physical modeling of MGFs from first a single magnetar and then a population within the local Universe. We develop tools for using the observed constraints discussed in \S \ref{sec:MGFsGRB} to constrain the underlying physical properties. In \S \ref{sec:GW} we discuss the prospects of detecting the contribution of MGFs to the stochastic GW background and in \S \ref{sec:theoreticalneutrino} their neutrino emission. We discuss implications and directions for future studies in \S \ref{sec:discuss} and conclude in \S \ref{sec:conclusion}.

\section{MGFs within the sGRB population - Previous inferences, new extragalactic candidates and observational constraints}
\label{sec:MGFsGRB}

\subsection{Summary of previous studies}
There have been multiple searches for a population of extragalactic MGFs. Many of these studies used spatial distributions of galaxies to constrain the fraction of sGRBs that have a MGF origin. Prompted by the detection of GRB~041227 from SGR 1806--20, \cite{Palmer+05} estimated the number of MGFs in the population of sGRB within 40~Mpc to be $<5\%$. Similarly, using spectral comparisons between MGFs and sGRBs, \cite{2005MNRAS.362L...8L} estimated the population to be $<4\%$. \cite{2005Natur.438..991T} looked at simulated data and real data from BATSE to build a correlation function to connect between short bursts of energy in space and the positions of galaxies within $\sim$100\,Mpc , finding that about $\sim10-25\%$ of low redshift bursts (z<0.025) bursts seem to be connected to the positions of galaxies. They further infer that the fraction of sGRBs that are potentially MGFs to be $\sim(8\pm6)\%$ (95\% confidence). A study by \citet{2006MNRAS.365..885P} looked for MGFs in nearby star-forming galaxies using data from BATSE, in particular focusing on four galaxies that are most likely to host a MGF. No convincing detections were found, leading to an upper limit on the galactic rate of MGFs with an energy release in the initial spike above $0.5\times10^{44}\,\rm{erg}$ of less than 1/30~$yr^{-1}$ in the Milky Way. Using data for short-hard GRBs  in IPN data, \citet{2006ApJ...640..849N} and \cite{2007ApJ...659..339O} both attempted to associate a sample of well localized sGRBs with local galaxies. \cite{2006ApJ...640..849N} searched for galaxies within $\sim$100\,Mpc, looking at a sample of six short-hard GRBs. Finding no credible galaxies in these well-localized error boxes, they estimated the lower limit of their energy output to be at least $10^{49}\rm{ergs}$, making some of them potentially detectable at distances greater than 1\,Gpc. This suggests that these bursts are part of a cosmological population of short-hard GRBs, consistent with previous observations. The energy estimations were at least two orders of magnitude higher than that seen in GRB~041227, suggesting that less than 15\% of short-hard GRBs are similar to that MGF. \cite{2007ApJ...659..339O} looked at 47 sGRBs and checked if the error region of each burst overlapped with the apparent disk of any of 316 bright, star-forming galaxies within 20 Mpc. Using a limiting fluence above $\sim10^{-6}\rm{erg\,cm^{-2}}$, this study was able to estimate the fraction of MGFs among sGRBs to be $<16\%$ and to set a lower limit of 1\% based on the Galactic MGF rate. Further studies utilizing IPN localizations for GRB catalogs and lower limiting fluence values ($\sim10^{-7}\rm{\, erg\,cm^{-2}}$) find the fraction of MGFs in the sGRB sample to be $\sim7\%-8\%$ \citep{2010AstL...36..231T, 2015MNRAS.447.1028S}. More recent studies have constrained the intrinsic rates of MGFs to $<3\rm{yr^{-1}}$ within 11~Mpc \citep[$<4\rm{yr^{-1}}$ within 200~Mpc][]{2018Galax...6..130M} and $<1.3^{+1.7}_{-0.8}\rm{yr^{-1}}$ within 200~Mpc for a limiting fluence of $\lesssim2\times10^{-8}\rm{erg\,cm^{-2}}$ \citep{2020MNRAS.492.5011D}. The latest estimate of the MGF fraction is $>1.6\%$ based on a limiting fluence of $\sim10^{-7}\rm{erg\,cm^{-2}}$ and comes from \cite{2021ApJ...907L..28B}. This study identified four MGF candidates in a sample of 250 sGRBs (three previously identified and one new candidate) by quantifying how likely a burst of a given fluence at Earth is to have a MGF origin from a nearby star-forming galaxy. From this work, they were able to determine an intrinsic volumetric rate (above $4\times 10^{44}\mbox{erg}$) of $\mathcal{R}_{\rm MGF}^0 = 3.8^{+4.0}_{-3.1}\times10^5$ Gpc$^{-3}$ yr$^{-1}$ \citep{2021ApJ...907L..28B}. There is some possibility of misidentification of MGFs with this procedure \citep{2023arXiv230305922S}, but is mitigated by better future localization capability and sensitivity.   In a comprehensive search for extragalactic MGFs, \cite{Pacholski2024} analyzed data from the IBIS instrument on the INTEGRAL satellite, targeting the Virgo Cluster and nearby galaxies with high star formation rates. This study used nearly 35 Ms of data from the Virgo Cluster and additional exposures of seven star-forming galaxies. No MGFs were detected in the Virgo Cluster, but a candidate, GRB 231115A, was identified in galaxy M82. Based on these findings, the study sets a 90\% upper confidence limit on MGF rates, estimating that a flare with energy $>3\times10^{45} \rm{erg}$ might occur once every 500 years per magnetar. A lower limit on the rate of MGFs with $E>10^{45}\rm{erg}$ was found to be $\mathcal{R}_{\rm MGF}>4\times10^{-4}\rm{yr^{-1}\,magnetar^{-1}}$.

\subsection{Extragalactic MGF candidates}
\label{sec:EMGF cands}
The first three MGF detections, all occurring in the Milky Way and Large Magellanic Cloud \citep{mazets1979observations,Feroci-1999-ApJL,hurley1999giant,Mazets+99,Hurley+05,Palmer+05} and defining the characteristics of these phenomena, displayed a bright, millisecond-long spike followed by a decaying tail modulated by the rotational period of the neutron star from which it originated. The modulated tail seen in these bursts is considered the smoking-gun signature of a MGF. Unfortunately, even for the brightest MGFs, the luminosities of these tails, which decay quasi-exponentially over hundreds of seconds, can be several orders of magnitude lower than that of the main peak \citep{Palmer+05}. This means that, at extragalactic distances, this unambiguous signal indicative of a MGF falls below the sensitivity threshold of current X-ray monitors with large fields of view. Therefore, identifying MGF candidates from extragalactic sources requires localization within nearby star-forming galaxies. By using distances to these galaxies, an isotropic-equivalent energy can be determined for a given burst to assess if it falls within the expected range for a MGF, as cosmological sGRBs tend to have much higher isotropic-equivalent energies. Multi-band and multi-messenger follow-up observations of the host galaxy can then help rule out other possible progenitor classes

In Table \ref{tab:sources} we summarize the constraints on the main MGF population parameters reported by several studies. Over the last $\sim$20 years, six extragalactic MGF candidates have been identified. Four of these candidates, GRB\,051103, GRB\,070201, GRB\,200415A, and GRB\,231115A, were localized to the nearby star-forming galaxies M81, M31, NGC 253, and M82, respectively \citep{ofek2006short,frederiks2007AstL...33...19F,mazets2008giant,Ofek2008ApJ...681.1464O,svinkin2021bright,Roberts2021Natur.589..207R,mereghetti2023magnetar,2024arXiv240906056T}, soon after their initial detection. Various follow-up observations in the optical and UV bands, as well as gravitational wave searches, provided convincing evidence that these events were likely extragalactic MGF candidates. GRB\,070222, initially localized as a long arc with no galaxy association, was identified through the population search in \cite{2021ApJ...907L..28B}. The statistical method described therein also identified three of the other extragalactic MGF candidates, associating GRB\,070222 with the galaxy M83. A reanalysis of gravitational wave data from around the time of this burst ruled out the possibility of a NS-NS merger at the distance to M83. The remaining extragalactic MGF candidate, GRB\,180128A, was identified during a search of archival \textit{Fermi}-GBM data for MGFs masquerading as sGRBs \citep{Trigg2024AA...687A.173T}. During the initial search, which utilized the sharp rise-time and short peak interval to down-select the GRB sample prior to localization and comparison with star formation rate (SFR) for nearby galaxies, this burst stood out as a likely candidate. Galaxy association through IPN localization, along with further spectral and temporal analysis, provided convincing evidence supporting the classification of this burst as an MGF.

Analyzing the temporal and spectral characteristics of extragalactic MGF candidates in comparison with those of the known Galactic MGFs is crucial for verifying whether these candidates indeed have a MGF origin. The bottom half of Table\,\ref{tab:sources} lists all of the relevant information for the six extragalactic MGF candidates. The distance measurements to the associated host galaxies along with the flux and fluence measurements, allow for the calculation of the $L_{\rm{iso}}$ and $E_{\rm{iso}}$ values for these bursts. These values are consistent with those of the three Galactic MGFs. For nearly all associated host galaxies, the SFRs are higher than the value of $1.65\pm0.19\,\rm{M}_{\odot}\,yr^{-1}$ \citep{Licquia2015ApJ...806...96L} seen in the Milky Way. The statistical method in \citet{2021ApJ...907L..28B} for determining the chance alignment significance with the associated host galaxies, denoted as False Alarm Rate (FAR), depends on a linear weighting of the SFRs. Finally, various time-integrated spectral analyses of these MGF candidates all fit a Comptonized spectral model. This function displays a power-law characterized by an index $\Gamma$ with an exponential cutoff at a characteristic energy, $E_{\rm{p}}$ near the spectral peak. The values for $\Gamma$ vary from between -1 and 0, with the higher $E_{\rm{p}}$ values corresponding to $\Gamma\sim0.0$ and the lower $E_{\rm{p}}$ values corresponding to $\Gamma\sim1.0$, consistent with the theoretical model developed in \cite{Roberts2021Natur.589..207R} and \cite{Trigg2024AA...687A.173T}.

As previously mentioned, the definitive indicator of the MGF nature of a transient would be the detection of the pulsating tail. From the Galactic MGF sample, we can infer the typical duration of MGF tails to be several minutes and the intrinsic total energy to be about $\sim10^{44}$ erg. As recently highlighted in \cite{2024FrASS..1188953N}, an agile and/or sensitive X-ray mission could detect such signatures either through hyper-fast re-pointing or through the serendipitous detection of an event occurring in the field of view. The former scenario would require an automated communication system between wide-field monitors and sensitive pointing X-ray observatories, as a re-pointing time of about (or under) a minute from the trigger is necessary to assure the detection of pulsation out to $\sim$3.5 Mpc \citep{2024FrASS..1188953N,2024arXiv240906056T}. The latter scenario allows us to compromise on sensitivity in favor of a large instantaneous sky coverage.

\begin{table}
\footnotesize
\scriptsize
\setlength{\tabcolsep}{3pt}
\begin{center}
\begin{tabular}{l||cccccccccccc}
\textbf{MGF} & \textbf{Distance D} & \textbf{SFR} & \textbf{Principal} & \textbf{Significance }& $\boldsymbol{L_{\rm iso}}$ & $\boldsymbol{E_{\rm iso}}$ & $\boldsymbol{\Phi}$ & \textbf{Rise Time} & $\boldsymbol{\Gamma}$ & $\boldsymbol{{\cal E}_{\rm peak}}$ & $B_{\rm{dipole}}$ & $\boldsymbol{E_{\rm B_{\rm dip}}}$\\
 & [Mpc] & [$M_\odot$ yr$^{-1}$] & \textbf{Instrument} & [FAR $\sigma$] & [$10^{46}$ erg s$^{-1}$] & [$10^{45}$ erg] & [$10^{-5}$ erg cm$^{-2}$] & [ms] & (photon index) & [keV] & [$10^{14}$ G] &  [$10^{45}$ erg]\\
\hline \hline
790305 & 0.054 & 0.56 & Konus &  $\infty$ &0.36 & 0.16 & 45 & $\lesssim$2 & -- & 500 & 5.6 & $1.1$\\
980827 & 0.0125 & 1.65 & KW & $\infty$ & >0.04 & >0.07 & $6\times10^3$ & $\lesssim$4 & -- & 1200 & 7.0 & 0.9 \\
041227 & 0.0087 & 1.65 & RHESSI/KW &  $\infty$ & 35 & 23 & $9\times10^4$ & $\sim$1 & -0.7 & 850 & 20.0 & 47 \\
\hline
051103 & 3.6 & 7.1 & KW & 4.2 & 180 & 53 & 3 & $\lesssim$4 & -0.1 & 2690  \\
070201 & 0.78 & 0.4 & KW & 3.7 & 12 & 1.5 & 2 & $\sim$20 &  -0.6 & 280  \\
070222 & 4.5 & 4.2 & KW & 4.3 & 40 & 6.2 & 0.3 & $\sim$4 & -1.0 & 1290  \\
180128A & 3.5 & 4.9 & GBM & 3.3 & 0.4 & 0.6 & 0.2 & $\sim$2 & -0.6 & 290  \\
200415A & 3.5 & 4.9 & GBM/KW & 4.4 & 140 & 13 & 0.9 & 0.08 & 0.0 & 1080  \\
231115A & 3.5 & 7.1 & GBM/KW & 5.0 & 1.14 & 1.15 & 0.8 & $\sim$3 & -0.1 & 600  \\
\hline
\end{tabular}
\end{center}
\caption{Table adapted from \citealt{2021ApJ...907L..28B}, and \citealt{Trigg2024AA...687A.173T,2024arXiv240906056T}. $B_{\rm{dipole}}$ values for the Galactic MGFs taken from  \citealt{OlausenKaspi14}. Rise time values are taken from the literature (see below). They are calculated using different methods, utilizing observations from new and historical missions with varying temporal resolutions. See also \cite{mazets1979observations,1980ApJ...237L...1C,Mazets+99,hurley1999giant,Palmer+05,ofek2006short,mazets2008giant,Roberts2021Natur.589..207R,svinkin2021bright}. } 
\label{tab:sources}
\end{table}

\subsection{Model independent constraints on MGF fraction in observed sGRB population}
\label{sec:modelindconst}
Given that the convincing extragalactic MGF candidates are all associated to galaxies under 5\,Mpc, but that current instrumentation can detect these events to much further distances, it is important to understand if we are failing to identify the most distance MGFs in the archival sample. To investigate this possibility we inject a set of GRBs into the method described in \citet{2021ApJ...907L..28B}. Here we assume a source galaxy with an integral SFR of 5\,M$_\odot$\,yr$^{-1}$, corresponding to a starburst galaxy, being an optimistic but reasonable case. To study the various identification capabilities of active instruments and networks, as well as future ones, we vary the GRB fluence in order of magnitude steps from $1.0\times10^{-9}$ to $1.0\times10^{-5}$\,erg\,cm$^{-2}$, 1-sigma circular-equivalent radii uncertainties of 0.03, 0.1, 0.3, 1, 3, 10, and 30\,deg, and place the fiducial host galaxy over a range of distances. The placement of the center of the GRB localization is randomized based on the angular extent of the galaxy (utilizing M82 as the baseline, scaled by distance) and the GRB uncertainty.

Swift-BAT and INTEGRAL have a capability of localizations with 0.03\,deg uncertainty or better and can detect bursts down to as low as a few $10^{-8}$\,erg\,cm$^{-2}$ although may not generally attain this sensitivity for bursts as hard as MGFs. With the full assumptions in \citet{2021ApJ...907L..28B}, we would identify the correct starburst host galaxy at $\sim$3$\sigma$ confidence to $\sim$30\,Mpc and would flag a candidate event to $\sim$100\,Mpc. The IPN is capable of performing triangulation for bursts above $1.0\times10^{-6}$\,erg\,s$^{-1}$\,cm$^{-2}$, from arcminute scale accuracy to several tens of square deg. For IPN localizations with 90\% containment area of $<1.0$\,deg$^2$ we would identify the real host galaxy at $\sim$3$\sigma$ confidence to 10\,Mpc, and with some ambiguity at 15-20\,Mpc. As there is some relation between brightness at Earth and IPN localization capability the exact maximum association distance depends on the intrinsic brightness. 

One possible solution as to why this may not occur is these events would have the highest E$_{\rm iso}$ and that may correspond to an intrinsically harder spectrum, giving fewer photons for a fixed energy, and preventing detection in some cases. However, this explanation may not be required. There is a local excess of SFR within 5-10\,Mpc (see Fig. \ref{fig:overdensity}), which may explain the lack of identification of MGFs from galaxies beyond 5\,Mpc because these starforming galaxies do not exist. To quantify this we use the full assumptions in \citet{2021ApJ...907L..28B} to rank the galaxies by their likelihood of producing a detectable MGF. All extragalactic MGF candidates are associated to galaxies in the top 10. The Milky Way, NGC 253, and M82, which each have two identified MGFs, are three of the top four galaxies. The fourth is M77 with a SFR of $\sim$32.5\,$M_{\odot}$\,yr$^{-1}$ at a distance of 12.3\,Mpc (these values and subsequent from \citealt{2019ApJS..244...24L}). The remaining galaxies in the top 10 without associated MGFs are IC 342 ($\sim$1.9\,$M_{\odot}$\,yr$^{-1}$, 2.3\,Mpc), the Circinus Galaxy ($\sim$3.9\,$M_{\odot}$\,yr$^{-1}$, 4.2\,Mpc), and NGC~6946~($\sim$6.1\,$M_{\odot}$\,yr$^{-1}$, 7.7\,Mpc). Two of these are under 5\,Mpc, one between 5-10\,Mpc, and one beyond. So far, the lack of identified MGFs beyond 5\,Mpc is consistent with Poisson variation in the observed sample.

\citet{2021ApJ...907L..28B} utilized HEALPix \citep{2005ApJ...622..759G} for discrete representation of sGRB skymaps, integral SFRs from modern galaxy catalogs and an approximate model of the intrinsic MGF energetics function to answer the question of how many MGFs can we confidently identify in the detected population of sGRBs. We utilize this same framework to instead answer a different question: what fraction of the detected sGRB population could have a MGF origin?

With the method in \citet{2021ApJ...907L..28B}, the maximum distance to which a MGF can be associated to a given host galaxy depends on both the localization precision and MGF flux. Thus, it is instrument-specific. It also depends on the property of a given host, so we get representative values (i.e., assuming they are star-forming galaxies similar to M82 and NGC\,253). For Swift-BAT and INTEGRAL IBIS, with localization precisions of a few arcminutes and sensitivity limits on the order of $10^{-8}$~erg~s$^{-1}$~cm$^{-2}$, the maximal host association distance is $\sim$25-30~Mpc. For reasonably precise IPN localizations, i.e, 0.1~deg$^2$ or less, the maximal host association distance is $\sim$20\,Mpc. For localizations up to 10-100~deg$^2$ the maximal association distance is $\sim$10\,Mpc. For worse localizations, no individual host association is robust.

We utilize the same approach as in \citet{2021ApJ...907L..28B}, with some modification. First, we minimize the number of assumptions by dropping the assumed intrinsic energetics function for MGFs, in effect treating the distance to other galaxies as separate from their likelihood to produce a MGF of a given brightness at Earth. We consider only galaxies within 10\,Mpc. Second, rather than identifying confident MGFs, we consider an event to be a viable MGF if the spatial Bayes factor defined in \citet{2021ApJ...907L..28B} exceeds 0.1, i.e. if the GRB localization is aligned with nearby galaxies to the extent that a MGF origin is favored by an odds ratio of at least 1 in 10. Note that because MGFs are a subdominant component of the detected sGRB population we expect many of these events to be cosmological sGRBs \citep[presumably from distant neutron star mergers,][]{2020LRR....23....4B}. We maintain the weighting of nearby galaxies by either the integral SFR or the total stellar mass as measured by \citet{2019ApJS..244...24L} and utilize the original 250 sGRB sample in \citet{2021ApJ...907L..28B}.

\begin{table}
\centering
\begin{tabular}{c||c|c|c}
\hline
\textbf{Max. Dist. [Mpc]} & \textbf{Weighting} & \textbf{Count} & \textbf{90\% Limit} \\ \hline \hline
20 & SFR           & 9  & $<$5.7\%   \\ 
   & Stellar Mass  & 9  & $<$5.7\%   \\ 
   & Inclusive     & 11 & $<$6.6\%   \\ \hline
30 & SFR           & 21 & $<$11.2\%  \\ 
   & Stellar Mass  & 18 & $<$9.9\%   \\ 
   & Inclusive     & 22 & $<$11.7\%  \\ \hline
\end{tabular}
\caption{The 90\% upper limit of the $N_{\rm MGF}/N_{\rm sGRB}$ occurrence ratio from the observed sGRB sample, according to the Gehrels statistic, for 6 combinations of assumptions. This is derived from IPN observations over 30~years, being a sample of 250 short GRBs with reasonably precise localizations. We allow the maximum detection distance to be either 20 or 30~Mpc. If magnetars arise predominantly from CCSN then the galaxies would be weighted by SFR. If instead they arise from a delayed channel the weighting would be total stellar mass. If multiple channels contribute, then the inclusive set of bursts from these two options are allowed. }
\label{tab:upper_limits}
\end{table}

We calculate both a lower limit and upper limit on the fraction of observed short GRBs which are due to MGFs. Out of the 250 GRBs there are 4 events (GRBs 051103, 070201, 070222, and 202415A). The 90\% lower limit using the Gehrels statistic \citep{1986ApJ...303..336G} is 1.745, or $>$0.7\%. The upper limit is more complex and depends on i) the assumed maximal detection distance and ii) the assumed weighting scheme. The 90\% upper limit from these combinations are shown in Table\,\ref{tab:upper_limits}, ranging from $<$5.7\%-$<$11.7\%.

\subsection{Over-density of Galaxies within 30 Mpc}
\label{sec:overdensity}
\begin{figure}
\centering
\includegraphics[scale=0.49]{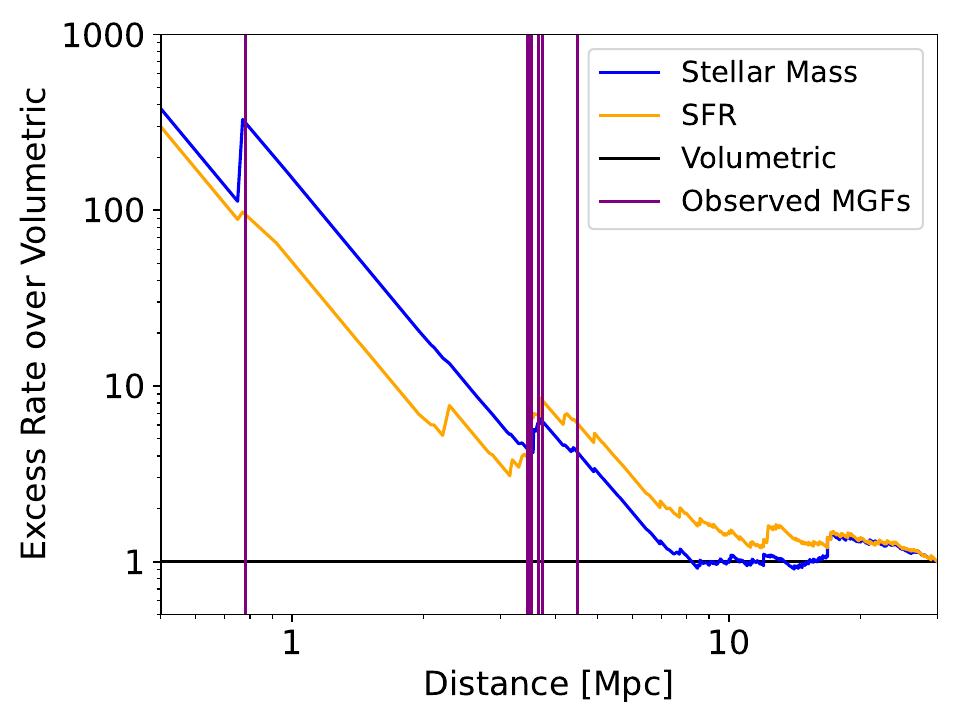}
\includegraphics[scale=0.49]{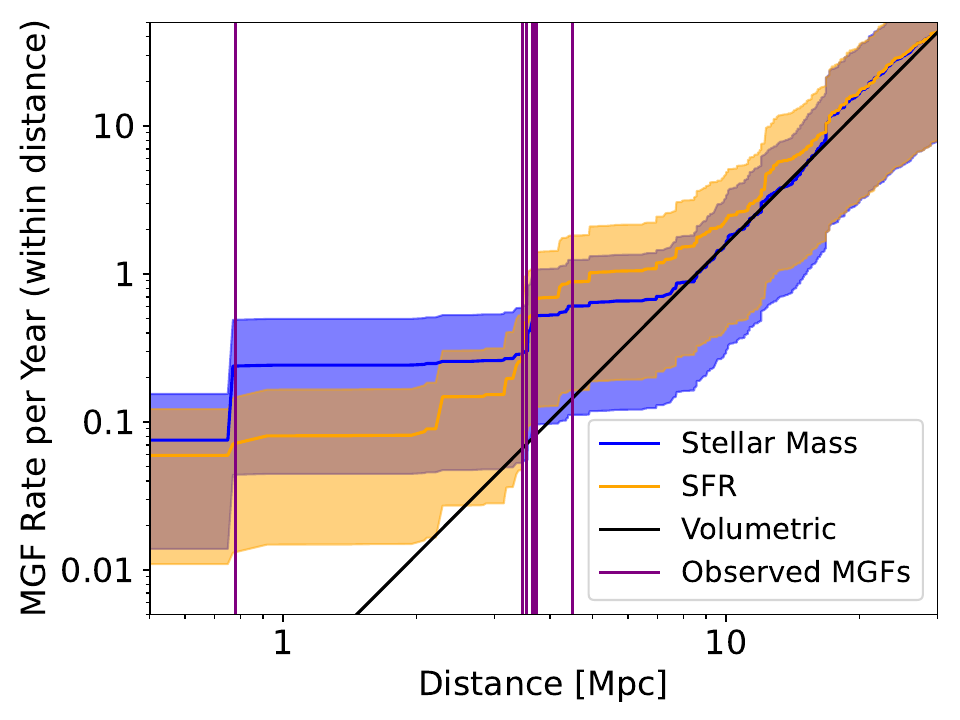}
\caption{The local galaxy distribution results in an overdensity of matter, as compared to the volumetric average on scales where the universe is isotropic and homogeneous. The normalization was set at 30~Mpc, which is roughly the scale where the Universe is homogeneous. Vertical lines denote the distances to MGF candidates listed in Table \ref{tab:sources}. Left: excess density factor ($\Delta(r)=n(r)/n(30\mbox{Mpc})$ in \S \ref{sec:overdensity}) for stellar mass and SFR following populations over this homogenized average as a function of distance in the local universe, normalized to 30\,Mpc, based on the z=0 Multiwavelength Galaxy Synthesis Catalog \citep{2019ApJS..244...24L}. Right: Effect on the cumulative rate within a given distance. }
\label{fig:overdensity}
\end{figure}

On the scales of $\gtrsim$30\,Mpc the universe is relatively uniform, appearing both homogeneous and isotropic. On smaller scales the local effects of gravity have caused galaxies to form and attract each other, causing overdensities for observers in galaxies, including us in the Milky Way. As MGFs are currently predominantly only detectable within 30~Mpc we must consider the local overdensity for detection and intrinsic rates. We note that this is a benefit to us in detecting MGFs, as well as other transients possibly associated with magnetars (e.g. FRBs). The local overdensity is shown in Fig.~\ref{fig:overdensity}. 
For the purposes of modeling, we denote by $\Delta(r)$ the ratio of densities of sources between a case in which sources follow the local rate of star-formation (or alternatively galaxy stellar mass) and the case in which they follow a homogeneous distribution that is normalized by the same quantity at distances $\gtrsim 30$\,Mpc, i.e. $\Delta(r)\equiv n_{\rm mag}(r)/n_{\rm mag}(r\approx \mbox{30Mpc})$. The rate of events from a distance $<r$ (ignoring cosmological corrections which are very small for these distances) is then $\dot{N}\propto \int \Delta(r) r^2dr$.

\section{Modeling the MGF and GRB populations}
\label{sec:analysis}
We seek to tie the magnetar burst energy properties to their intrinsic properties. We begin, in \S \ref{sec:IndMag}, with modeling individual magnetars. We consider their magnetic energy reservoir and temporal evolution and connect those to their observed burst energy distribution. In \ref{sec:MGFpop} we then extend this analysis to a population of magnetars. We account for detectability constraints as well as for the non-uniform distribution of sources in the near Universe. Since MGFs are hidden as a small sub-population of the sGRB population, one has to account for the latter in order to infer the properties of the former. To this end, we model the sGRB population in \S \ref{sec:sGRB}.
Finally, in \S \ref{sec:combanalyticresult} we present how different observed samples can be used to inform the most important parameters in the modeling of the two populations. In particular we use our model independent constraints derived in \S \ref{sec:modelindconst} to constrain the most important parameters in our modeling.

\subsection{Burst Energy-Age Distributions from Individual Magnetars}
\label{sec:IndMag}

\begin{table}
\setlength{\tabcolsep}{4pt}
\begin{center}
\begin{tabular}{c||c|l}
\textbf{Parameter} & \textbf{Range} & \textbf{Description} \\
\hline \hline
$s$ & $\sim 1.7$ & Universal power-law index for burst event size distribution   \\
$b_c$ & $\sim 1$ & Energy distribution cutoff steepness parameter   \\
$f_E$ & $[0.03,0.3]$ & Ratio between maximum (beaming corrected) MGF energy and magnetar's dipolar magnetic energy \\
$f_{\rm dip}$ & $[0.1,1]$ & Ratio between magnetar's dipolar magnetic energy and free magnetic energy\\
$f_{b}$ & $[0.1,1]$ & Beaming fraction, i.e. $E_{\rm t}/E$\\
$f_{\rm bol}$ & $[0.3,1]$ & Ratio between observed X-ray + $\gamma$-ray and bolometric emitted energy (isotropic equivalent)\\
$f_{\rm fl}$ & $[0.1,1]$ & Fraction of magnetic energy decay channeled into flares (note $f_{\rm fl}>f_E f_{\rm dip}$)\\
$f_{\rm mag}$ & $[0.15,1]$ & Fraction of core-collapse SNe resulting in magnetars\\
$\mathcal{R}_{\rm CCSN}(z=0)$ & $\sim 10^5\,\,\mbox{Gpc}^{-3}\mbox{\,yr}^{-1}$ & Local core-collapse SNe rate\\
$\tau_{\rm d,0}$ & $[10^{3},10^{4}]$\,yr & Magnetar's initial internal magnetic field decay time$\approx1800\mbox{yr}/f_{\rm mag}$\\
$\alpha$ & $[-1,1]$ & Index for internal field decay, $\dot{B}\propto B^{1+\alpha}$ \\
$B_0$ & $[10^{14.5},10^{16}]$\,G & Magnetar's initial total magnetic field (determining initial magnetic energy $E_{\rm B,0}$)\\
$B_{\rm min}$ & - & Magnetar's minimum initial total magnetic field (assuming a distribution between sources)\\
$B_{\rm max}$ & - & Magnetar's maximum initial total magnetic field (assuming a distribution between sources)\\
$\beta$ & - & PL index for initial magnetic field distribution\\
$\eta_{\rm kin}$ & $\sim 1$ & ratio between MGF kinetic and EM energy\\
$\eta_{\rm GW}$ & - & ratio between MGF GW and kinetic energy\\
\hline
$E_{\rm min}$ & $\sim 1.5\times 10^{49}$\,erg & Minimum (isotropic equivalent) gamma-ray energy radiated along the core of a sGRB jet\\
$E_{*}$ & $\sim 6\times 10^{51}$\,erg & Break scale of the (isotropic equivalent) gamma-ray energy radiated along the core of a sGRB jet\\
$\alpha_{E}$ & $\sim 0.95$ & PL index for (isotropic equivalent) core gamma-ray energy distribution at $E_{\rm min}<E_{\rm c}<E_*$\\
$\beta_{E}$ & $\sim 2$ & PL index for (isotropic equivalent) core gamma-ray energy distribution at $E_{\rm c}>E_*$\\
$\theta_{\rm c}$ & $\sim 0.1$ & Opening angle of sGRB jet core\\
$\theta_{\rm max}$ & $\sim 1$ & Maximum angle to which sGRB PL energy angular profile extends\\
$a$ & $\sim 4.5$ & PL index of (isotropic equivalent) kinetic energy angular profile\\
$\tilde{a}$ & $\sim 6$ & PL index of (isotropic equivalent) gamma-ray energy angular profile\\
$\mathcal{R}_{\rm BNS}(z=0)$ & $\sim 320\,\,\mbox{Gpc}^{-3}\,\mbox{yr}^{-1}$ & Local BNS merger rate\\
$f_{\rm sGRB}$ & $\sim 1$ & Fraction of BNS mergers resulting in sGRBs\\
\hline
\end{tabular}
\end{center}
\caption{Table of Magnetar (above the horizontal line) and sGRB (below the line) energy distribution parameters used in this work. } 
\label{tab:paras}
\end{table}

For each individual magnetar, the differential distribution function of (collimation-corrected) energies is
\begin{equation}
\frac{\partial {\cal N}}{\partial E_{\rm t}}\propto E_{\rm t}^{-s}\exp\left[-\left(\frac{E_{\rm t}}{E_{\rm c,t}}\right)^{b_c}\right]\ ,\quad
E_{\rm c,t} = f_{E} f_{\rm dip} E_{B}(\tau)\ ,
\label{eq:dndesimp}
\end{equation}
where here and elsewhere in the paper $E_{\rm t}$ to denotes true (collimation-corrected) energies, $E_{\rm c,t}$ is the true (collimation-corrected) cutoff energy scale, $b_c$ parametrizes the cutoff steepness, $E_{B}(\tau)$ the free energy or magnetic helicity available to power bursts and $\tau$ is the age of the magnetar. In addition, the fraction $f_{\rm dip}=E_{B_{\rm dip}}(\tau)/E_{\rm B}(\tau)<1$ is the fraction of the magnetar's magnetic energy that resides in the dipole component (observations of Galactic magnetars suggest that $0.1\lesssim f_{\rm dip}\lesssim 0.3$, \citealt{Dall'Osso2012D}) and $f_{E}$ is a scaling factor between the Dipole magnetic energy and the maximum MGF energy at an age $\tau$ (which in principle could be either smaller or larger than unity). The actual magnetic energy dissipated to power the MGF likely resides in toroidal components, as the dipole field does not change much from before to after the MGF, but still the dissipated energy appears to be a fraction of the dipole field energy, $f_E\lesssim1$ (see Table \ref{tab:sources}). Indeed it is remarkable that the three Galactic magnetars that have had an observed MGF are also among the top four with the largest magnetic dipole field strengths. The index $s \sim 1.7$ is the universal index observed for lower-energy short bursts across the population of local magnetars, and across temporally-distant burst episodes in individual magnetars \citep{1996Natur.382..518C,Gogus1999,1999ApJ...519L.139W,Gogus2000,2004ApJ...607..959G,Gotz2006A&A...445..313G,2010A&A...510A..77S,2011ApJ...739...94S,2012ApJ...755....1P,2012ApJ...749..122V,2015ApJS..218...11C,2021ApJ...907L..28B}. Such a universality is analogous to magnitude distributions in earthquakes, and other driven dissipative or complex adaptive systems exhibiting self-organized criticality. A summary of the parameters used in our modeling is given in Table \ref{tab:paras}. In the derivation below, we will generally assume that $s<2$, such that the MGF energy distribution is dominated by the more energetic bursts (in the opposite limit, the results depend instead on the minimum burst energy for which Eq. \ref{eq:dndesimp} holds). As described above, the assumption $s<2$ is consistent with observational constraints (see also \S \ref{sec:discuss}).

We note that the dependence on $f_{\rm dip}, f_{E}$ enters solely through the combination $f_{E}f_{\rm dip}$ and the individual values of these two factors cannot be constrained by our analysis. Nonetheless, it is useful to introduce this division because: (i) the dipole energy reservoir can be observationally inferred in magnetars through measurements of $P,\dot{P}$ and (ii) MGF energies in known Galactic magnetars are an order unity fraction of their inferred dipole energy reservoirs. 

It is convenient to use the isotropic-equivalent energy, $E$, as it is directly\footnote{When comparing with observations, the observed energy in the X-ray/gamma-ray band could be somewhat lower than the bolometric isotropic equivalent energy release, i.e. $E_{\rm obs}=f_{\rm bol}E$ with $f_{\rm bol}<1$. For specific events one can use the observed spectral shape to estimate $f_{\rm bol}$ and convert from $E_{\rm obs}$ to $E$.} related to observed fluence.
We define the beaming fraction converting between true and isotropic bolometric energies, $f_b \equiv E_{\rm t}/E$. This corresponds to the solid angle fraction that is larger between the physical collimation of the MGF and the relativistic beaming cone, i.e. $f_b\approx  [1-\min(\cos \theta_0,\beta)]$ (for a double sided outflow). $f_b$ could, in principle, vary with the bursts' energy output. Indeed, less energetic short bursts are quasi-thermal and likely isotropic, while MGFs are expected to be more collimated (with outflows) owing to their supra-Eddington luminosities and Comptonized fireball photon pressure exceeding the local magnetic confinement pressure (in contrast to short bursts). However, we note that: (1) our analysis in this paper focuses primarily on MGFs (this is because only such energetic bursts are viewable beyond our Galaxy. Furthermore, since $s<2$, the total magnetar's energy release in the form of bursts is guaranteed to be dominated by the more energetic events) and (2) in the best studied case to date, MGF~041227, the inferred beaming is relatively modest, with $f_b\sim 0.15$ \citep{Granot+06}. For these reasons we focus below on the case in which $f_b$ is energy independent. We also assume that the orientation of individual bursts from each source is randomly drawn from an isotropic distribution.
With these definitions we have an observed energy distribution
\begin{equation}
\label{eq:dNdEiso}
\frac{\partial {\cal N}_{\rm obs}}{\partial E}= \frac{\partial {\cal N}_{\rm obs}}{\partial {\cal N}} \frac{\partial {\cal N}}{\partial E_{\rm t}} \frac{\partial E_{\rm t}}{\partial E}=f_b^2 \frac{\partial {\cal N}}{\partial E_{\rm t}} \propto f_b^{2-s} E^{-s}\exp\left[-\left(\frac{E}{ E_{\rm c}}\right)^{b_c}\right]\ ,\quad
E_{\rm c} = E_{\rm c,t}/f_b\ ,
\end{equation}
where ${\cal N}_{\rm obs}$ here only accounts for the cases in which the observer is within the beaming cone of the bursts, and (at this stage) no energy cutoff is imposed.

Eq.~(\ref{eq:dndesimp}) can be expanded to include an age $\tau$-dependent rate. For simplicity, we assume the rate and energy distribution are separable functions. This is likely a secure assumption since the self-organized critical process producing the power-law distribution is universal, while the energy injection process has the only temporal scales dimensionally accessible in the problem. Accordingly,
\begin{equation}
\left(\frac{\partial^2 {\cal N}}{\partial E_{\rm t} \partial\tau} \right) =A(\tau) E_{\rm t}^{-s}\exp\left[-\left(\frac{E_{\rm t}}{E_{\rm c,t}}\right)^{b_c}\right]\ ,\quad
E_{\rm c,t} = f_{E} f_{\rm dip} E_{B}(\tau)\ ,
\label{eq:dndev2}
\end{equation}
where $A(\tau)$ is a scaling factor that is determined by the (average) rate of magnetic energy loss that is deposited in bursts, $\dot{E}_{\rm fl,t}(\tau)$
\begin{equation}
\label{eq:Edotfl}
\int_0^\infty dE_{\rm t} E_{\rm t}\left(\frac{\partial^2 {\cal N}}{\partial E_{\rm t} \partial\tau} \right) = \dot{E}_{\rm fl,t}(\tau)\equiv f_{\rm fl} |\dot{E}_{B}(\tau)|\ ,
\end{equation}
where $f_{\rm fl}$ is the fraction of the magnetic energy decay channeled into bursts. The condition $f_{\rm fl}>f_{E}f_{\rm dip}$ is necessary to guarantee that one MGF doesn't remove more energy from the magnetar than restricted by $f_{\rm fl}$. Observational constraints imply that $0.1<f_{\rm fl}<1$ (see \S \ref{sec:ObsEnergy} for details). The isotropic equivalent observed energy release rate in the form of a bursts, $\dot{E}_{\rm fl}(\tau)$, can be directly constrained from observations. Note that this quantity is independent of $f_b$ so long as the total energy channeled into bursts remains fixed. This is because by virtue of Eq. \ref{eq:dNdEiso},
\begin{equation}
\dot{E}_{\rm fl}=\int dE E \left(\frac{\partial^2 {\cal N}_{\rm obs}}{\partial E \partial\tau} \right) =\int_0^\infty d(E_{\rm t}/f_b) (E_{\rm t}/f_b) f_b^2\left(\frac{\partial^2 {\cal N}}{\partial E_{\rm t} \partial\tau} \right)=\dot{E}_{\rm fl,t}.
\end{equation}
In other words, due to beaming, the recorded energy from each burst is increased by a factor $f_b^{-1}$ while the observed rate is decreased by a factor $f_b$. Overall, the average energy release remains the same as in the case with no beaming. This remains true also when $f_b$ is energy dependent.

Plugging Eq. \ref{eq:dndev2} into Eq. \ref{eq:Edotfl} we get
\begin{equation}
\label{eq:Atau}
A(\tau) = f_{\rm fl}\frac{b_c E_{\rm c,t}^{s-2}(\tau)}{\Gamma\left(\frac{2-s}{b_c}\right)} |\dot{E}_B(\tau)|\ .
\end{equation}
where $\Gamma(x)$ here is the Gamma function. Thus $A(\tau)$ decreases in direct proportion to $|\dot{E}_B(\tau)|$, with a weak dependence on $E_{\rm c,t}$. Note that for a sharp cutoff, i.e. $b_c\rightarrow \infty$, we have $b_c/\Gamma[(2-s)/b_c] \rightarrow 2-s $.
We assume that the available magnetic free energy in MHD equilibria (given by a component with non-zero helicity) is proportional to and indeed comparable to the curl-free component's field energy, i.e. $E_{B} \propto B^2 R^3$.
For $|\dot{B}| = B/\tau_d(B)\propto B^{1+\alpha}$ (where $\tau_d(B)\propto B^{-\alpha}$ is the field decay rate, see, e.g. \citealt{Colpi2000}) and for $\alpha\neq 0$ we have
\begin{equation}
\dot{B}(\tau) = -\frac{B(\tau)}{\tau_{d,0}+\alpha \tau}\ ,\quad  \dot{E}_B(\tau) = - \frac{2E_B(\tau)}{\tau_{d,0}+\alpha \tau}\ ,
\end{equation}
where $\tau_{d,0}$ is the initial magnetic field decay rate. Explicitly, for $\alpha \neq 0$, 
\begin{equation}
B(\tau) = B_0 (1+\alpha \tau/\tau_{d,0})^{-1/\alpha}\ ,\quad  E_B(\tau) = E_{B,0} (1+\alpha \tau/\tau_{d,0})^{-2/\alpha}\ ,
\end{equation}
where $E_{B,0}\equiv E_B(\tau=0)$. The rate of bursts with energy greater than $E$ is $\partial N(>E_t)/\partial \tau=\int_{E_t} dE_t\partial^2 {\cal N}/(\partial E_{\rm t} \partial\tau)$. It's inverse, $(\partial N(>E_t)/\partial \tau)^{-1}$, is the average waiting time between bursts of energy $>E_{t}$.
While the source of energy of bursts is assumed to be the central magnetic field, it is possible that energy builds up slowly in the crust as an intermediate step before bursts. This will then act as a bottleneck to getting temporally close energetic events from a single source. In the context of our formalism, sufficient energy supply is guaranteed by construction, and such a situation will simply correspond to an effectively larger value of $\tau_{\rm d,0}$. Moreover, since $\tau_{\rm d,0}$ is inferred from observations, our quantitative results are largely independent of such assumptions, while the physical interpretation may vary. In other words, the approach presented here can be used to constrain the underlying physical conditions for scenarios with / without a slow process of energy build-up in the crust before bursts.

\begin{figure}
\centering
\includegraphics[scale=0.15]{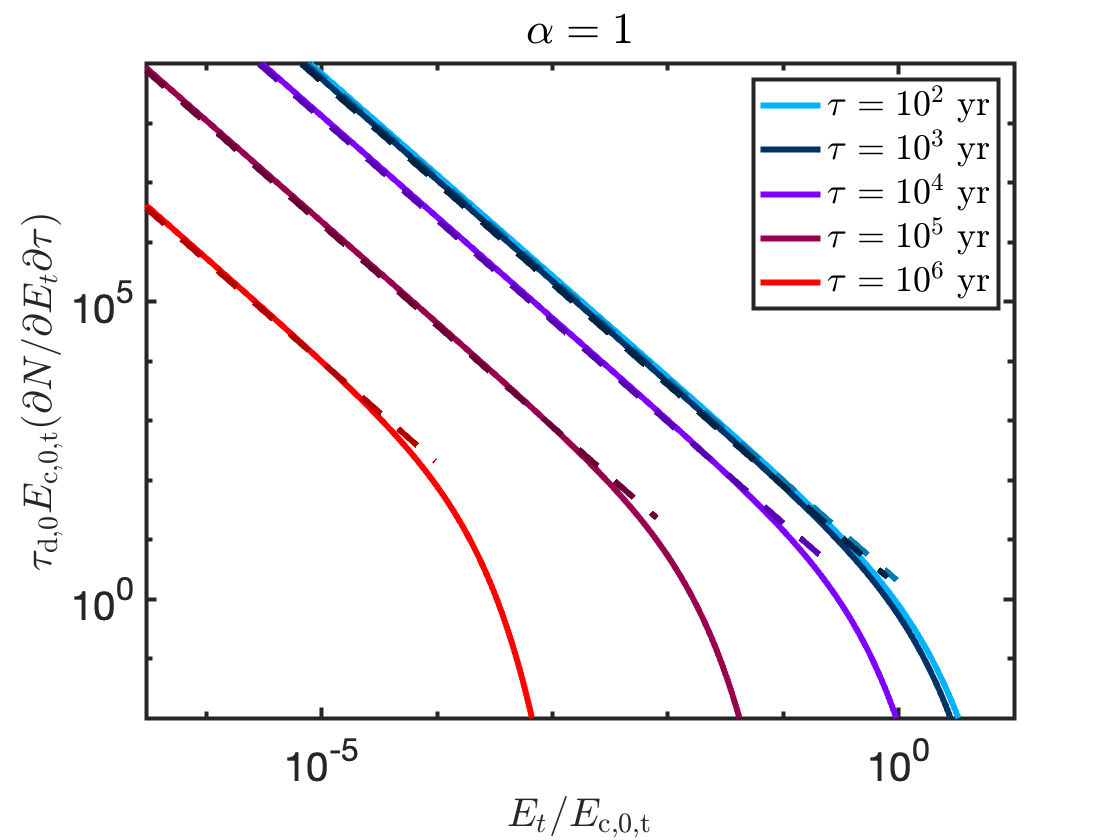}
\includegraphics[scale=0.15]{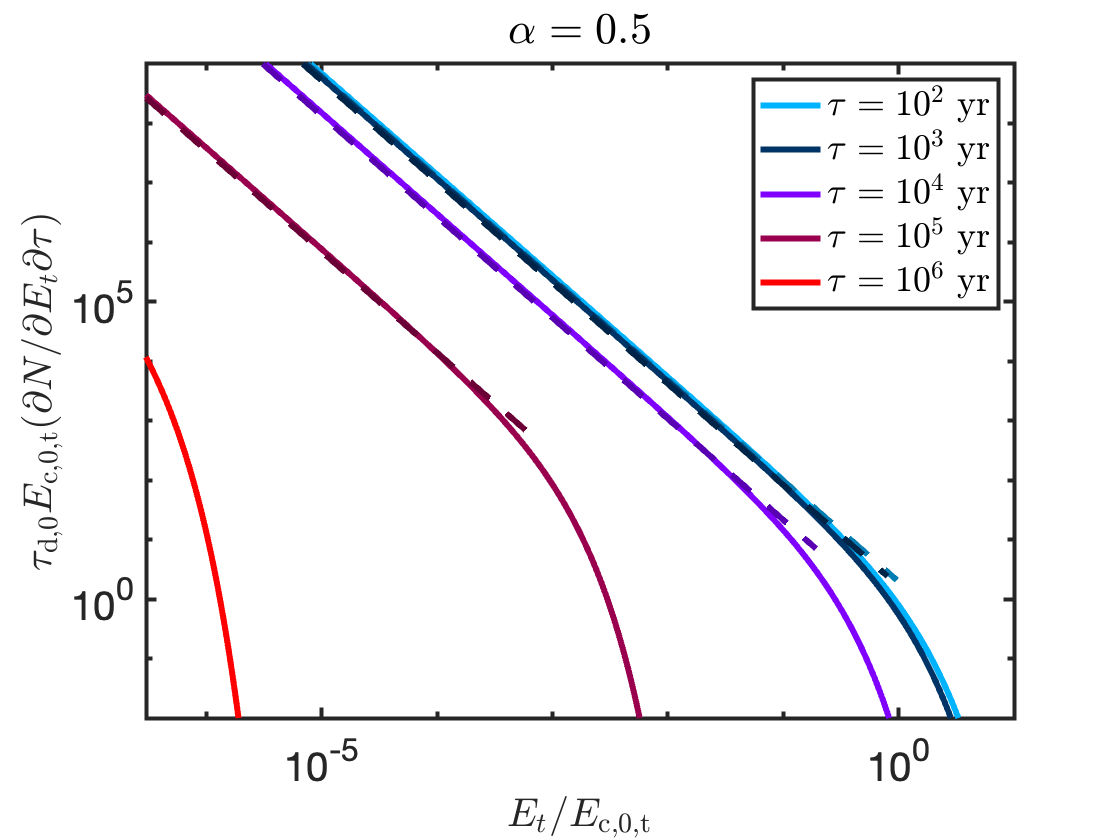}
\includegraphics[scale=0.15]{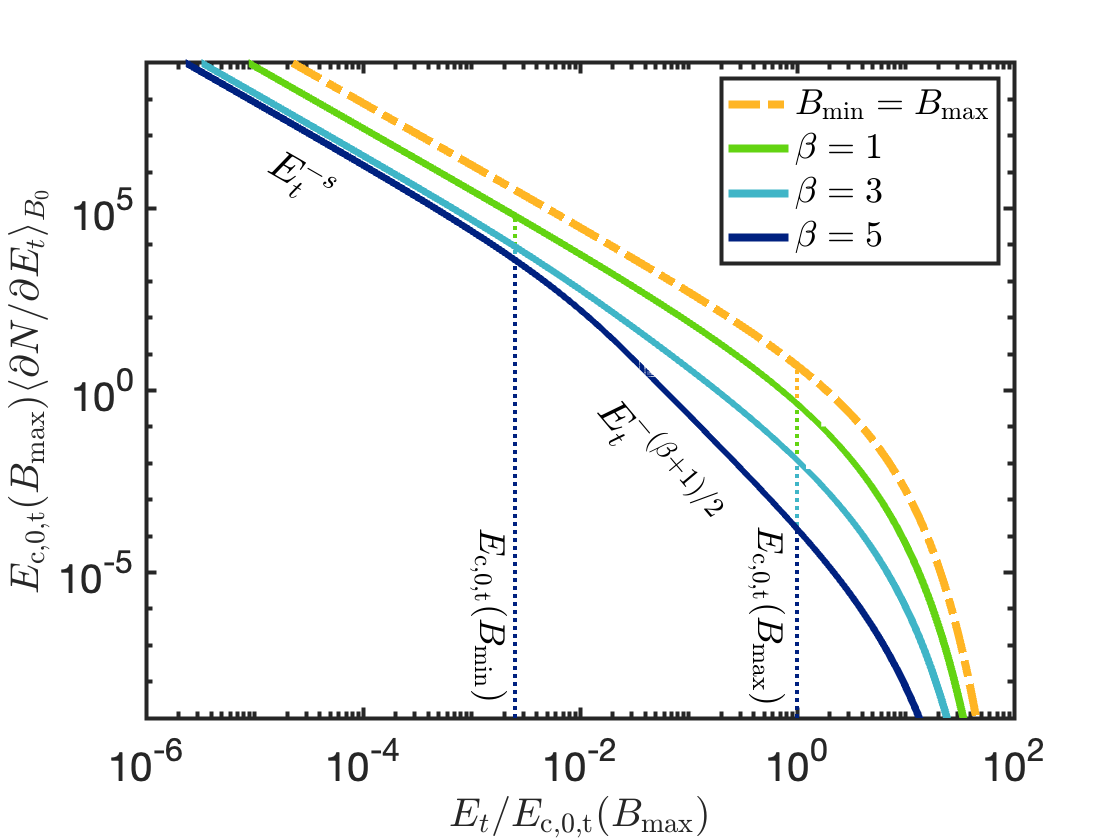}
\caption{Left and center: distribution of magnetar burst energies, $\frac{\partial^2 {\cal N}}{\partial E_{\rm t} \partial\tau}$ (solid lines; see Eq. \ref{eq:dndev2}), at different stages in a magnetar's evolution. The cut-off power-law approximation given by Eq. \ref{eq:dndedtau} is shown in dashed lines. The magnetar parameters used for the plot are $s=1.7, b_c=1, f_{\rm fl}=f_E=f_{\rm dip}=0.3, \tau_{\rm d,0}=10^4$ yr as well as two different values of $\alpha$ ($\alpha =1$ on the left and $\alpha=0.5$ on the right) and $\tau \in \{10^2,...,10^6\}$ yr in one decade increments. As is clear from this figure, the energy release in bursts is almost independent of $\alpha$ at $\tau\lesssim \tau_{\rm d,0}$. At greater values of $\tau$, $\alpha$ strongly affects the burst energy distribution. However in all cases, the instantaneous energy release at $\tau>\tau_{\rm d,0}$ is significantly reduced (compared to lower $\tau$) and even the integrated energy release from the age range $\tau \in [\tau_{\rm d,0},\infty]$ is at most comparable to that at earlier times ($\tau \in [0,\tau_{\rm d,0}]$). Thus $\alpha$ has a minimal effect on the overall energy release in bursts during a magnetar's lifetime. Right: The burst energy distribution for an individual magnetar (averaged over the distribution of birth magnetic fields). The top line presents the case of a delta function distribution of magnetar birth fields, i.e. $B_{\rm min}=B_{\rm max}$. Other lines, are for distributions with $B_{\rm min}=10^{-1.3}B_{\rm max}\ll B_{\rm max}$ and varying values of $\beta$. Once $\beta>2s-1$, a new power-law with $\langle \frac{\partial {\cal N}}{\partial E_t}\rangle_{B_0}\propto E_t^{-(\beta+1)/2}$ appears between $E_{c,0,t}(B_{\rm min})$ and $E_{c,0,t}(B_{\rm max})$. Other parameters assumed here are: $s=1.7, b_c=1, f_{\rm fl}=f_E=f_{\rm dip}=0.3, \alpha=1$.
}
\label{fig:dndev2}
\end{figure}


We consider the burst energy distribution resulting from a single magnetar, and integrate it over its lifetime.
For $E_{\rm t}<E_{\rm c,0,t}\equiv E_{\rm c,t}(\tau=0)$, the desired distribution can be formulated as 
\begin{equation}
	\label{eq:dNdEint}
	\frac{\partial {\cal N}}{\partial E_{\rm t}}=\int_0^{\tau_E} d\tau \frac{\partial^2 {\cal N}}{\partial E_{\rm t} \partial\tau} 
\end{equation}
where $\partial^2 {\cal N}/(\partial E_{\rm t} \partial\tau)$ is given by Eq. \ref{eq:dndev2} and
\begin{equation}
	\label{eq:tauE}
	\tau_{E}=\frac{\tau_{\rm d,0}}{\alpha}\left( \left(\frac{E_{\rm t}}{E_{\rm c,0,t}}\right)^{-\alpha/2}-1\right)
\end{equation}
is the time it takes for the magnetar's field to sufficiently decay as to reach $E_{\rm t}=E_{\rm c,t}(\tau)\equiv f_{E}f_{\rm dip} E_{\rm B}(\tau)$ (after this time the magnetar effectively stops forming bursts with energy $E_{\rm t}$).

As long as $E_{\rm t}\ll E_{\rm c,t}(\tau)$, Eq. \ref{eq:dndev2} can be approximated as
\begin{eqnarray}
&& 	\frac{\partial^2 {\cal N}}{\partial E_{\rm t} \partial\tau} \approx f_{\rm fl} \dot{E}_{\rm B,0} E_{\rm c,0,t}^{s-2} (2-s)E_{\rm t}^{-s}\left(1+\frac{\alpha \tau}{\tau_{\rm d,0}}\right)^{\frac{2-\alpha-2s}{\alpha}} \Theta(f_E f_{\rm dip}E_{B}(\tau)-E_{\rm t}) \nonumber \\
&& =\frac{2 f_{\rm fl}}{\tau_{\rm d,0}} \left(f_E f_{\rm dip}\right)^{s-2} E_{\rm B,0}^{s-1} (2-s)E_{\rm t}^{-s}\left(1+\frac{\alpha \tau}{\tau_{\rm d,0}}\right)^{\frac{2-\alpha-2s}{\alpha}}\Theta(f_E f_{\rm dip}E_{B}(\tau)-E_{\rm t})
	\label{eq:dndedtau}
\end{eqnarray}
where $\Theta(x)$ is the Heaviside function.
Plugging this into Eq. \ref{eq:dNdEint} we see that $\frac{\partial {\cal N}}{\partial E_{\rm t}}\propto (1+\alpha \tau/\tau_{\rm d,0})^{\frac{2-2s}{\alpha}}$. For $s>1, \alpha>0$\footnote{If $\alpha<0$ the field formally decays to 0 at a finite time, $-\tau_{\rm d,0}/\alpha$. Even in that case, the contribution at times after which the field has declined substantially (say to $E_{\rm B,0}/2$) is negligible compared to earlier times, as long as $s>1$.}, this PL is negative, meaning that the contribution to Eq. \ref{eq:dNdEint} at $\tau\gg \tau_{\rm d,0}$ is suppressed. This is consistent with results from simulations of magnetar field evolution and stresses building in the crust \citep[e.g.,][]{2011ApJ...727L..51P,2023ApJ...947L..16L}, suggesting that young magnetars (with $\tau\lesssim 10^3\mbox{ yr}$) burst significantly more often than older ones.
The result is that the effective lifetime of MGF sources is given by the initial poloidal field decay time and Eq. \ref{eq:dNdEint} can be integrated up to $\tau=\infty$ with no loss of generality, yielding
\begin{eqnarray}
	&& 		\frac{\partial {\cal N}}{\partial E_{\rm t}}=f_{\rm fl} \frac{2-s}{s-1} \left(f_E f_{\rm dip}\right)^{s-2} E_{\rm B,0}^{s-1} E_{\rm t}^{-s} ,\quad
\mbox{ for }E_{\rm t}<E_{\rm c,0,t} .
	\label{eq:dndev3}
\end{eqnarray}
In Fig.~\ref{fig:dndev2} (left and center) we plot the distribution of burst energies per time as defined by Eq.~\ref{eq:dndev2} and compare with the cut-off PL approximation given by Eq. \ref{eq:dndedtau}. As anticipated in Eq. \ref{eq:dndev3} the overall energy release in bursts during a magnetar's lifetime has almost no dependence on $\alpha, \tau_{\rm d,0}$ (these parameters are only important for determining the rate of bursts from a given magnetar). 

Both the maximum burst energy $E_{\rm c,0}=E_{\rm c,0,t}/f_b$ and the number of observable bursts with that energy ${\cal N}_{\rm obs}(E_{\rm c,0})=f_b{\cal N}(E_{\rm c,0,t})\sim f_b E_{\rm c,0,t} \frac{\partial {\cal N}}{\partial E_{\rm t}}|_{\rm E_{\rm c,0,t}}$ can, under circumstances discussed below, be constrained by observations. The latter can be approximated by (ignoring factors of order unity)
\begin{equation}
	{\cal N}_{\rm obs}(E_{\rm c,0})\sim \frac{f_{\rm fl}f_b}{f_E f_{\rm dip}}
 \label{eq:dndev4}
\end{equation}
Therefore, the combinations of two quantities $E_{\rm c,0}\sim (f_{E}f_{\rm dip}/f_b) E_{\rm B,0}$ and ${\cal N}_{\rm obs}(E_{\rm c,0})  = f_{\rm fl}f_b/(f_E f_{\rm dip})$ can both be deduced directly from high-energy observations of a given magnetar. Note that $E_{\rm c,0} {\cal N}_{\rm obs}(E_{\rm c,0}) = f_{\rm fl} E_{\rm B,0}$.

When considering a given magnetar, it is easiest to observe the number of MGFs per object per unit time. An analytic approximation for this is given by Eq. \ref{eq:dndedtau}. Evaluating this at $E_{\rm c,0}$, we can get a simple approximate expression, $\dot{\cal N}_{\rm obs}(E_{\rm c,0})\sim \frac{f_{\rm fl}f_b}{f_E f_{\rm dip}\tau_{\rm d,0}}$. At first glance, this appears to depend explicitly on $\tau_{\rm d,0}$. However, one should consider that the total number of Galactic magnetars $N_{\rm mag}^{\rm MW}\approx 30$ is approximately $N_{\rm mag}^{\rm MW}\sim \tau_{\rm d,0} \dot{N}_{\rm mag}^{\rm MW}\sim \tau_{\rm d,0} f_{\rm mag} \dot{N}_{\rm CCSN}^{\rm MW}$ where we have defined $f_{\rm mag}$ as the ratio between the magnetar birth and CCSN rate, which is convenient as the rate of CCSN in the Galaxy, $\dot{N}_{\rm CCSN}^{\rm MW}\approx 17\mbox{ kyr}^{-1}$ is well constrained by different observational inferences \citep{Beniamini2019}. All-together, we see that $\tau_{\rm d,0}\approx 1800\mbox{yr}/f_{\rm mag}$ and therefore that
\begin{equation}
\label{eq:Nobsdot}
\dot{\cal N}_{\rm obs}(E_{\rm c,0})\sim \frac{f_{\rm fl}f_bf_{\rm mag}}{1800f_E f_{\rm dip}} \sim 10^{-4} \left(\frac{f_{\rm fl}f_{\rm mag}}{0.06}\right) \left(\frac{0.3 f_b}{f_E f_{\rm dip}}\right)\mbox{ yr}^{-1}\mbox{ magnetar}^{-1}.
\end{equation}
Where the motivation for this split of terms and the normalization adopted for each will be discussed in detail in \S \ref{sec:combanalyticresult}.

\subsubsection{Distribution of birth magnetic fields}
\label{sec:changeB0}
A natural extension of the burst energy distribution calculation is to allow for a situation in which the birth magnetic field varies between magnetars. Lacking concrete knowledge about the distribution of $B_0$, an illustrative case is to consider a PL probability distribution of $B_0$,
\begin{equation}
    \frac{d\mbox{Pr}}{dB_0}=C_B \left(\frac{B_0}{B_{\rm max}}\right)^{-\beta} \mbox{ for } B_{\rm min}<B_0<B_{\rm max}
\end{equation}
where $C_B$ is taken such that the probability is normalized $\int dB_0(d\mbox{Pr}/dB_0)\!=\!1$. We calculate the average burst energy distribution per magnetar, $\langle \frac{\partial {\cal N}}{\partial E_t}\rangle_{B_0}$ where the average is according to the probability distribution $dP/dB_0$ and over $B_0$. Using Eq. \ref{eq:dndev3}, 
\begin{eqnarray}
\label{eq:B0variationEdist}
 &   \langle \frac{\partial {\cal N}}{\partial E_t}\rangle_{B_0}=\int_{B_{\rm min}}^{B_{\rm max}} dB_0 \frac{dP}{dB_0}  \left. \frac{\partial {\cal N}}{\partial E_t}\right|_{B_0}\approx f_{\rm fl} \frac{2-s}{s-1} \left(f_E f_{\rm dip}\right)^{s-2} E_{\rm B,0,max}^{s-1}E_{\rm t}^{-s}C_B \int_{\max(B_{\rm cr},B_{\rm min})}^{B_{\rm max}}  \left(\frac{B_0}{B_{\rm max}}\right)^{2s-2-\beta} dB_0 \\ & \approx  f_{\rm fl} \frac{2-s}{s-1} \left(f_E f_{\rm dip}\right)^{s-2} E_{\rm B,0,max}^{s-1}\frac{(1-\beta)\Theta(E_{c,0,t}(B_{\rm max})-E_t)}{(2s-1-\beta)[1-(B_{\rm max}/B_{\rm min})^{1-\beta}]} E_{\rm t}^{-s}\left\{ \begin{array}{ll}  \frac{(B_{\rm min}/B_{\rm max})^{2s-1-\beta}}{\beta+1-2s} & \mbox{ for }\beta>2s-1, E_t<E_{c,0,t}(B_{\rm min}) \\ (6E_t/(f_Ef_{\rm dip}B_{\rm max}^2R^3))^{\frac{2s-1-\beta}{2}} & \mbox{ for }\beta>2s-1, E_t>E_{c,0,t}(B_{\rm min})\\
     \frac{1}{2s-1-\beta} & \mbox{ for } \beta<2s-1
\end{array} \right.   \nonumber
\end{eqnarray}
%
where $E_{\rm B,0,max}$ is the initial magnetic energy of a magnetar with a birth magnetic field $B_{\rm max}$ and $B_{\rm cr}=[6E_t/(f_Ef_{\rm dip}R^3)]^{1/2}$ is the magnetic field of a magnetar for which the initial MGF energy is $E_t$ (i.e. $f_Ef_{\rm dip} E_{\rm B,cr}=E_t$). Eq. \ref{eq:B0variationEdist} shows that as long as the birth magnetic field distribution is not very steeply declining ($\beta<2s-1\approx 2.4$) then the average burst energy distribution is (up to a normalization constant) the same as for a single birth field (i.e. $\langle \frac{\partial {\cal N}}{\partial E_t}\rangle_{B_0}\propto E_t^{-s}$ as in Eq. \ref{eq:dndev3}) with strength $B_{\rm max}$. For steeper birth field distributions, as possibly realized in the tail of a Gaussian, the burst energy distribution is dominated by magnetars with $B_0=B_{\rm min}$ up to a MGF energy of $E_{c,0,t}(B_{\rm min})$. Above this energy, a new (softer) power-law develops, which represents contributions from (increasingly rare) magnetars with gradually larger birth magnetic field values ($\langle \frac{\partial {\cal N}}{\partial E_t}\rangle_{B_0}\propto E_t^{-(\beta+1)/2}$).
We plot $\langle \frac{\partial {\cal N}}{\partial E_t}\rangle_{B_0}$ for different values of $\beta$ in the right panel of Fig. \ref{fig:dndev2}.

\subsection{Burst Distributions from an Extragalactic Magnetar Population}
\label{sec:MGFpop}

With some assumptions (detailed below), the fluence distribution from a population of magnetars can be calculated by integrating the rate of bursts per unit energy from a single magnetar, discussed in \S \ref{sec:IndMag}, over the volume from which bursts can be detected. For clarity, we focus on the analytic expressions below primarily on a Euclidean geometry\footnote{Cosmological distances/volume as well as redshift corrections for the rates, energetics, etc. are accounted for self-consistently in the results shown in the figures - see \S \ref{sec:stoch} for an explicit example of these corrections in the context of the gravitational wave background.}. The validity of this approximation can be easily checked by comparison to the maximal distance from which bursts of a given energy are detectable. Assuming the condition for burst detectability is dominated by their fluence, it is useful to define $E_{\rm \phi,t}(r)\equiv 4\pi f_b r^2 \phi$, the corresponding burst energy for a detection fluence $\phi$.
We define also the volumetric density of magnetars (averaged over volumes larger than $[30 \rm \, Mpc]^3$), $n_{\rm mag}$, which is directly related to the volumetric magnetar formation rate $\mathcal{R}_{\rm mag}\approx n_{\rm mag}/\tau_{\rm d,0}$ (with a constant of proportionality depending slightly on $\alpha$). As above we parameterize the magnetar formation rate, as a fraction $f_{\rm mag}$ of the CCSN rate, $\mathcal{R}_{\rm mag}\equiv f_{\rm mag}\mathcal{R}_{\rm CCSN}$.
For simplicity, we initially assume also that all magnetars have identical properties (we readdress this point below) but were born at different times. As long as $\mathcal{R}_{\rm mag}$ evolves on timescales that are long compared to $\tau_{\rm d,0}$ the age distribution of magnetars at a given distance $r$ is uniform.
Under these assumptions, the number of detected bursts per unit time above a fluence $\phi$ is
\begin{equation}
\label{eq:Ndotphi0}
    \dot{N}_{\rm pop}(>\phi)=\int_0^{\infty} \frac{d\tau}{\tau_{\rm d,0}} \int_0^{r_{\rm max}} dr \, 4\pi r^2 n_{\rm mag} f_b \frac{\partial {\cal N} (E_{\rm t}>E_{\rm \phi,t}(r))}{\partial \tau}\approx \frac{4\pi f_{\rm mag}\mathcal{R}_{\rm CCSN}(r=0) }{3} f_b\int_0^{r_{\rm max}} \Delta(r) d(r^3) {\cal N}(E_{\rm t}>E_{\rm \phi,t}(r))
\end{equation}
where $r_{\rm max}(\phi)=[E_{\rm c,0}/(4\pi\phi)]^{1/2}=[E_{\rm c,0,t}/(4\pi\phi f_b)]^{1/2}$ is the maximum distance from which the highest energy magnetar bursts (with isotropic equivalent energy $E_{\rm c,0}=E_{\rm c,0,t}/f_b$) may be detected, considering the limiting fluence. As explained in \S \ref{sec:IndMag}, it is useful to calculate the contribution of each magnetar at $\tau=0$ since the maximum MGF energy is highest then and it evolves only by an order unity factor by $\tau_{\rm d,0}$ ($\tau\lesssim \tau_{\rm d,0}$ dominates the burst contribution per magnetar, see Fig. \ref{fig:dndev2}). The term $\Delta(r)\equiv n_{\rm mag}(r)/n_{\rm mag}(r\approx \mbox{30Mpc})$ is useful to define in case there is a deviation from a uniform density within the region extending up to $r_{\rm max}$. This is useful when considering typical distances of order $\lesssim 30$\,Mpc, for which local environment of our galaxy cannot be neglected, as shown in Fig.\,\ref{fig:overdensity} and discussed in \S \ref{sec:overdensity}. For such distances, the SFR and stellar density distributions at a given $r$ deviate from the volume averaged densities up to $r$. 
Using Eqns. \ref{eq:dndev3} and \ref{eq:Ndotphi0}, we get an approximation for $ \dot{N}_{\rm pop}(>\phi)$,
\begin{eqnarray}
\label{eq:Ndotphi}
     \dot{N}_{\rm pop}(>\phi)=\frac{(2-s)f_{\rm mag}\mathcal{R}_{\rm CCSN}(r=0)f_{\rm fl}(f_E f_{\rm dip})^{1/2} E_{B,0}^{3/2}}{(s-1)^2 (5-2s)(4\pi f_b)^{1/2}}\phi^{-3/2}\int_0^{y_{\rm max}}dy(5-2s) y^{5-2s}\Delta(y)
\end{eqnarray}
where $y \equiv r/r_{\rm max}$ and for the isotropic case ($\Delta(y)=1$), the dimensionless $y$ integral is unity.
Eq. \ref{eq:Ndotphi} reproduces the well known `log N - log S' relation $\dot{N}_{\rm pop}\propto \phi^{-3/2}$ and provides the normalization in terms of the magnetar volumetric density and the MGF properties. Writing $\dot{N}_{\rm pop}$ in terms of $\{ E_{\rm c,0,t}\,, r_{\rm max} \}$, we see that Eq. \ref{eq:Ndotphi} results in $\dot{N}_{\rm pop}(>\phi)\propto f_b$. This is in accordance with the results for individual sources, as shown in Eq. \ref{eq:dndev4}.

In practice, if $r_{\rm max}(\phi)$ gets too large, then it is no longer possible to reliably identify the host galaxy and therefore to infer the associated distance and energy scales. In such a situation, while the MGF may be detected, it may not be identified as such. Indeed, a significant, $\sim 5-20\%$ fraction of the sGRB population may consist of such MGF interlopers, which are difficult to realize as such (see \S \ref{sec:MGFsGRB} and references therein). We therefore define an additional limiting radius, $r_{\rm cc}$, which is the distance above which the probability of chance coincidence (i.e. false alarm) of a true physical association with the nearest projected host on the sky is too large for a confident association. For a given limiting fluence, $\phi$, the limiting distance for detection and confirmation as a MGF is $r_{\rm lim}=\min(r_{\rm max}(\phi),r_{\rm cc})$\footnote{In general $r_{\rm lim}$ is instrument response dependent and also determined by the intrinsic spectral properties of MGFs.}.  We can therefore define a critical fluence, $\phi_{\rm c,0}=E_{\rm c,0}/(4\pi r_{\rm cc}^2)$ below which the $\dot{N}_{\rm pop}(>\phi)$ distribution becomes {\it volume} limited (i.e. deficit of active objects) rather than {\it energy} limited (i.e. common events not bright enough). Overall, we have
\begin{equation}
 \dot{N}_{\rm pop}(>\phi)\propto \!\left\{ \begin{array}{ll}  \phi^{1-s} & \phi<\phi_{\rm c,0}\mbox{ (volume limited)}\ ,\\ \phi^{-3/2} & \phi>\phi_{\rm c,0}\mbox{ (energy limited)}\ ,
\end{array} \right.   
\end{equation}
At $\phi>\phi_{\rm c,0}$ the observed distribution is dominated by bursts with $E\sim E_{\rm c,0}$, while at $\phi<\phi_{\rm c,0}$ the distribution becomes dominated by gradually decreasing burst energies.

The same information can be equivalently conveyed in terms of $d\dot{N}_{\rm pop}/dE$, the differential number distribution of bursts per isotropic-equivalent energy interval (for a particular instrument with limiting fluence $\phi_{\rm lim}$). Carrying out the integral in Eq. \ref{eq:Ndotphi} for a fixed $E$ and defining $E_{\rm lim}\equiv 4\pi r_{\rm cc}^2 \phi_{\rm lim}$, we get (assuming homogeneity,  $\Delta(y)=1$ for clarity)
\begin{eqnarray}
\label{eq:dNMGFpopdE}
     \frac{d\dot{N}_{\rm pop}}{dE}=\frac{4\pi (2\!-\!s)f_b^{2-s}f_{\rm mag}\mathcal{R}_{\rm CCSN}f_{\rm fl}(f_E f_{\rm dip})^{s-2} E_{B,0}^{s-1}}{3 (s-1)}\!\left\{ \begin{array}{ll}  (4\pi \phi_{\rm lim})^{-3/2} E^{3/2-s} & E<E_{\rm lim}\ \textrm{ (sensitivity limited)}
     \\ 
     r_{\rm cc}^3 E^{-s} & E>E_{\rm lim}\ \textrm{ (volume limited)}
\end{array} \right.   
\end{eqnarray}
We see that below the critical energy, $E_{\rm lim}$ the observed energy distribution, $d\dot{N}_{\rm pop}/dE\propto E^{1.5-s}$, is flatter than the original distribution, owing to the fact that larger energy bursts can be seen to greater distances. This is the sensitivity limited regime. In the other limit, $E>E_{\rm lim}$, all bursts are energetic enough to be seen up to the maximum distance $r_{\rm cc}$ and the intrinsic distribution $\frac{\partial^2 {\cal N}}{\partial E_{\rm t} \partial\tau}\propto E_{\rm t}^{-s}$ is reproduced in the observed one. This is the volume limited regime (notice that the ordering of the two regimes flips when presented as a function of energy instead of fluence). As above, when writing Eq. \ref{eq:dNMGFpopdE} in terms of collimation corrected energies, we find $\left(E_t \frac{d\dot{N}_{\rm pop}}{dE_t}\right)\propto f_b$ in the volume limited regime, which is consistent with Eqns. \ref{eq:dndev4}, \ref{eq:Ndotphi}.

Consider a survey with both a limiting fluence $\phi_{\rm lim}$ and a limiting radius, $r_{\rm cc}$. Integrating Eq. \ref{eq:dNMGFpopdE} over $E$ we get the total (all-sky) rate of MGFs detected by such a survey, $\dot{N}_{\rm pop}= \int dE\frac{d\dot{N}_{\rm pop}}{dE}$. In particular, we see that if $E_{\rm lim}<E_{\rm c,0}$, then most of the contribution to $\dot{N}_{\rm pop}$ comes from $E\approx E_{\rm lim}$, whereas if $E_{\rm lim}>E_{\rm c,0}$ then, while $s<2$, most of the contribution comes from $E\approx E_{\rm c,0}$.
Put together, we have
\begin{eqnarray}
    \dot{N}_{\rm pop}\approx \frac{2\pi (2\!-\!s)\mathcal{R}_{\rm CCSN}f_{\rm mag}f_{\rm fl}}{(5/2-s)}\!\left\{ \begin{array}{ll}  \left(4\pi \phi_{\rm lim}\right)^{1-s} r_{\rm cc}^{5-2s}E_{B,0}^{s-1}\left(f_E f_{\rm dip}/f_b\right)^{s-2} & E_{\rm lim}<E_{\rm c,0}
     \\ 
     \frac{2}{3(s-1)}\left(4\pi \phi_{\rm lim}\right)^{-3/2}E_{B,0}^{3/2}\left(f_E f_{\rm dip}/f_b\right)^{1/2} & E_{\rm lim}>E_{\rm c,0}
\end{array} \right.   
\label{eq:Ndotpop}
\end{eqnarray}

Finally, if the distance $r$ can be measured (e.g., by host localizations) for many MGFs, it is possible to compare the observed data to the number of bursts per unit logarithmic energy (or fluence) and per logarithmic unit of distance
\begin{equation}
\label{eq:dNdEMGFpop}
    \frac{\partial^2\dot{N}_{\rm pop}}{\partial\log E \partial\log r}=4\pi r^3 f_{\rm mag}f_{\rm fl}\mathcal{R}_{\rm CCSN}\frac{2-s}{s-1}(f_E f_{\rm dip}/f_b)^{s-2} E_{\rm B,0}^{s-1}E^{1-s} \Theta(f_E f_{\rm dip}E_{B,0}/f_b-E) .
\end{equation}
Since $\phi$ is linearly proportional to $E$, we have $\frac{\partial^2\dot{N}_{\rm pop}}{\partial\log \phi \partial\log r}=\frac{\partial^2\dot{N}_{\rm pop}}{\partial\log E \partial\log r}$.

We extend the results discussed in this sub-section to a population with varying birth fields. This is done by simply replacing $\frac{\partial {\cal N}}{\partial E_t}$ with $\langle \frac{\partial {\cal N}}{\partial E_t}\rangle_{B_0}$ (see \S \ref{sec:changeB0}) in Eqns. \ref{eq:Ndotphi}, \ref{eq:dNMGFpopdE}, \ref{eq:Ndotpop}.

\subsection{The sGRB Population}
\label{sec:sGRB}

Classical sGRBs may be confused with MGFs, as discussed in \S\ref{sec:modelindconst}. sGRBs involve relativistic jets, and relativistic beaming implies that gamma-rays are only observable from material moving at small angles relative to the line of sight. While the angular structure and collimation of sGRB jets is still not fully understood \citep{Duffell+18,Matsumoto-Masada-19,Lazzati-Perna-19,Hamidani-Ioka-21,Gottlieb+21,Beniamini2022,Salafia2022}, the GW-triggered GRB 170817A provides very useful constraints. Despite the event being much nearer than other localized sGRBs, it was orders of magnitude fainter than those bursts.
The coincident trigger by GWs made it possible to associate GRB 170817A to GW170817 despite this faintness, which was ultimately shown to be the result of our misaligned line of sight towards the GRB jet.

Cosmological sGRBs are likely viewed mostly from angles close to or within their jet cores \citep{BN2019,Gill+20,O'Connor2024}.  Therefore the inferred gamma-ray energy distribution from this population, likely probes the (isotropic equivalent) core energy (denoted here as $E_{\rm c}$) distribution. This distribution is typically modeled as a broken power-law above $E_{\rm min}$ and with a break energy $E_*$
\begin{eqnarray}
	\label{eq:PEc}
	\frac{d\mbox{Pr}}{d\log E_{\rm c}}\propto \!\left\{ \begin{array}{ll}  \left(\frac{E_{\rm c}}{E_{\rm *}}\right)^{-\alpha_E} & E_{\rm min}<E_{\rm c}<E_*\ ,\\  \left(\frac{E_{\rm c}}{E_*}\right)^{-\beta_E} & E_{\rm c}>E_*\ .
	\end{array} \right.   
\end{eqnarray}
\cite{WP2015} find $\alpha_E=0.95, \beta_E=2, E_*\approx 6\times 10^{51}\mbox{ erg}, E_{\rm min}\approx 1.5\times 10^{49}\mbox{ erg}$\footnote{\cite{WP2015} derive their results for the sGRB (isotropic equivalent) luminosity function. Here we have converted from $L_{\gamma}$ to $E_{\gamma}$ using a typical sGRB duration of $T\approx 0.3$\,s.}. 

The angular profile of (isotropic equivalent) kinetic energy is typically modeled by a broken PL approximation\footnote{Some authors favour a Gaussian angular energy profile \citep[e.g.,][]{Rossi02,Saleem2020,Cunningham2020}. Observationally, the situation is not strongly constrained as there is only a single sGRB (GRB 170817A), which is undoubtedly observed off-axis. For this reason, and for the sake of clarity of the analytic results, we focus on PL models in the following but note that an extension to Gaussian jets is straight-forward.}, such that
\begin{equation}
	\label{eq:Ektheta}
	E_{\rm k}(\theta)=E_{\rm k,c}\left[\Theta(\theta_c-\theta)+ \Theta(\theta-\theta_c) \Theta(\theta_{\rm max}-\theta) \left(\frac{\theta}{\theta_c}\right)^{-a}\right]
\end{equation}
where $\theta_{\rm c}$ is the core jet angle, $a$ is the PL index of the energy decline above the core, $\theta_{\rm max}$ is the maximum angle to which this jet extends and $E_{\rm k,c}$ is the isotropic equivalent kinetic energy of the jet's core. For the sake of simplicity and concreteness (and since this option is consistent with currently available data), we assume in what follows that the structure of sGRBs is universal in the sense that only the core energy changes, but all other parameters remain roughly constant between jets. Taking $\theta_c=0.087\mbox{ rad}, a=4.5, E_{\rm k,c}=10^{53}\mbox{ erg}$ (all angles here and elsewhere in the text are in radians) as well as an observer viewing angle of $\theta_{\rm obs}=0.47\mbox{ rad}$, the structure represented by Eq. \ref{eq:Ektheta} matches well the observed afterglow data from GRB 170817A (see \citealt{BGG2020} and references therein). This is related to the isotropic equivalent gamma-ray energy, $E(\theta)$,via the efficiency, $\epsilon(\theta)$, of converting the initial energy reservoir to gamma-rays, $	E(\theta)=\frac{\epsilon(\theta)}{1-\epsilon(\theta)} E_{\rm k}(\theta)\approx \epsilon(\theta) E_{k}(\theta)$ (where we have assumed $\epsilon(\theta)\ll1$)\footnote{This relation between $E(\theta)$ and $E_{\rm k}(\theta)$ implicitly assumes that the observed energy at any given line-of-sight is dominated by the material moving approximately in the same direction (rather than by more energetic material that is closer to the core but, the emission of which is deboosted relative to the observer). As shown in, e.g., \cite{BN2019}, for values of $\tilde{a}$ discussed below, this is typically a good approximation for practically all observation angles.}. 
$\epsilon(\theta)$ is not well constrained by observations. However, evidence from cosmological sGRBs, suggests that at the core $\epsilon(\theta<\theta_c)\approx 0.15$ \citep{Beniamini2015}, while the prompt emission of GRB 170817A requires $\epsilon(\theta_{\rm obs})\approx 6\times 10^{-3}$ (see \citealt{BPBG2019}). We therefore model $\epsilon(\theta)$ as a broken PL, with a functional form similar to that described in Eq. \ref{eq:Ektheta},
\begin{equation}
	\label{eq:epsilontheta}
	\epsilon(\theta)=\epsilon_{\rm c}\left[\Theta(\theta_c-\theta)+ \Theta(\theta-\theta_c)\Theta(\theta_{\rm max}-\theta)  \left(\frac{\theta}{\theta_c}\right)^{-\delta}\right]
\end{equation}
with $\epsilon_{\rm c}=0.15$ and $\delta=1.9$. Under these approximations $E(\theta)$ will also have the same form,
\begin{eqnarray}
	E(\theta)\approx E_{\rm c} \left[\Theta(\theta_c-\theta)+ \Theta(\theta-\theta_c) \Theta(\theta_{\rm max}-\theta) \left(\frac{\theta}{\theta_c}\right)^{-\tilde{a}}\right]
\end{eqnarray}
with modified values of the core energy and the PL index of the energy profile, $E_{\rm c}\approx \epsilon_{\rm c}E_{\rm k,c}$, $\tilde{a}=a+\delta$.

For a given  isotropic equivalent core gamma-ray energy, $E_{\rm c}$ and for a given viewing angle, $\theta$, the differential probability of a GRB having an energy $E$ is (see also \citealt{Guetta2005}),
\begin{equation}
	\label{eq:dPdEatEctheta}
	\left.\frac{d\mbox{Pr}}{dE}\right\vert_{\theta, E_{\rm c}}=\Theta(\theta_c-\theta)\delta(E-E_c)+\Theta(\theta-\theta_{\rm max})\delta(E) +\Theta(\theta-\theta_c)\Theta(\theta_{\rm max}-\theta)\delta\left(E-E_{\rm c} (\theta/\theta_c)^{-\tilde{a}}\right)
\end{equation}
where $\delta(x)$ is the Dirac function.
The probability of a event with unknown values of $\theta, E_{\rm c}$ having an energy $E$ along the line of sight is obtained by averaging Eq. \ref{eq:dPdEatEctheta} over $\theta$ (assumed to be distributed isotropically relative to the jet axis, i.e. $d\mbox{Pr}/d\theta=\sin \theta$) and over the distribution of $E_c$ (given by Eq. \ref{eq:PEc})
\begin{eqnarray}
&	\frac{d\mbox{Pr}}{dE}=\int \sin \theta d\theta \int  \frac{d\mbox{Pr}}{dE_{\rm c}}(E_{\rm c}) dE_{\rm c}  \left.\frac{d\mbox{Pr}}{dE}\right\vert_{\theta, E_{\rm c}}(\theta,E_{\rm c})\\ & \approx  \frac{\theta_c^2}{\tilde{a}}E^{-1-2/\tilde{a}}\left[E_*^{\alpha_E+1} \frac{E_2^{2/\tilde{a}-\alpha_E}-E_1^{2/\tilde{a}-\alpha_E}}{2/\tilde{a}-\alpha_E}\Theta(E_2-E_1)+E_*^{\beta_E+1} \frac{E_4^{2/\tilde{a}-\alpha_E}-E_3^{2/\tilde{a}-\alpha_E}}{2/\tilde{a}-\alpha_E}\Theta(E_4-E_3)\right]+\nonumber \\&\frac{\theta_c^2}{2}C\left[ \left(\frac{\theta}{\theta_*}\right)^{-\alpha_E-1}\Theta(E_*-E)\Theta(E-E_{\rm min}) + \left(\frac{\theta}{\theta_*}\right)^{-\beta_E-1}\Theta(E-E_*)\right]+\delta(E)\cos \theta_{\rm max} \nonumber
\end{eqnarray}
where $E_1\!\equiv \!\max(E_{\rm min},E)$, $E_2\!\equiv\! \min(E_{*},E (\theta_{\rm max}/\theta_c)^{\tilde{a}})$, $E_3\!\equiv \!\max(E,E_*)$, $E_4\!\equiv\! E (\theta_{\rm max}/\theta_c)^{\tilde{a}}$ and $C=\left[ (\beta_E^{-1}-\alpha_E^{-1}+\alpha_E^{-1} (E_{\rm min}/E_*)^{-\alpha_E})E_*\right]^{-1}$.
In particular, for $2/\tilde{a}<\alpha_E$ (which is the case for our canonical parameter values), we see that the angular structure affects only the low energy tail of the energy distribution \citep{BPBG2019}, i.e. 
\begin{equation}
	\frac{d{\rm Pr}}{d \log E}\propto  \!\left\{ \begin{array}{ll}  E^{-2/\tilde{a}} & E_{\rm min}(\theta_{\rm max}/\theta_c)^{-\tilde{a}}<E<E_{\rm min}\ ,\\ E^{-\alpha_E} & E_{\rm min}<E<E_*\ ,\\  E^{-\beta_E} & E>E_*\ .
	\end{array} \right.   
\end{equation}

Using the angle and core energy averaged energy distribution, we estimate next the observed rate of GRBs with a given energy.
\begin{equation}
\label{eq:ratesGRB}
	\frac{d\dot{N}_{\rm sGRB}}{dE}=\int_0^{z_{\rm max}} dz \frac{dV}{dz} \frac{\mathcal{R}_{\rm sGRB}(z)}{1+z} \frac{d{\rm Pr}}{dE}
\end{equation}  
where $z_{\rm max}$ is the maximum redshift to which a sGRB with energy $E$ can be detected, $dV/dz$ is the change in cosmological comoving volume with redshift, the comoving sGRB rate is $\mathcal{R}_{\rm sGRB}(z)=f_{\rm sGRB}(z) \mathcal{R}_{\rm BNS}(z)$ (this is reduced by a factor of $1+z$ in the observer frame due to cosmological time dilation) and $f_{\rm sGRB}(z)$ is the fraction of BNS mergers that result in sGRB jets that successfully break-out of the ejecta, which is assumed here to be 1 (see \citealt{BPBG2019}).

When considering the local population of sGRBs, we can ignore cosmological effects and simplify Eq. \ref{eq:ratesGRB} to
\begin{equation}
	\frac{d\dot{N}_{\rm sGRB}}{dE}=\int_0^{r_{\rm max}} dr 4\pi r^2 \mathcal{R}_{\rm sGRB}(0) \frac{d{\rm Pr}}{dE}
\end{equation}  
where the local sGRB rate is $\mathcal{R}_{\rm sGRB}(0)=f_{\rm sGRB}(0) \mathcal{R}_{\rm BNS}(0)\approx f_{\rm sGRB}(0) 320\mbox{ Gpc}^{-3}\mbox{ yr}^{-1}$ \citep{Mandel2022}.

The low energy tail of the sGRB distribution $d\dot{N}_{\rm sGRB}/dE\propto E^{-2/\tilde{a}}$, that arises due to bursts observed progressively off-axis, is visible as long as $E_{\rm lim}<E<E_{\rm min}$. This happens for
\begin{equation}
\label{eq:PhilimGRB}
    \Phi_{\rm lim}<\Phi_{\rm min}\equiv \frac{E_{\rm min}(1+z(d_{\rm L,cc}))}{4\pi d_{\rm L,cc}^2}=8.8\times 10^{-8}E_{\rm min,49} (1+z(d_{\rm L,cc})) d_{\rm L,cc,Gpc}^{-2} \mbox{erg cm}^{-2}
\end{equation}
where $d_{\rm, L,cc}$ is the imposed limit on the sample's luminosity distance.
If $E_{\rm lim}>E_{\rm min}$, the off-axis sGRB emission is visible only in the sensitivity limited regime and the observed distribution becomes steeper by a factor of $E^{3/2}$ (as in Eq. \ref{eq:dNMGFpopdE} for the MGFs, see also Fig. \ref{fig:MGFsGRBPhi}).
Similarly, if $E_{\rm lim}>E_{*}$ ($E_{\rm lim}>E_{\rm max}$), then the segment of the distribution dominated by the $E^{-\alpha_E}$ ($E^{-\beta_E}$) become suppressed by a factor of $E^{3/2}$.

In an analogy with \S \ref{sec:MGFpop}, we can define a limiting fluence, $\Phi_*\equiv E_*(1+z(d_{\rm L,cc}))/(4\pi d_{\rm L,cc}^2)$ ($\Phi_{\rm min}\equiv E_{\rm min}(1+z(d_{\rm L,cc}))/[4\pi d_{\rm L,cc}^2$]), below which events with $E_*$ ($E_{\rm min}$) can be seen within the entire volume up to $d_{\rm L,cc}$. As discussed in \S \ref{sec:modelindconst}, for the purpose of identifying a sub-population of MGFs within the sGRB population one typically imposes a limiting distance only on the MGF (which are intrinsically more abundant but fainter due to their smaller energies) and not the sGRB population (which due to their large energies and low rates are typically seen from much larger distances). Even if one imposes no specific limiting distance on the sGRB population, there is an effective limiting distance $d_{\rm L,cc}$ which corresponds to the luminosity distance beyond which the volume of the Universe within $d_{\rm L,cc}$ grows significantly slower than the Euclidean expectations $V(<d_{\rm L,cc})\propto d_{\rm L,cc}^3$. We find that a good approximation is to take $d_{\rm L,cc}\approx 5$\,Gpc (or $z\approx 0.8$).

\subsection{Mixing the MGF and sGRB populations}
\label{sec:combanalyticresult}

\begin{figure}
\centering
\includegraphics[scale=0.22]{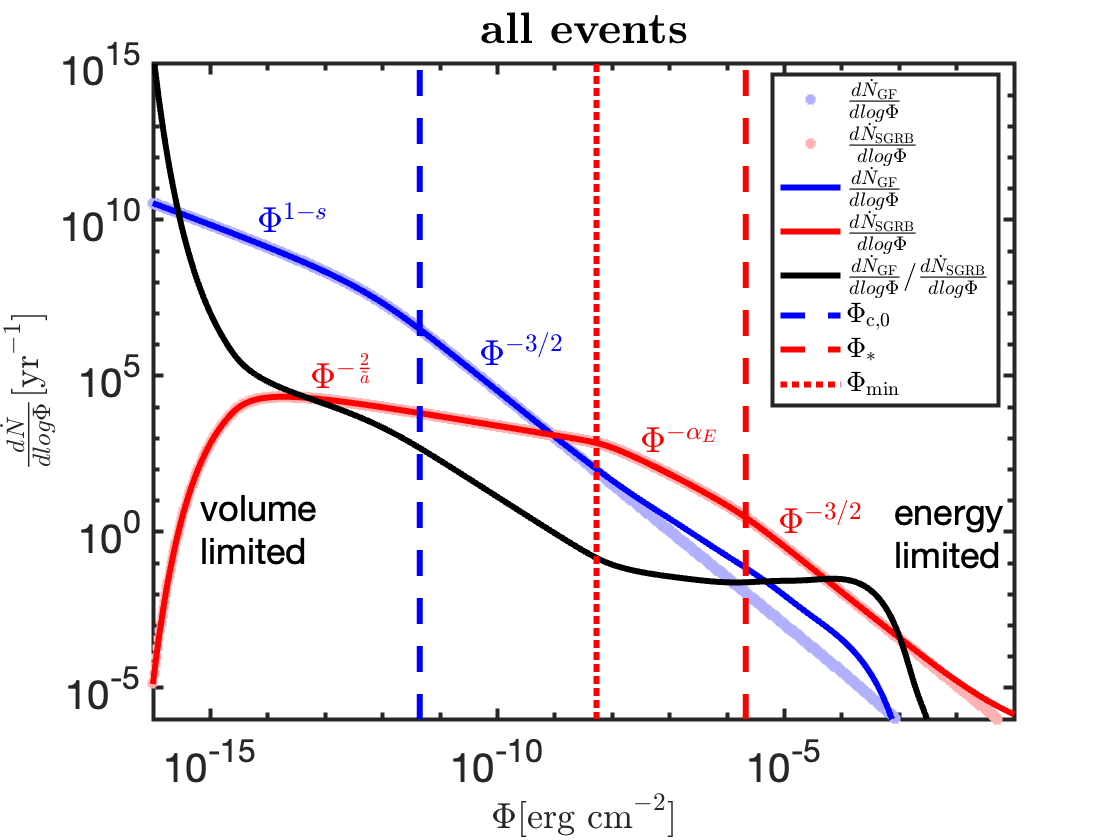}
\includegraphics[scale=0.22]{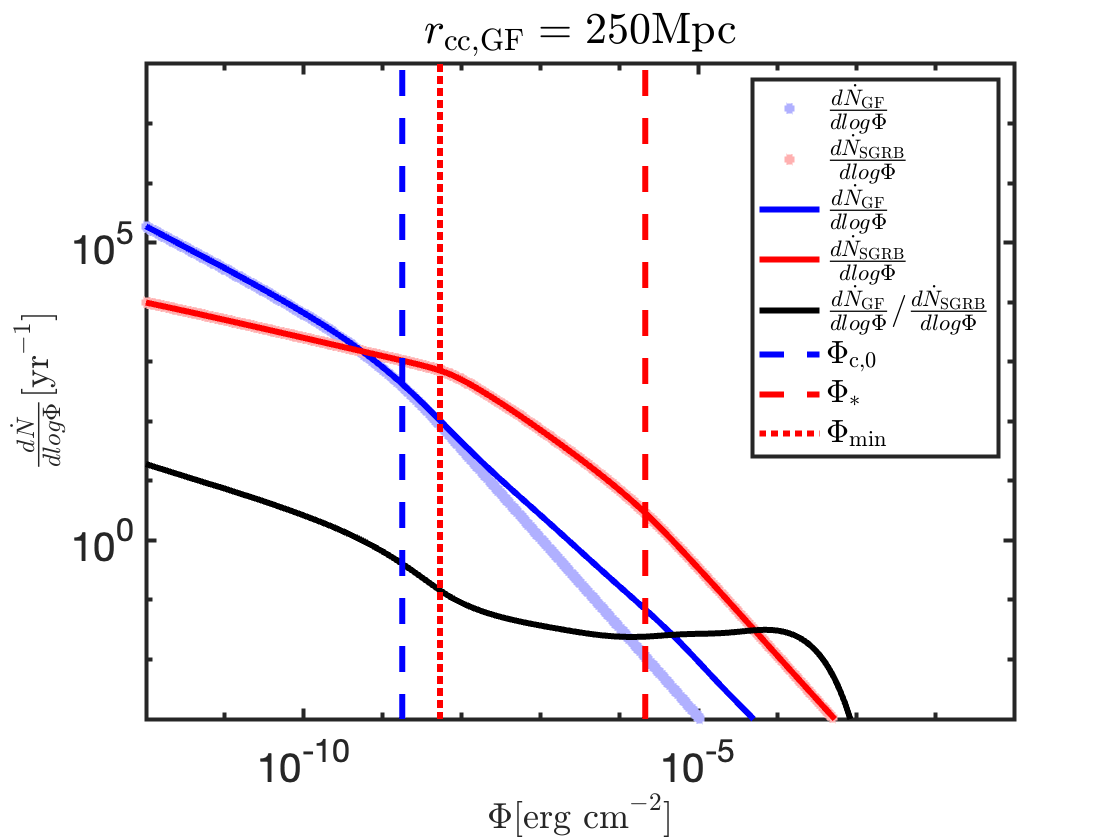}\\
\includegraphics[scale=0.22]{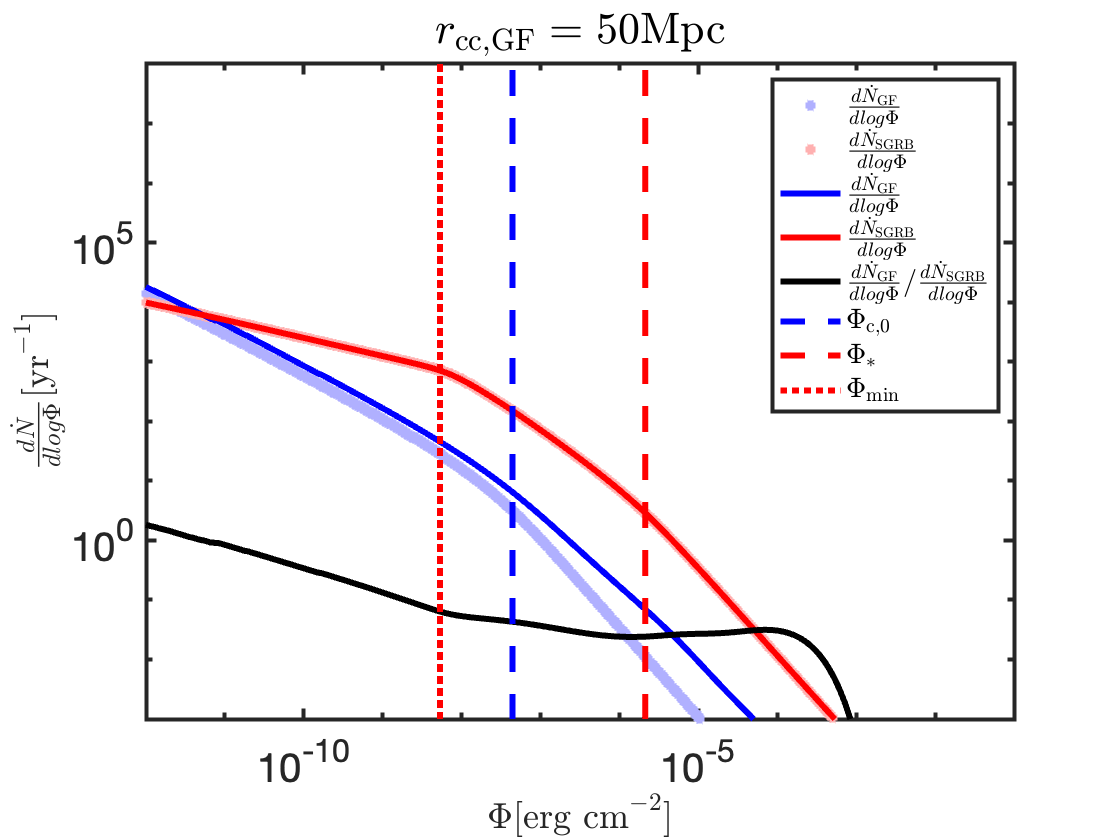}
\includegraphics[scale=0.22]{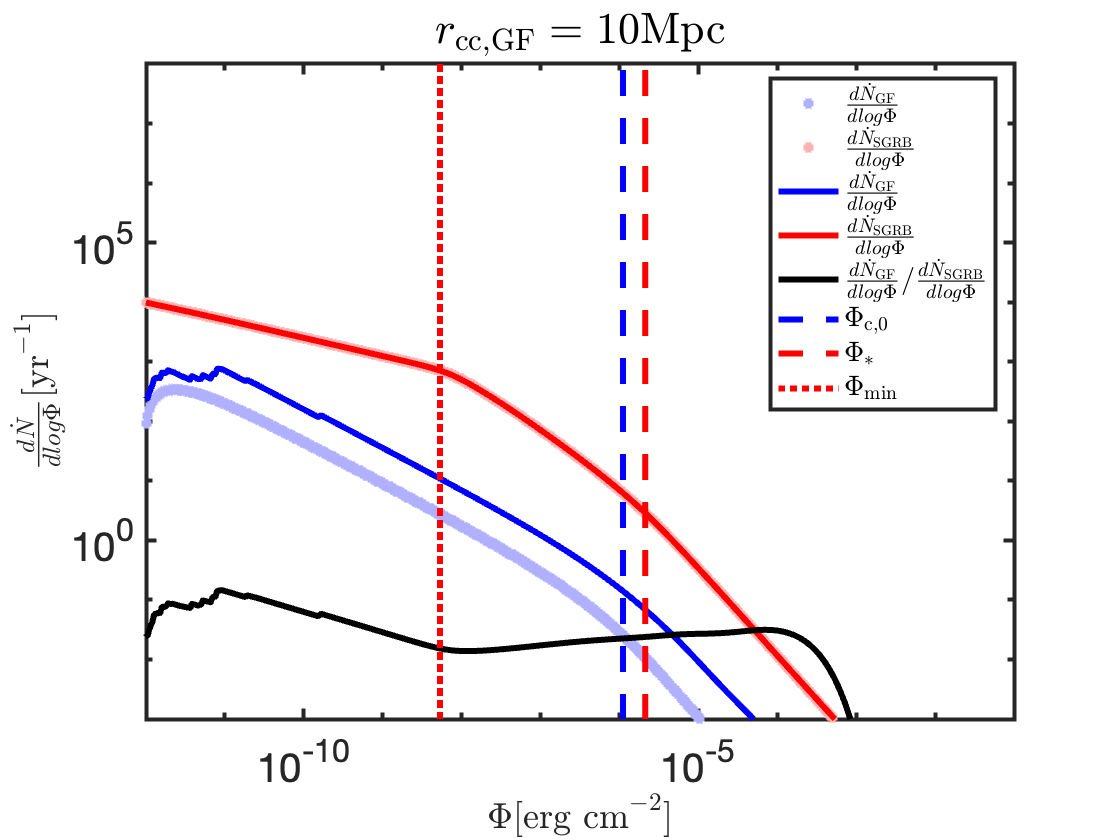}
\caption{Rate of events per logarithmic unit of fluence for different assumptions on the limiting discrimination distance of the MGF sample (no similar limit is imposed on the sGRB population). Results are shown for MGFs (blue), sGRBs (red) and for the ratio of the two (black). The light blue and red shaded wide bands depict the results for $\Delta=1$ (i.e. uniform distribution of sources), while the thin lines show the modification due to the local over-density of sources in the vicinity of the Milky Way (where $\Delta(r)$ is assumed to track star formation for MGFs and galaxy mass for sGRBs). Since sGRBs are seen to cosmological distances, the effect of $\Delta(r)$ is only apparent for that population at the highest fluences which probe increasingly closer events. This can be seen at the top left panel where the red line and the light red shaded region begin to diverge from each other.
The parameters used for this calculation are: $\mathcal{R}_{\rm CCSN}=10^5\mbox{Gpc}^{-3}\mbox{ yr}^{-1},f_{\rm mag}=0.2, s=1.7, f_{fl}=f_E=f_{\rm dip}=f_b=0.3, B_0=5\times 10^{14}\mbox{ G}, b=1, \alpha=1$ and: $\mathcal{R}_{\rm BNS}=320\mbox{Gpc}^{-3}\mbox{ yr}^{-1},f_{\rm sGRB}=1, \theta_{\rm c}=0.1, \theta_{\rm max}=1, \tilde{a}=6.5, \alpha_E=0.95, \beta_E=2, E_*=6\times 10^{51}\mbox{ erg}, E_{\rm min}=1.5\times 10^{49}\mbox{ erg}$. The distributions agree well with the derived asymptotic PLs in \S \ref{sec:MGFpop},\ref{sec:sGRB}.}
\label{fig:MGFsGRBPhi}
\end{figure}

In order to explore the joint distribution of MGFs and sGRBs, we begin by looking at the number of events detected per year and per logarithmic interval of fluence. The result is shown in Fig. \ref{fig:MGFsGRBPhi}, for all events, as well as for different values of the limiting distance to which MGF host galaxies can be identified, $r_{\rm cc,MGF}$. The distributions of both MGFs and sGRBs are volume limited at small values of $\Phi$, and energy limited at large values as anticipated in Eqns. \ref{eq:Ndotphi}, \ref{eq:PhilimGRB}. In addition, the asymptotic PLs in those regimes are well-reproduced by the numerical results. We explore the modification due to the local over-density of sources in the Milky Way vicinity (where $\Delta(r)$ is assumed to track star formation for MGFs and galaxy mass for sGRBs). This, especially at low $r_{\rm cc,MGF}$, leads to a modification from the results in the uniform source density case, and in particular, lead to an excess (over homogeneous expectations) of nearby and thus high fluence sources (approximately leading to $d\dot{N}/d\log \Phi$ being flatter by $1/4$ compared to the isotropic case, $d\dot{N}/d\log \Phi\propto \Phi^{-3/2}$).

\begin{figure}
\centering
\includegraphics[scale=0.22]{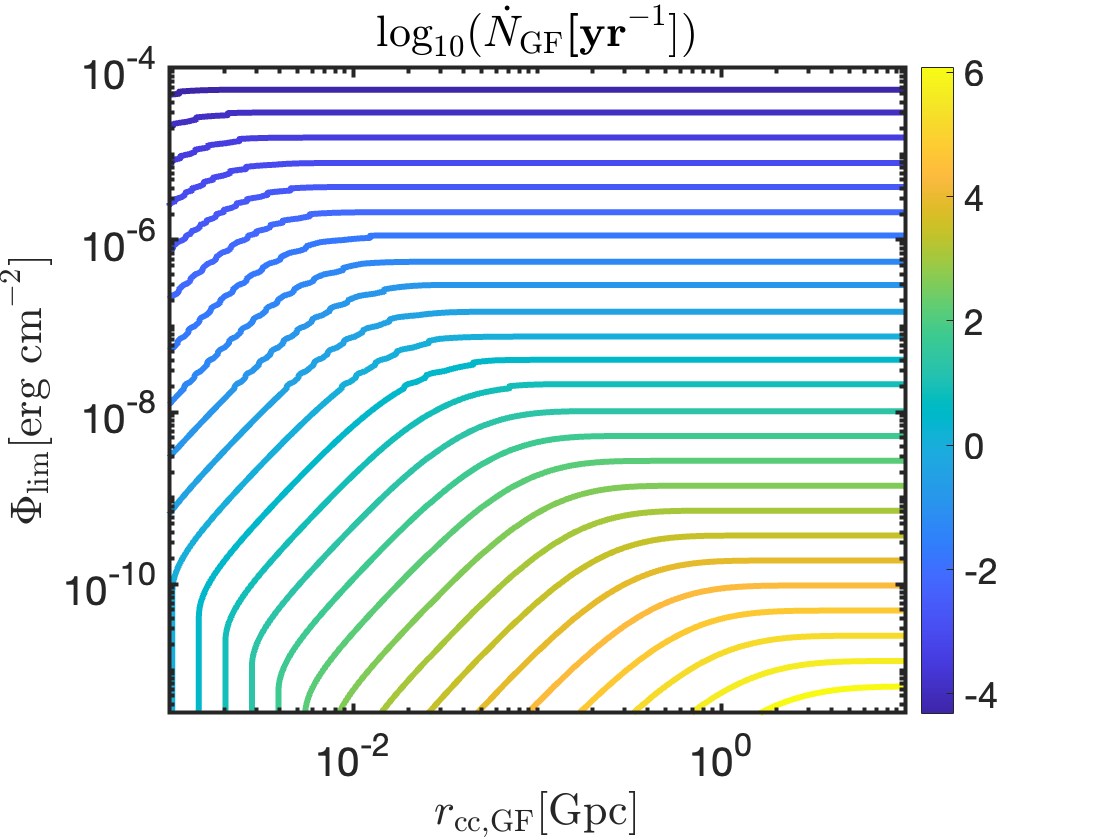}
\includegraphics[scale=0.22]{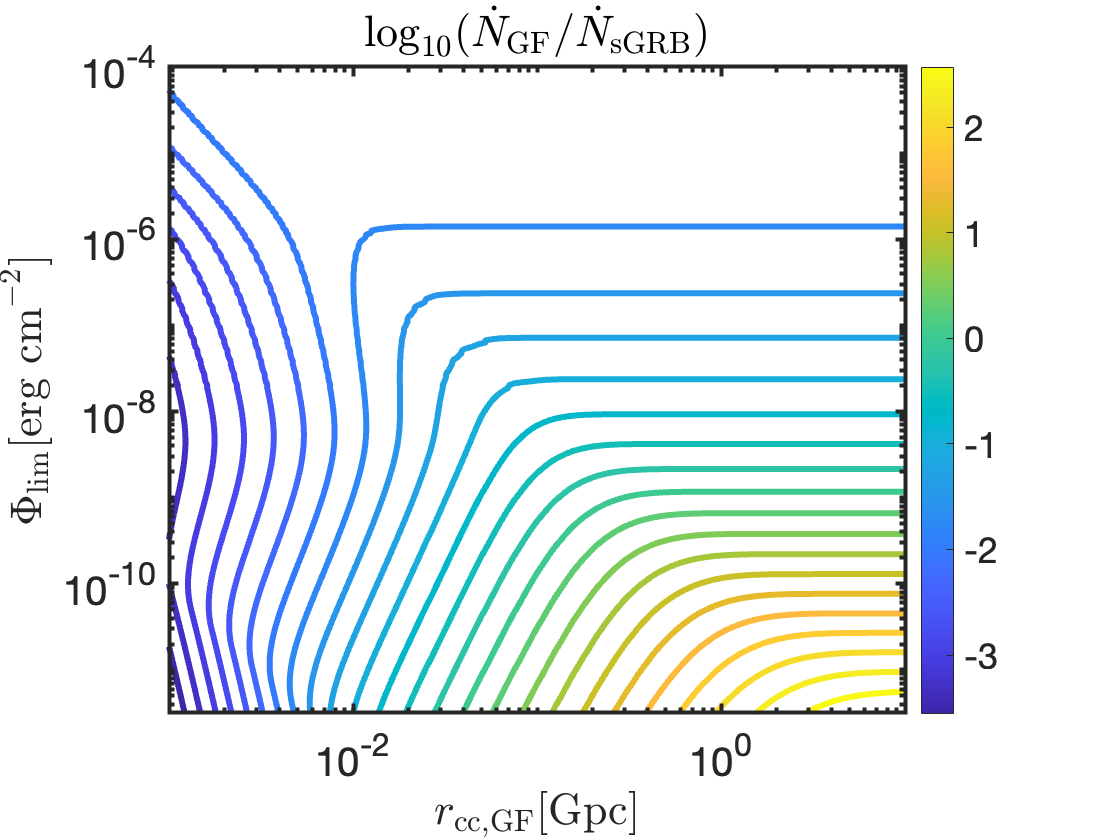}
\includegraphics[scale=0.22]{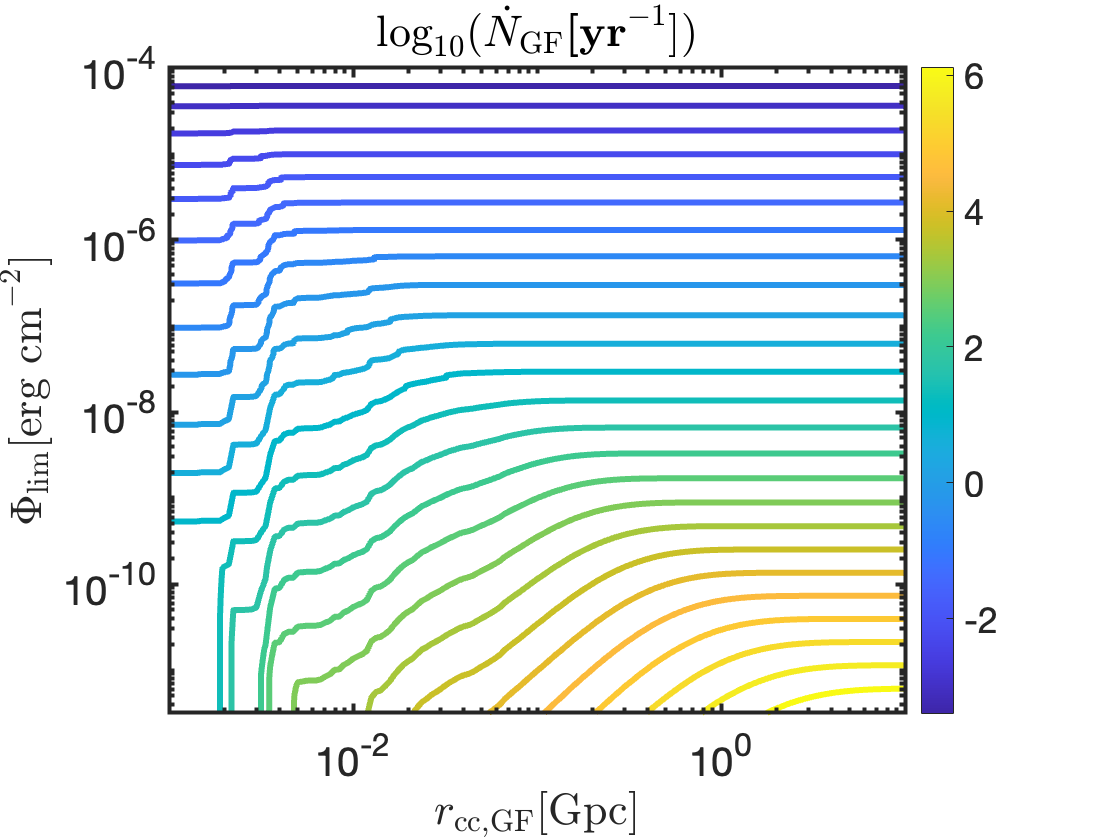}
\includegraphics[scale=0.22]{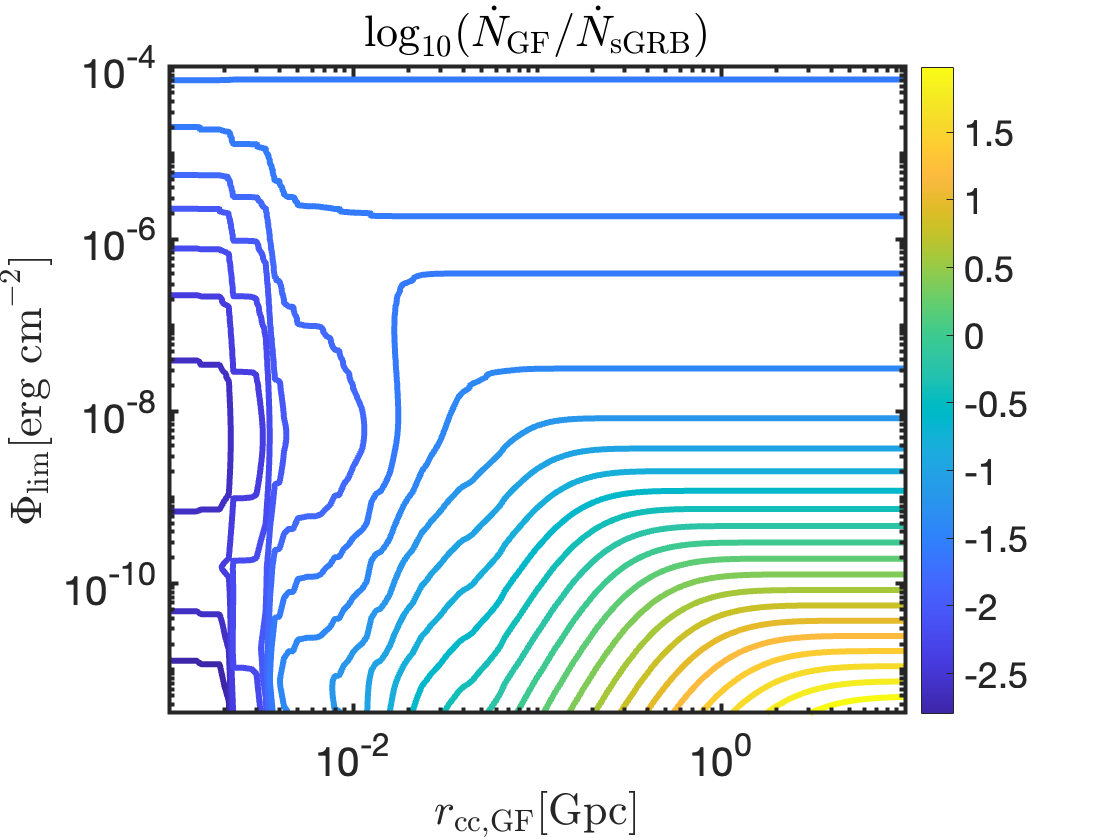}
\caption{Left: All-sky rate of MGFs above a given limiting fluence $\Phi>\Phi_{\rm lim}$ and below a certain luminosity distance $d_{\rm L}<r_{\rm cc,GF}$. Right: ratio between MGFs with $\Phi>\Phi_{\rm lim}$ and $d_{\rm L}<r_{\rm cc,G}$ to sGRBs with $\Phi>\Phi_{\rm lim}$ (and no additional limiting distance). Results are shown for $\Delta=1$ on the top panels and accounting for overdensity at small distances on the bottom panels.}
\label{fig:MGFsGRBratiocontour}
\end{figure}

\begin{figure}
\centering
\includegraphics[scale=0.22]{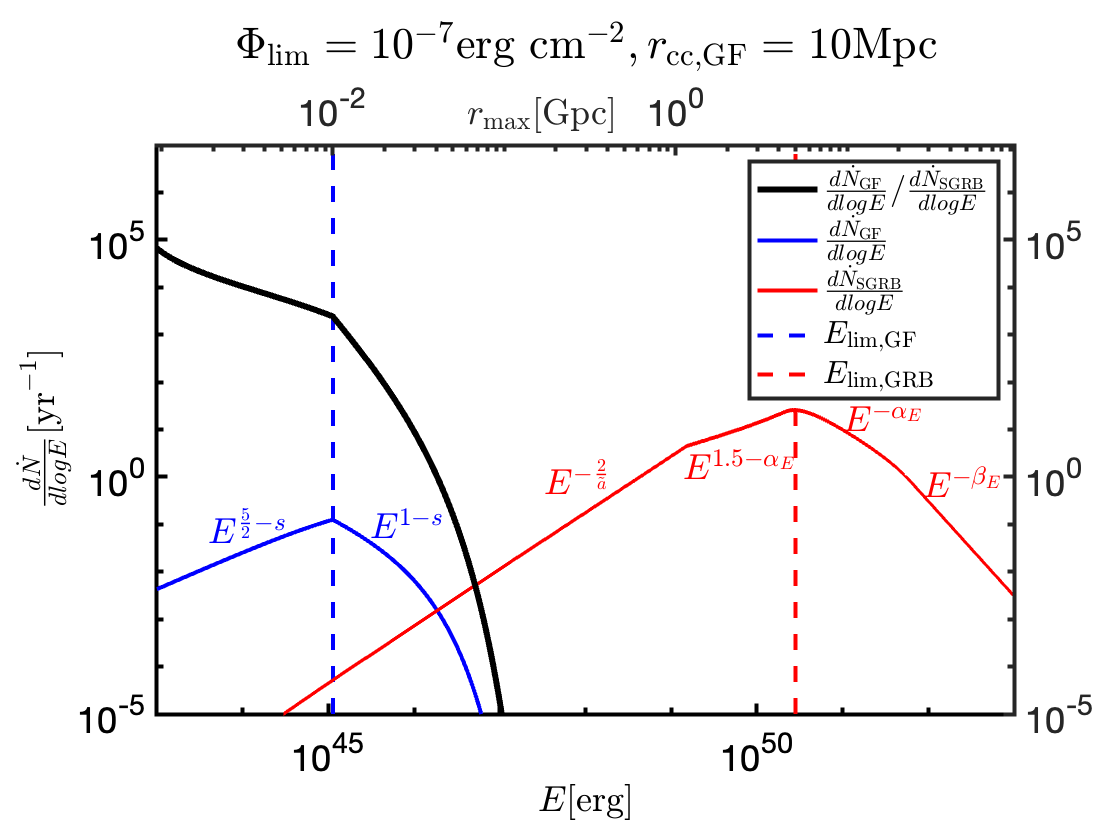}
\includegraphics[scale=0.22]{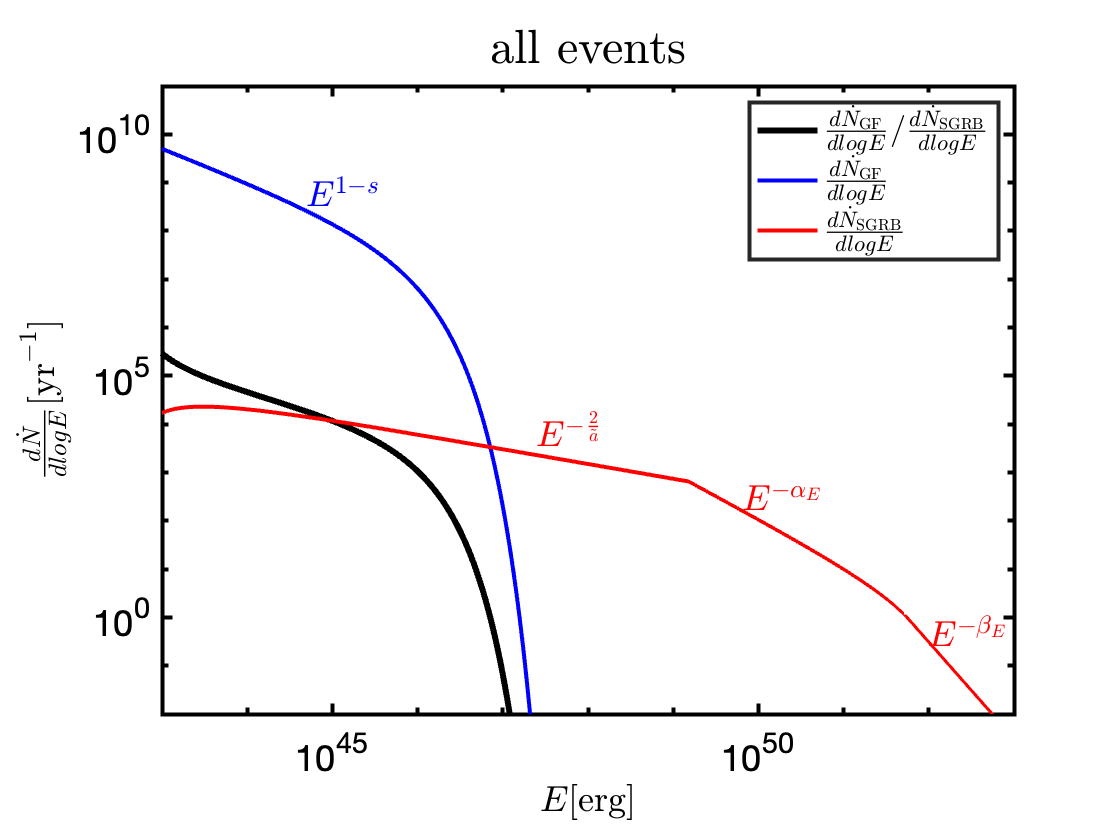}\\
\includegraphics[scale=0.22]{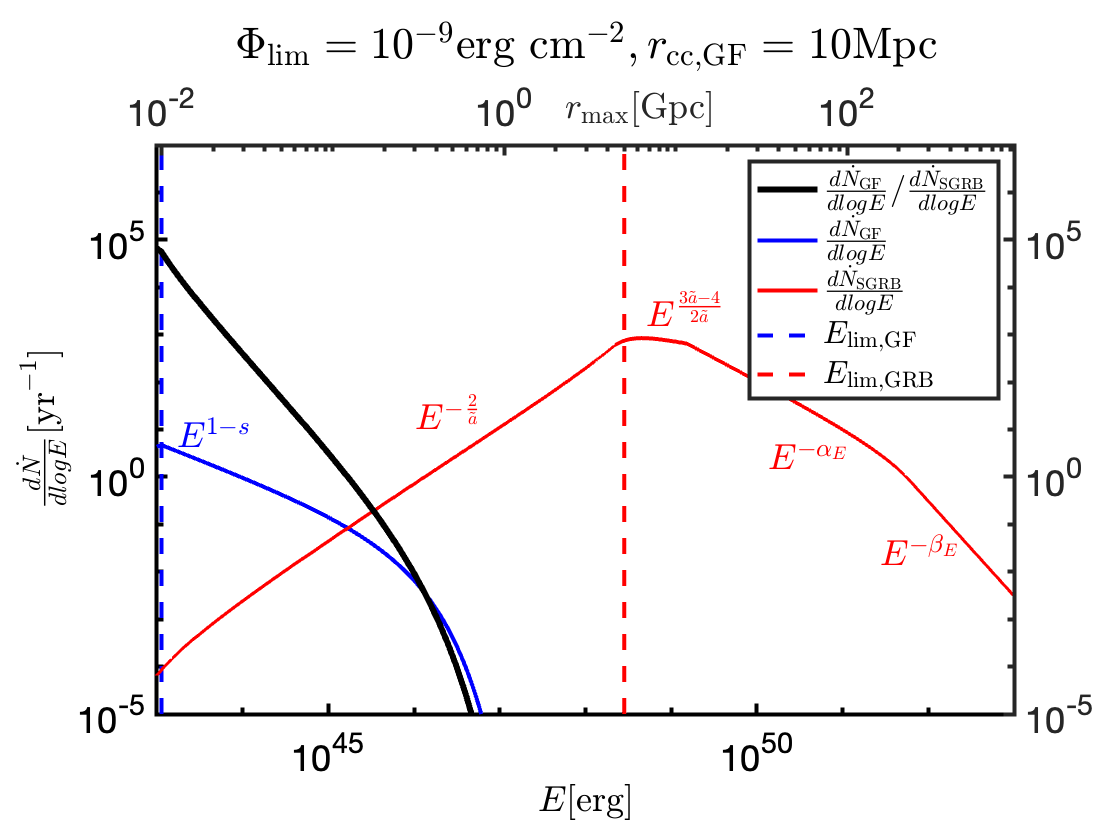}
\includegraphics[scale=0.22]{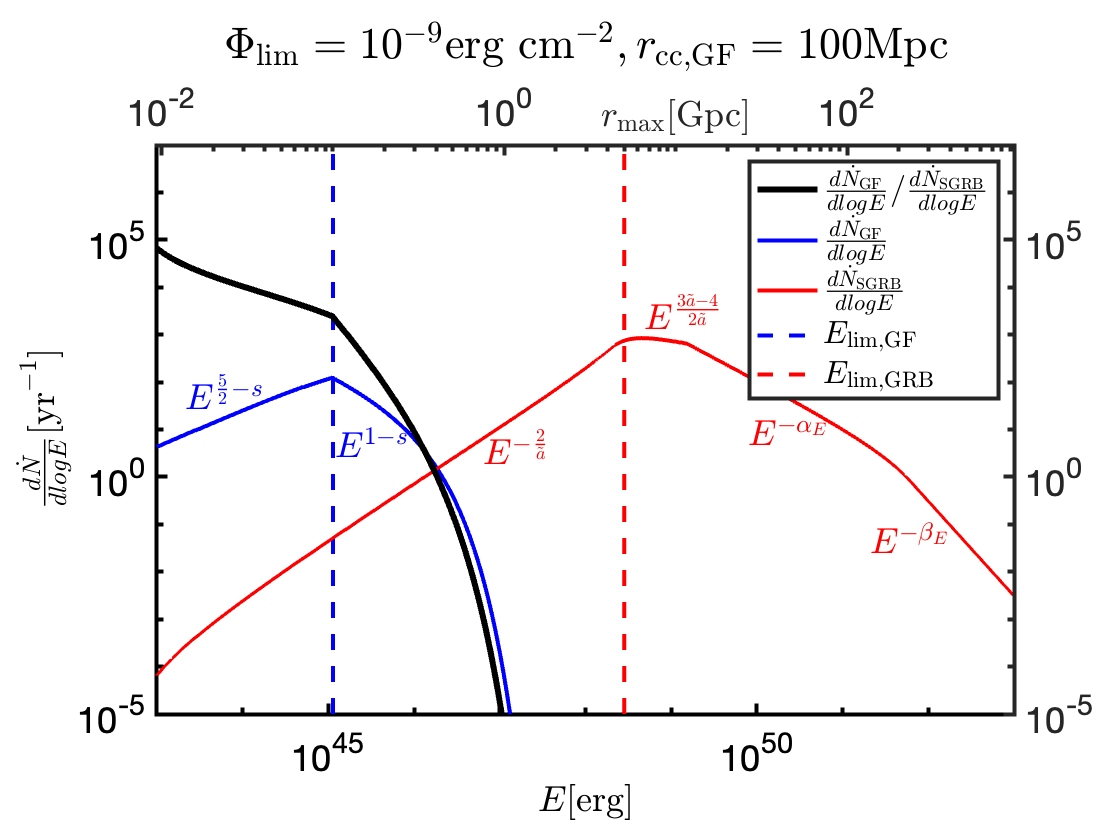}
\caption{Rate of events per logarithmic unit of isotropic equivalent energy, assuming different fluence thresholds, $\Phi_{\rm lim}$, and limiting identification distances for MGFs, $r_{\rm cc,GF}$, as well as (top right) no observational cut on the distributions. Results are shown separately for MGFs (blue) and sGRBs (red). We also plot the ratio of the two in black. The top X axis in each case, depicts the equivalent value of $r_{\rm max}=[E/(4\pi \Phi_{\rm lim})]^{1/2}$. The parameters assumed are the same as in Fig. \ref{fig:MGFsGRBPhi}. The distributions agree well with the derived PLs in \S \ref{sec:MGFpop},\ref{sec:sGRB}. The sensitivity limit regime for MGFs (sGRBs), $r_{\rm max}<r_{\rm cc}$, resides to the left of the dashed blue (red) line, and the volume limited regime, $r_{\rm max}>r_{\rm cc}$, to the right. For clarity and comparison with the analytic scalings, the results are depicted for the case with a uniform distribution of sources, $\Delta=1$ (for deviation from this assumption see Fig.s \ref{fig:MGFsGRBPhi}, \ref{fig:MGFsGRBratiocontour}).}
\label{fig:MGFsGRB}
\end{figure}

The observed number of sGRBs dominates over that of MGFs, except for low $\Phi_{\rm lim}$ (and to a lesser extent large $r_{\rm cc,GF}$). This is readily apparent in Fig. \ref{fig:MGFsGRBratiocontour}, which depicts the rate of observed MGFs and the ratio of MGFs to sGRBs as a function of $r_{\rm cc,GF},\Phi_{\rm lim}$. In particular, for $r_{\rm cc,GF}=10$Mpc, simply increasing sensitivity by reducing the limiting fluence without better localization capability will actually lead to a greater increase in the number of detected sGRBs than in that of MGFs. This trend persists until a limiting fluence of $\sim 5\times 10^{-9}\mbox{ erg cm}^{-2}$.
Fig. \ref{fig:MGFsGRB} presents a complimentary view of the observable parameter space, considering the rate of events per unit isotropic energy, for varying assumptions on the limiting fluence and distance. 

\begin{figure}
\centering
\includegraphics[scale=0.22]{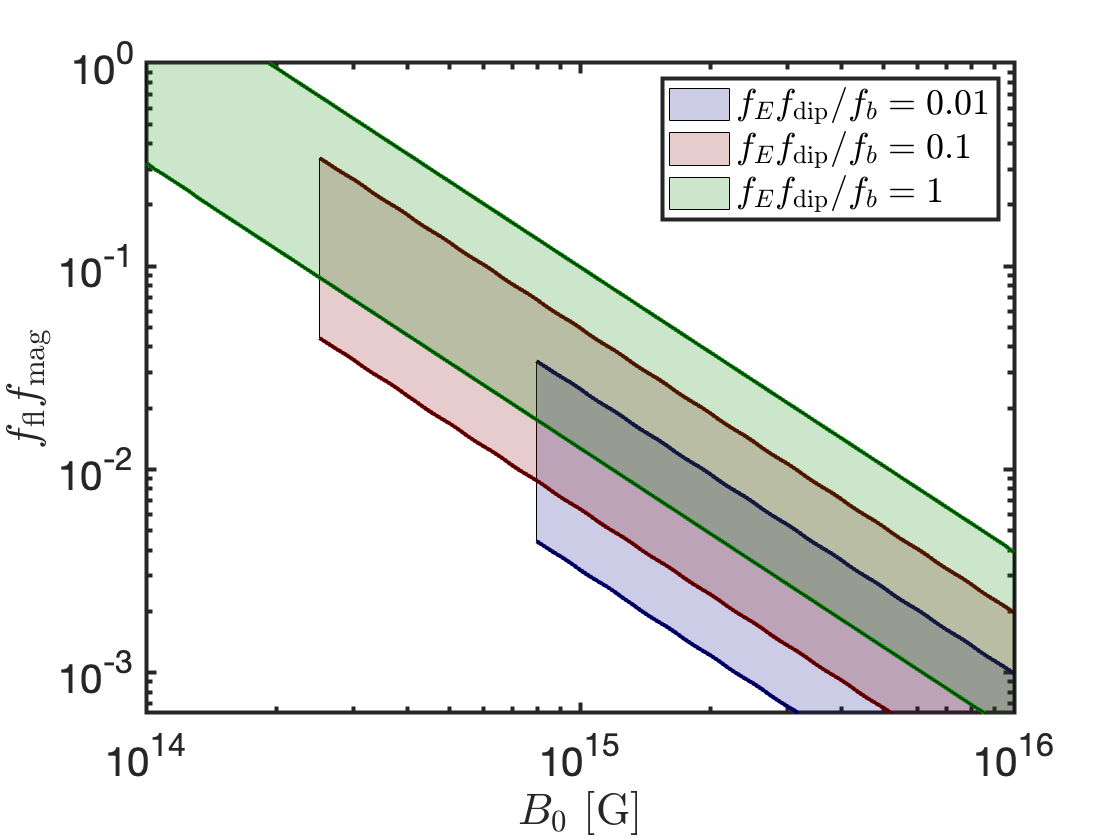}
\includegraphics[scale=0.22]{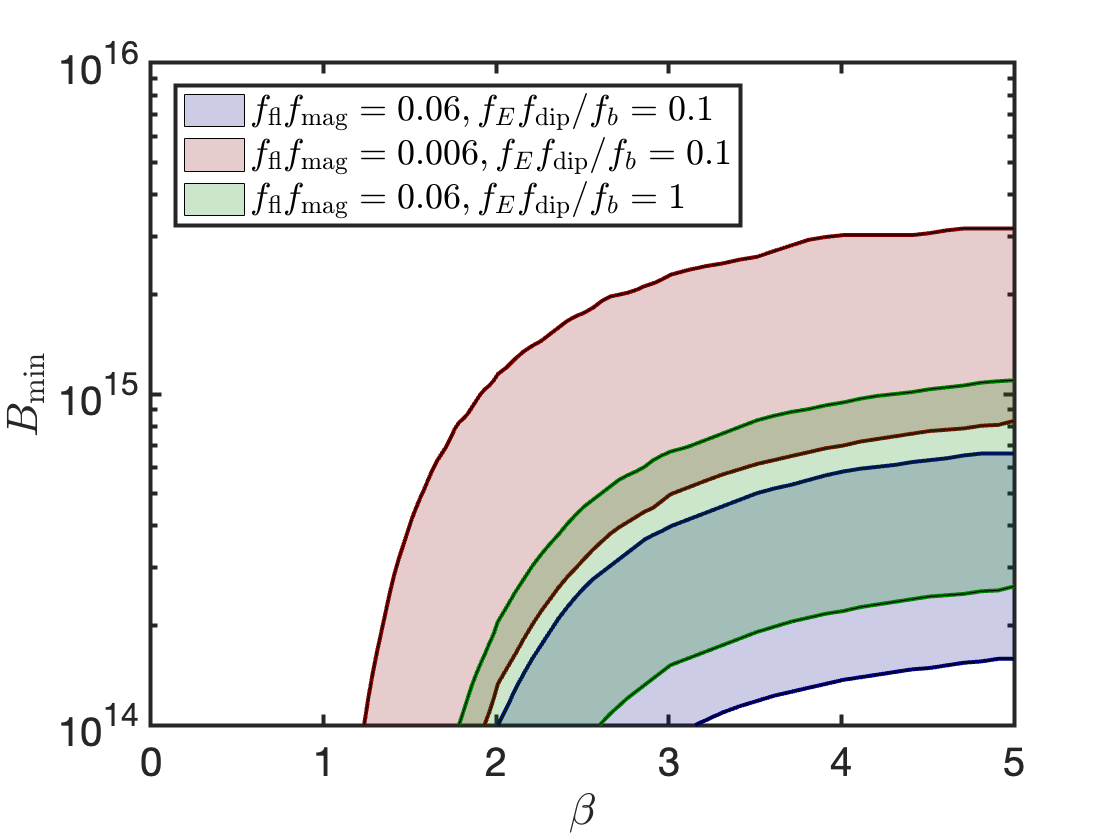}
\caption{Allowed parameter space for model parameters considering observational limits on MGF occurrence rates within the sGRB population. The allowed region is shown for the case where all magnetars are born identical (left) or assuming a power-law distribution of birth magnetic fields above a varying $B_{\rm min}$ and up to $B_{\rm max}=10^{16}$\,G with $d{\rm Pr}/dB_0\propto B_0^{-\beta}$ (right). }
\label{fig:AllowedParamSpace}
\end{figure}

A comparison of our modeling with observational bounds can significantly constrain the allowed parameter space for MGF rates and energies.
As discussed in \S \ref{sec:modelindconst}, using the sGRB sample of \citet{2021ApJ...907L..28B}, the ratio of MGFs/sGRBs is $0.007<\dot{N}_{\rm MGF}/\dot{N}_{\rm GRB}<0.057$ at a $90\%$ confidence limit and for $\Phi_{\rm lim}\approx 2\times 10^{-6}, r_{\rm cc,GF}=10\mbox{ Mpc}$, corresponding to $E_{\rm lim}=2\times 10^{46}\mbox{ erg}$ (meaning bursts above this energy can be detected in the entire volume up to $r_{\rm cc,GF}$). As clear from Eqns. \ref{eq:Nobsdot}, \ref{eq:Ndotpop}, the parameters $f_{\rm fl},f_{\rm mag}$ enter this ratio only via their product. Similarly, as explained in \S \ref{sec:IndMag}, the parameters $f_{E},f_{\rm dip}, f_b$ enter the observed MGF rate only through the specific combination $f_{E}f_{\rm dip}/f_b$. The result is that the MGF/sGRB ratio above is primarily a function of three parameters: (i) $B_0$, (ii) $f_{\rm fl}f_{\rm mag}$ and (iii)  $f_{E}f_{\rm dip}/f_b$. The allowed parameter space is depicted in Fig. \ref{fig:AllowedParamSpace}.
It is apparent that $f_{E}f_{\rm dip}/f_b$ has a weak effect on the allowed parameter space, as long as the maximum MGF energy per magnetar, $E_{\rm c,0}=f_E f_{\rm dip} E_{\rm B,0}/f_b$, is large enough to account for the observed MGF energetics (in what follows we impose a rather modest constraint of $E_{\rm c,0}>10^{45}\mbox{ erg}$). The reason for this weak dependence is that, as shown in Eq. \ref{eq:Ndotpop}, $\dot{N}_{\rm MGF}$ is proportional to this combination to the power of $s-2=-0.3$ when $E_{\rm lim}<E_{\rm c,0}$ and to the power of $0.5$ when $E_{\rm lim}>E_{\rm c,0}$. As both powers are weak, and they straddle 0 when transitioning between the two energy regimes, the results are largely insensitive to $f_{E}f_{\rm dip}/f_b$. Eq. \ref{eq:Ndotpop} also shows that for a fixed rate of detected events, $f_{\rm fl}f_{\rm mag}\propto B_0^{2-2s}=B_0^{-1.4}$ when $E_{\rm lim}<E_{\rm c,0}$ and $f_{\rm fl}f_{\rm mag}\propto B_0^{-3}$ when $E_{\rm lim}>E_{\rm c,0}$. Considering that for the chosen sample $E_{\rm lim}=2\times 10^{46}\mbox{ erg}$ and that we impose the condition $E_{\rm c,0}>10^{45}\mbox{ erg}$, the former limit is applicable within most of the parameter space. This matches well with the anti-correlation shown between these parameters in Fig. \ref{fig:AllowedParamSpace}. 
Our results can be approximately summarized as 
\begin{equation}
\label{eq:Bconstraint}
    B_0\approx 4.5 \times 10^{14}\left(\frac{f_{\rm fl}f_{\rm mag}}{0.06}\right)^{-5/7} \mbox{ G} \quad \& \quad B_0\gtrsim 1.4\times 10^{14} \left(\frac{0.3 f_b}{f_E f_{\rm dip}}\right)^{1/2}\mbox{ G}.
\end{equation}
Plugging Eq. \ref{eq:Bconstraint} back into Eq. \ref{eq:Nobsdot},
we get
\begin{equation}
    \dot{\mathcal{N}}_{\rm obs}(E_{\rm c,0})\!\approx \!9\times 10^{-5} \left(\frac{B_0}{5\times 10^{14\mbox{ G}}}\right)^{-7/5}\left(\frac{0.3 f_b}{f_E f_{\rm dip}}\right)\mbox{ yr}^{-1}\mbox{ magnetar}^{-1} \quad \& \quad E_{\rm c,0}\!\approx \!1.3\times 10^{46} \left(\frac{f_E f_{\rm dip}}{0.3 f_b}\right)\left(\frac{B_0}{5\times 10^{14}\mbox{ G}}\right)^2\mbox{ erg}.
\end{equation}
The rate of MGFs per magnetar can be compared with recent, independent observational constraints, based on long-term searches for MGFs in the Virgo cluster by {\it{INTEGRAL}} \citep{Pacholski2024}. Using $\dot{\mathcal{N}}_{\rm obs}(E)=\dot{\mathcal{N}}_{\rm obs}(E_{\rm c,0})(E/E_{\rm c,0})^{1-s}$, we estimate for the same choice of canonical parameters as above, $\dot{\mathcal{N}}_{\rm obs}(3\times 10^{45}\mbox{ erg})\approx 2.5\times 10^{-4}  \mbox{ yr}^{-1}\mbox{ magnetar}^{-1}$, $\dot{\mathcal{N}}_{\rm obs}(10^{45}\mbox{ erg})\approx 5.4\times 10^{-4}  \mbox{ yr}^{-1}\mbox{ magnetar}^{-1}$. Both numbers are consistent with those from \cite{Pacholski2024} that find $\dot{\mathcal{N}}_{\rm obs}(3\times 10^{45}\mbox{ erg})\lesssim 2\times 10^{-3}  \mbox{ yr}^{-1}\mbox{ magnetar}^{-1}$
 and $\dot{\mathcal{N}}_{\rm obs}(10^{45}\mbox{ erg})\gtrsim 4\times 10^{-4}  \mbox{ yr}^{-1}\mbox{ magnetar}^{-1}$ (their analysis is less constraining at both lower energies - as they are too faint to be seen up to the Virgo cluster, and higher energies - considering the observation duration and field of view).
 
Applying the conditions $f_E f_{\rm dip}/f_b\lesssim 0.3$, $f_{\rm mag}\approx 0.2$ and $f_{\rm fl}\gtrsim 0.1$ (see \S \ref{sec:IndMag} and \S \ref{sec:ObsEnergy}), the allowed parameter range becomes $B_0 \approx 2\times 10^{14} - 2\times 10^{15}$\,G.
The results in Eq. \ref{eq:Bconstraint} can be generalized to a situation in which there is a broad distribution of magnetar initial field strengths, as discussed in \S \ref{sec:changeB0}. Fixing the maximum $B_{\rm max}=10^{16}\mbox{ G}$, Fig. \ref{fig:AllowedParamSpace} shows the constraints on $B_{\rm min},\beta$ for various values of $f_{\rm fl}f_{\rm mag}$ and $f_E f_{\rm dip}/f_b$. Since for $\beta>1$, the population is dominated by magnetars with $B_0\approx B_{\rm min}$, we see that the results match closely with those for a fixed $B_0$ distribution when substituting $B_0 \to \langle B_0\rangle\approx 2^{1\over (\beta-1)} B_{\rm min}$ (where the last transition holds for $\beta>1, B_{\rm max}\gg B_{\rm min}$). The possibility of an initial field distribution can be probed with future data, which could provide meaningful constraints not only on the overall fraction of MGFs in the sGRB population, but also on their magnetic energy distribution.

\section{Prospects for Burst Gravitational Waves and the Stochastic Gravitational Waves Background from Magnetars}
\label{sec:GW}

We turn to investigate the detectability of burst GWs and of a stochastic background of gravitational waves (SGWB) generated by the population of MGFs discussed above. Numerous previous works have considered GWs from magnetars from various standpoints, empirically \citep[e.g.,][]{2007PhRvD..76f2003A,2008PhRvL.101u1102A,2009ApJ...701L..68A,2011ApJ...734L..35A,2019ApJ...874..163A,2024ApJ...966..137A} and in terms of forecasts or mechanisms \citep[e.g.,][]{2001MNRAS.327..639I,2011PhRvD..83j4014C,2011MNRAS.418..659L,2012PhRvD..85b4030Z,2021ApJ...918...80M,2022PhRvD.106f3007K,2024MNRAS.527.2379Y,2024MNRAS.tmp.1988B}. Other papers made different assumptions from MGFs: for example, \cite{2011MNRAS.411.2549M} considered GWs emitted by young, fast spinning magnetars distorted by a very strong magnetic field. The magnetar formation rate is estimated to be a large fraction of the CCSNe rate, and thus magnetars should be extremely common throughout cosmic history contributing to a SGWB at relevant (see below) frequencies. CCSNe themselves may source GWs during the proto-NS stage through excitation of fluid modes, particular f-like modes of the proto-NS. However, neutrinos in the CCSNe and hot proto-NS run away with most of the explosion energy, limiting the efficacy of GW production. In contrast, neutrino losses are not dominant in MGFs. Magnetars are also far more compact than the stage where GWs are radiated from the proto-NSs created in CCSNe. Over their lifetime, magnetars may contribute to GWs significantly as the rate of MGFs implies multiple MGFs occur in the lifetime of any given magnetar.

We consider here that some fraction of the energy in a MGF is released in the form gravitational waves. For simplicity and definiteness, we assume a linear scaling between the electromagnetic isotropic-equivalent energy and the gravitational wave production,
\begin{equation}
E_{\rm GW}^{\rm MGF} = \eta_{\rm gw} E_{\rm EM}^{\rm MGF}.
\label{eq:efficiency}
\end{equation}
The proportionality factor $\eta_{\rm gw}$ may vary considerably between different sources, or events within the same source depending on the details and idiosyncrasies of burst dynamics. However, a-priori we have no strong theoretical or observational justification yet for a deviation from a constant proportionality. If the trigger of MGFs is internal to the NS and the efficiency of MGF electromagnetic emission is not high, it is not even a-priori demanded that $\eta_{\rm gw} \leq 1$, but we assume so for definiteness.
As shown in Eq. \ref{eq:Edotfl} the total energy channeled into bursts is independent of $f_b$. Given our assumptions above, the same holds true also for the GWSB contributed by MGFs.

We explore below two physical mechanisms for the generation of gravitational waves in a MGF, (A) outflows or jets from MGFs and (B) excitation of fundamental global oscillation modes of the NS. Scenario (A) has never been considered before in the MGF context, while (B) was proposed as a possible SGWB source in \cite{2021ApJ...907L..28B} and in considered in \citet{2022PhRvD.106f3007K} with comparable assumptions. The method for calculating SGWB in each case is presented in \S \ref{sec:stoch}.

\subsection{Scenario A: GWs from Acceleration of Relativistic Outflows in MGFs}
\label{sec:GWoutflow}

MGFs involve super-Eddington luminosities and are energetic enough to overcome magnetic confinement in the magnetosphere and to blow open field lines near polar locales. This picture was realized soon after the 1979 March 5 event \citep{1980Natur.287..122R,1981Natur.292..319L,1982Natur.299..321L,1984Natur.310..121L,1988MNRAS.235...79B,Paczynski1992}. They ought to generically lead to baryonic outflows involving variable mass loadings, perhaps even leading to r-process enrichment \citep{2024MNRAS.528.5323C}. Additionally, some MGFs could involve trapped charged plasma of significant mass in the magnetosphere, sourcing possible continuous GWs \citep{2024MNRAS.527.2379Y}. Following an impulsive energy injection (corroborated by the short rise time of MGFs), such relativistic ejecta must originate and be accelerated relatively close to the neutron star in a short timescale, forming a baryon-loaded optically thick expanding and accelerating collimated fireball. The pulsating tails of MGFs are also highly indicative of coherent bulk acceleration of baryon-rich plasma \citep{2016MNRAS.461..877V}. Generically, fast acceleration of a blob of compact matter ought to lead to GWs, which we examine below.

An outflow scenario for burst GWs has, to our knowledge, not been previously considered in a magnetar context, although it has been considered for outflows of classical GRBs \citep{1987Natur.327..123B,2001PhRvD..64f4018S,2004PhRvD..70j4012S,2013PhRvD..87l3007B,2021PhRvD.104j4002L,2022arXiv221002740P}. 
While the energetics of classical GRBs are orders of magnitude higher than MGFs, we argue that this mechanism is promising for gravitational wave generation in MGFs for several reasons:
\begin{itemize}
    \item The recent GeV afterglow of MGF 200415A in the Sculptor galaxy \citep{2021NatAs...5..385F} confirms that ultrarelativistic overflows are indeed possible for MGFs, over more mildly relativistic ones detected in Galactic MGFs of SGR 1806-20 and SGR 1900+14 \citep{1999Natur.398..127F,2005Natur.434.1104G,Granot+06,2007Ap&SS.308...39G}. The mass associated with such outflows is significant, possibly up to $10^{-8}-10^{-6} M_\odot$ \citep[e.g.,][]{2007Ap&SS.308...39G} with kinetic energy comparable to the MGF prompt electromagnetic emission energetics. Outflows of significant mass are also possibly suggested by large timing anomalies of SGR 1935+2154 without MGFs, so may be common in some magnetars over their lifetime \citep{2023NatAs...7..339Y,2024Natur.626..500H}. 
    \item The $\gamma$-ray variability and rise times of MGFs are much shorter than classical GRBs, implying extremely fast bulk Comptonization and acceleration conducive to more efficient generation of GWs. $10 \,\mu$s variability and rise times have been detected in MGF 200415A \citep{Roberts2021Natur.589..207R}.
    \item MGFs are far more common than classical GRBs, which do not exceed a fraction of a percent of the CCSNe rate \citep{WP2015,Ghirlanda2016}, and therefore are statistically nearer.
\end{itemize}

In the classical GRB case, the maximum strain amplitude of burst GWs is of order $h_{\rm GRB, max} \sim 3\times 10^{-25}$ at 100 Mpc distance \citep{2021PhRvD.104j4002L}. The MGF case is similar at $\sim10$ kpc (see below). 

The relativistic outflow that could be caused by a MGF can be described in a simplified way as a point mass with $m_{\rm outflow}$, that accelerates uniformly from rest to a final speed $v_{\rm final}$ within the MGF rise time, $t_{\rm rise} \ll 1 {\rm\, ms}$. If this accelerated motion has a kinetic energy comparable to that of the observed electromagnetic MGF, that is, $10^{45}-10^{49}$ erg, and assuming that it needs to reach at least the neutron star escape velocity ($v_{\rm escape} \sim 0.7c$) in order to become unbound, we have at most $m_{\rm outflow} \sim 10^{-9}-10^{-5}M_{\odot}$.
The well-known GW quadrupole formula is not strictly valid for relativistic motion, and the relativistic treatment results in interesting effects such as anti-beaming of the waveform, beaming of the energy emission and a lingering memory effect. For example, \cite{2021PhRvD.104j4002L} estimates that 50\% of the radiated GW energy will be beamed in a cone with opening angle $\sqrt{2/\Gamma} \approx 70^{\circ}$ for our example\footnote{Note that the detection of the 2020 MGF in the Sculptur galaxy suggests $\Gamma \lesssim 100$ \citep{2021NatAs...5..385F} while the outflow of MGF~041227 suggests an initial $\Gamma\approx 1.5$ \cite{Granot+06}.}. For simplicity, we choose to neglect relativistic corrections to obtain an order of magnitude estimate for the radiated energy in gravitational waves and efficiency parameter defined in Eq. (\ref{eq:efficiency}), which we will use for calculating the resulting SGWB. Such relativistic effects largely only impact beaming of GWs and do not impact the total energy emitted in GWs  \citep[within factors of order unity, see e.g., Eq.~8 in][]{2021PhRvD.104j4002L}. The SGWB calculation is also unaffected except for a possible shift in the observed frequency spectrum of GWs than from a non-relativistic calculation; this phenomenologically, however, is degenerate with the unknown acceleration time. These beaming effects could enhance detectability of individual burst GWs from MGF outflows.

Without loss of generality, we stipulate that the motion happens in the $z$-direction. Then, the only nonzero component of the quadrupole moment is $M_{zz} = m_{\rm outflow}z^2(t)$, where $z(t) = (v_{\rm final}/t_{\rm rise})t^2/2$ for $0 \le t \le t_{\rm rise}$ and $z(t) = v_{\rm final}t$ for $t > t_{\rm rise}$. The gravitational wave strain $h_{+}$ at a distance $R$ from the source and at an angle $\theta$ from the direction of motion is then given by
\begin{equation}
h_{+}(t, R, \theta,\phi) = \frac{G}{Rc^4}\frac{d^2M_{zz}}{dt^2}\sin^2\theta = \frac{3G}{Rc^4}m_{\rm outflow}(v_{\rm final}/t_{\rm rise})^2t^2\sin^2\theta\,, \ {\rm for} \ 0 \le t \le t_{\rm rise}\,,
\end{equation}
and $h_{+}(t, R, \theta,\phi) = 0$ otherwise. At $R \sim 10$ kpc, $\theta = \pi/2$ and $t = t_{\rm rise}$, with $v_{\rm final} = v_{\rm escape}$, the peak burst GW strain is 
\begin{equation}
 h_{+}(t, R, \theta,\phi)\sim 10^{-24} \left(\frac{m_{\rm outflow}}{10^{-7} M_\odot} \right)\left( \frac{10 {\, \rm kpc}}{R} \right) .
 \label{eq:outflow_hmax}
\end{equation}
This strain (maximum) amplitude is independent of rise time and is also the scale of GW memory. Detection of these GWs would occur in time intervals (and associated frequencies) commensurate with the rise time (see below). The gravitational wave luminosity can be estimated as
\begin{equation}
\frac{dE_{\rm GW}}{dt} = \frac{2G}{15c^5}\left(\frac{d^3M_{zz}}{dt^3}\right)^2 = \frac{2G}{15c^5}m^2_{\rm outflow}(v_{\rm final}/t_{\rm rise})^4t^2\,,  \ {\rm for} \ 0 \le t \le t_{\rm rise}\,,
\end{equation}
and $dE_{\rm GW}/dt = 0$ otherwise, so the total energy emitted in gravitational waves is 
\begin{eqnarray}
&E_{\rm GW}\! =\! \int^{t_{\rm rise}}_0 \frac{dE_{\rm GW}}{dt} dt \!=\! \frac{2G}{3 c^5} \frac{m_{\rm outflow}^2 v_{\rm final}^4}{t_{\rm rise}} \!\nonumber \\& \approx \!1.4\times 10^{39} \left(\frac{m_{\rm outflow}}{10^{-7} M_\odot} \right)^2 \left(\frac{v_{\rm final}}{0.7 c} \right)^4 \left(\frac{10 \, \mu \rm s}{t_{\rm rise}}\right)  \, \, \rm erg= 10^{37}\left(\frac{B_0}{5\times 10^{14} \mbox{ G}} \right)^4 \left(\frac{f_E f_{\rm dip} \eta_{\rm kin}}{0.1} \right)^2\left(\frac{10 \, \mu \rm s}{t_{\rm rise}}\right) \rm erg
\label{eq:Egwtotal}
\end{eqnarray}
where $\eta_{\rm kin}$ is the ratio between the MGF kinetic and EM energy. The GW efficiency  is high for short acceleration timescales, $\eta_{\rm GW} \sim E_{\rm GW}/(m_{\rm outflow} v_{\rm final}^2) \propto t_{\rm rise}^{-1}$. A typical rise time of $10 \, \mu$s gives an efficiency $\eta_{\rm GW} \sim 3\times 10^{-9}$ in the conversion of the MGF kinetic energy to $E_{\rm GW}$ in this simplistic picture. 

In the frequency domain, the energy spectrum, integrated over the solid angle, is
\begin{eqnarray}
    \frac{dE_{\rm GW}}{d\nu} &=& \frac{4 G}{15c^5}(2 \pi \nu)^6\tilde{M}_{zz}(2 \pi \nu)\tilde{M}^{*}_{zz}(2 \pi \nu) \quad {\rm for} \quad \omega_{\rm min} \le 2 \pi \nu \ll \omega_{\rm max}\,  \\
    & \sim & \frac{G m_{\rm outflow}^2 v_{\rm final}^4 c}{1500} \left(\frac{2 \pi t_{\rm rise} \nu}{c}\right)^6  \quad {\rm for} \quad \omega_{\rm min} \le 2 \pi \nu \ll \omega_{\rm max}\,
    \label{eq:outflowdGWdnu}
 \end{eqnarray}
where $\tilde{M} = \int_0^{t_{\rm rise}} M_{zz}(t)e^{2 \pi i \nu t}dt$. 
This expression was obtained in the quadrupole approximation with a non-relativistic equation of motion, and it is not formally applicable when $2 \pi \nu t_{\rm rise}  v_{\rm final}/c$ is not small (i.e. $2 \pi \nu \approx c/(v_{\rm final}t_{\rm rise})$ or higher). As Eq.~(\ref{eq:outflowdGWdnu}) diverges with $\nu$, we define $\omega_{\rm max}$ numerically by capping the corresponding integrated $E_{\rm GW}$ to the estimate given by Eq.~(\ref{eq:Egwtotal}). This regularization requires $\omega_{\rm max} \approx (1 + 14000\pi)^{1/7}/t_{\rm rise} \approx 4.6/t_{\rm rise} $ such that $\int d\nu dE_{\rm GW}/d\nu  \equiv  E_{\rm GW}$. 

\begin{figure}
\centering
\includegraphics[scale=0.32]{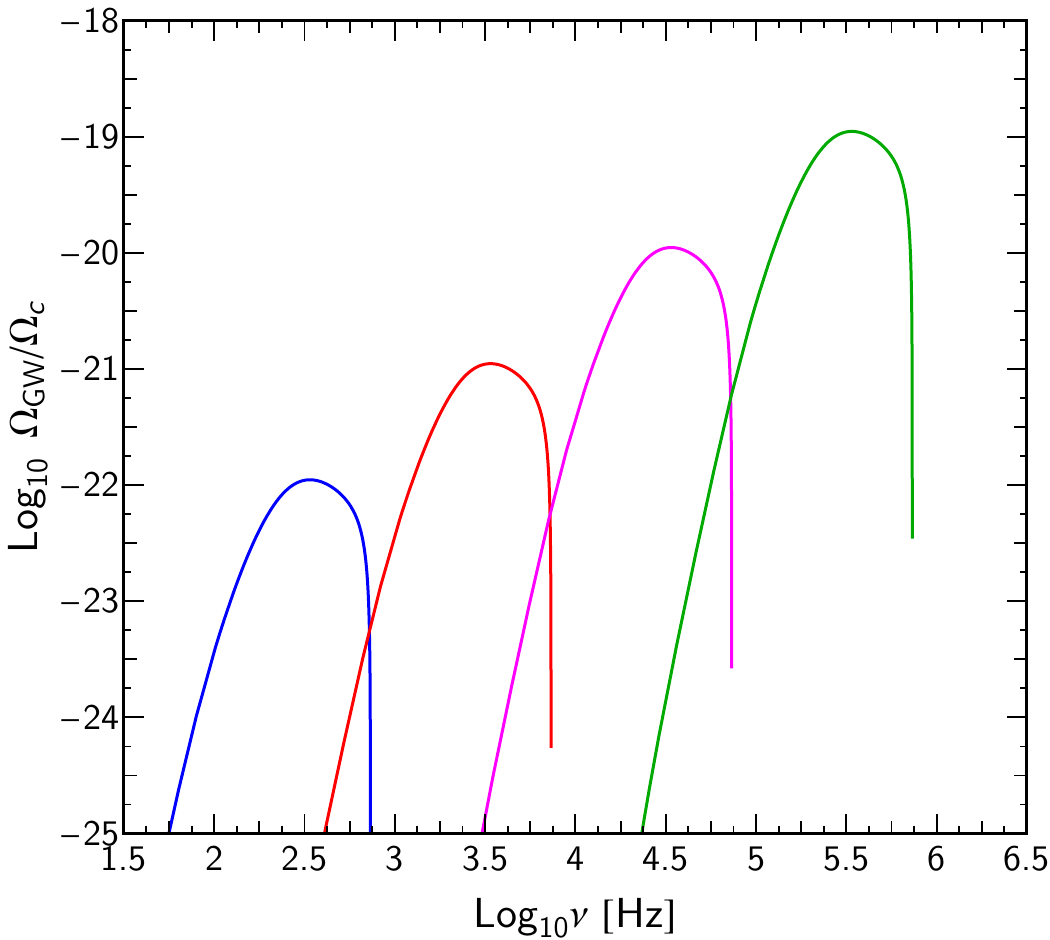}
\includegraphics[scale=0.32]{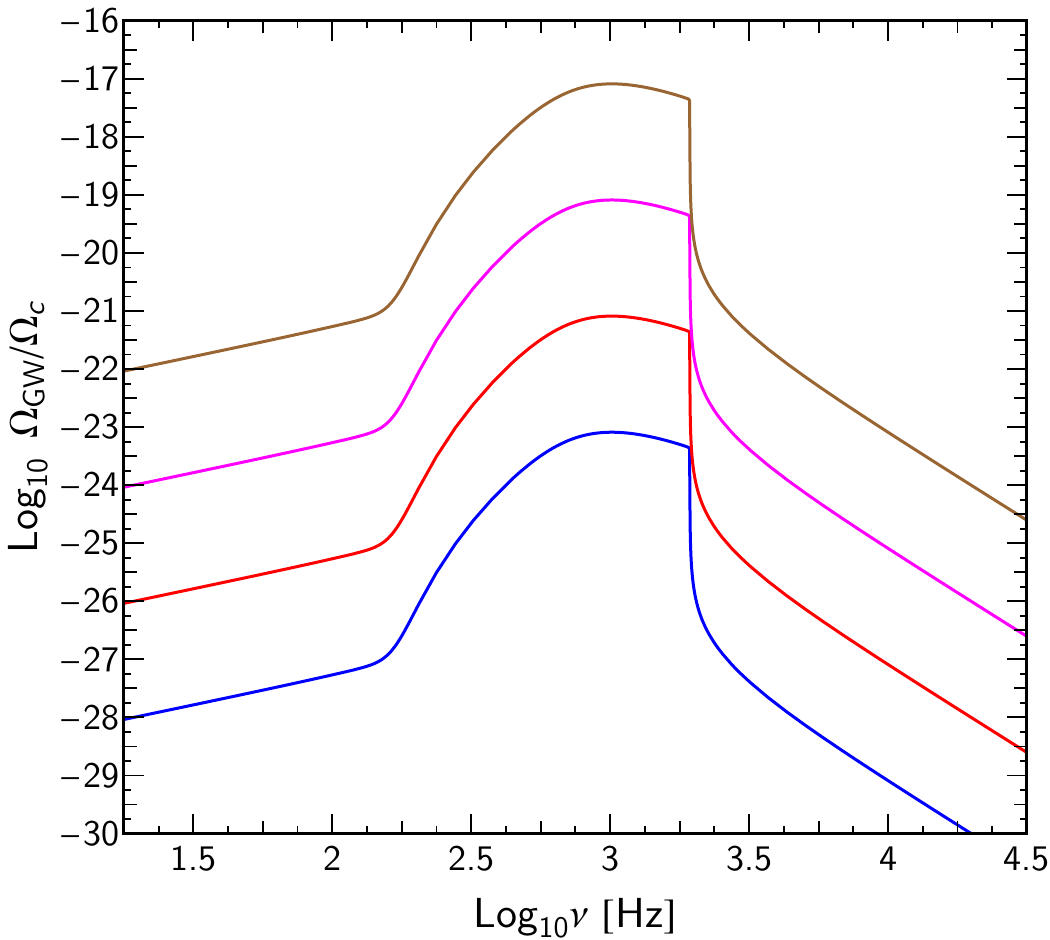}
\includegraphics[scale=0.32]{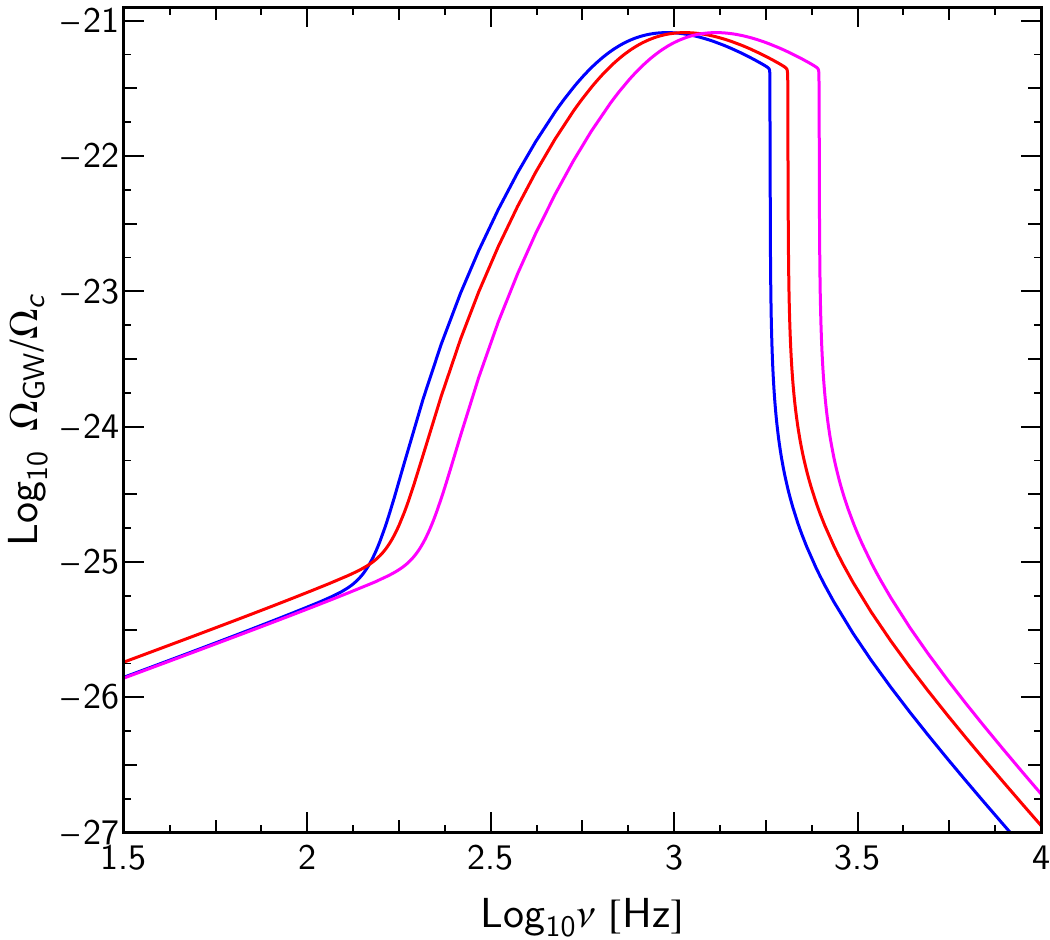}
\caption{Left: SGWB ($\Omega_{\rm GW}$) from outflows from magnetars in units of the critical energy density for a flat universe ($\Omega_{\rm c}$), varying the acceleration (rise) timescale from $10^{-3}$ s (blue) to $10^{-6}$ s (green). Here, $m_{\rm out} = 10^{-7} M_{\odot}$ with $v_{\rm final} = 0.7 c$. Depending on the radiation hydrodynamics during the MGF, rise times comparable to the light crossing timescale of a few microseconds may be realized for acceleration of matter close to the magnetar surface. This would imply a SGWB peaking well above LIGO's band, in the regime of high-frequency GW experiments. Center: SGWB from $f$-mode oscillations, shown for $E_{\rm GW} = \{10^{36},10^{38},10^{40},10^{42}\}$ erg with blue to brown curves, respectively. This assumes a magnetar mass of $1.4 M_\odot$ in the Sly4 EOS, corresponding to a mode of frequency 1.934 kHz and width  $\tau = 0.195$ s. Right: Same as center but with varying mass, $M = \{1.2,1.6,2.0\} M_\odot$ (blue, red and magenta colors respectively) in the Sly4 EOS (and associated damping times) for $E_{\rm GW} = 10^{38}$ erg. Varying masses thus shifts the SGWB frequency peak.}
\label{fig:GWoutflow}
\end{figure}

Observationally, the maximum frequency available in the data is the Nyquist frequency determined by the sampling rate of the detector (LIGO data are calibrated up to 5 kHz). The minimum frequency is determined by the inverse of $t_{\rm rise}$, i.e., $\omega_{\rm min} = 1$ kHz for $t_{\rm rise} = 1$ ms while $\sim 1\,\mu$s would be in the MHz regime. In Fig.~\ref{fig:GWoutflow} (left) we compute the SGWB for a simple scenario where $m_{\rm out} = 10^{-7} M_\odot$, $v_{\rm final} = 0.7c$, and $t_{\rm rise} \in \{10^{-6},10^{-5},10^{-4},10^{-3}\}$~s with green to blue curves, respectively. The peak for each curve occurs from the contribution at $z\sim 1$ from the maximum (Nyquist) frequency $\nu_{\rm peak, obs} \sim 4.6/t_{\rm rise}/(1+z)/(2\pi)$ of sources in that redshift range. Thus, $t_{\rm rise} \lesssim 10\,\mu$s will result in SGWB generally peaking beyond $\sim10$ kHz.  Galactic GW burst sources of this type could range from $v_{\rm obs} \sim 100$ Hz to $300$ kHz depending on the outflow radiation hydrodynamics and baryon loading.

\subsection{Scenario B: Excitation and GW Damping of NS Global Oscillation Modes}
\label{sec:fmode}

Alternatively, the main driver of GW emission by the MGF could be the excitation of fundamental fluid modes in the NS. These modes can be described as exponentially damped sinusoids, so that the strain amplitudes can be given by 
\begin{equation}
h_+(t) = Ae^{-t/\tau}\sin(\omega_{\rm R}t + \delta)\,,\ {\rm for}\, t > 0,
\end{equation}
where the amplitude $A$ is inversely proportional to the distance $R$ to the source, $\omega_{\rm R}$ is the angular oscillation frequency of the mode and $\tau$ is the mode damping time. The power spectrum can be approximately written as a Lorentzian
\begin{equation}
|\tilde{h}_+(\nu)|^2 = \frac{C}{(\Delta \nu)^2 + (\nu-\nu_0)^2}\,,
\end{equation}
where the centroid frequency $\nu_0$ and frequency width $\Delta \nu$ of the Lorentzian can be identified with the frequency and the inverse of the damping time of the mode to $O(\Delta \nu/\nu_0)^2$, and $C$ is proportional to the mode amplitude $A$  \citep{2019ApJ...871...95M}. The total energy emitted in gravitational waves can be calculated as
\begin{equation}
E_{\rm GW} = \frac{2\pi^2 R^2\nu_0^2c^3}{G}\int_{-\infty}^{+\infty}|\tilde{h}_+(\nu)|^2d\nu\,,
\end{equation}
see \cite{2011MNRAS.418..659L, 2022PhRvD.106f3007K,2024MNRAS.tmp.1988B}. This expression can be used to constrain the mode amplitude for a given assumed energy budget. From this, we obtain the normalized GW energy spectrum to leading order,
\begin{equation}
\frac{d E_{\rm GW}}{d\nu} \approx  \frac{2 E_{\rm GW} (1/\tau)}{4 \pi^2 (\nu-\nu_0)^2 +(1/\tau)^2}.
\label{eq:fmodedEgwdnu}
\end{equation}
Quasiperiodic oscillations observed in the tail of MGFs have been identified with torsional modes of the neutron star.  \cite{2011MNRAS.418..659L} find that these modes could be strongly excited. However, they are very weakly damped by the emission of gravitational waves, with estimated damping times $O(10^4 {\rm yr})$, and are expected to have a typical strain amplitude of approximately $10^{-28}$ at 10 kpc \citep{1983MNRAS.203..457S}. Therefore, gravitational wave detections of these modes are unlikely.

Arguably, the (fundamental) $f$-modes are the most likely to be excited with detectable amplitudes but how prolific and strong is their excitation in MGFs is debated. This debate stems from the uncertainty of the trigger of MGFs, the interior field structure of magnetars, and population characteristics of MGFs (i.e. whether strong mode excitation requires peculiar conditions or is more universal). \cite{2011MNRAS.418..659L} find that the energy in the $f$-modes is approximately $10^{-6} \times E_{\rm mag}$. \cite{2012PhRvD..85b4030Z} also discuss the detectability of GWs excited by MGFs and found that typically $E_{\rm GW} \sim 5 \times 10^{36}$ erg, with a strong dependence on the assumed magnetic field configuration. They assumed a catastrophic hydromagnetic instability mechanism for MGFs that generally may only occur once in the lifetime of a magnetar. \cite{2022PhRvD.106f3007K} consider the SGWB of $f$-modes excited by MGFs and conclude that they are probably undetectable by third generation GW detectors such as Cosmic Explorer \citep{2017CQGra..34d4001A} and Einstein Telescope \citep{2010CQGra..27s4002P}, with $E_{\rm GW} \sim 5 \times 10^{37}$ erg in a representative event. Here, we revisit this calculation, considering our estimates for the population of extragalactic MGFs. 

Using the formalism described in appendix (\S\ref{sec:stoch}), we compute the GW background for such an $f$-mode scenario, and the results are illustrated in the center and right panels of Fig. \ref{fig:GWoutflow}. We adopt a SLy4 equation of state \citep{2001A&A...380..151D} for the fundamental $n=0, l=2, m=0$ modes and their damping times \citep{2015PhRvD..91d4034C}. As the largest flares dominate the energetics of magnetars over their lifetime ($s<2$, see \S\ref{sec:IndMag}), the distribution of the largest event will also dominate in the SGWB if excitation mode amplitudes are linearly proportional to MGF electromagnetic counterpart energy. This is typically expected to be $\sim 3\times 10^{45}-3\times 10^{46}$\,erg (see \S \ref{sec:analysis}) and at the highest is bounded at $\sim10^{49}$ erg \citep{2011PhRvD..83j4014C}. Such a hugely energetic event requires an extremely strong magnetic field, which based on our analysis presented in this work, may only ever be realized for very rare magnetars (and even then will likely happen once in that object's lifetime).

\section{Theoretical considerations regarding neutrino emission}
\label{sec:theoreticalneutrino}

\subsection{ MeV neutrinos}

MeV neutrino cooling in the crust is likely a thermostat in magnetar outbursts over timescales of days to months. Direct external heating of the crust and reprocessing of emission  during a MGF may result in a MeV neutrino burst of $\gtrsim 10^{38}$ erg \citep{1995MNRAS.275..255T}. As MGFs possibly also involve significant heating and rearrangement of fields in the crust, additional transient MeV neutrino emission is expected given the strong temperature dependence of neutrino emissivities  \citep{2006NuPhA.777..497P}. However, neutrino luminosities depends on the nature of transient energy injection and heat transport within the crust and core. These interior timescales, however, are generally much slower compared to a MGF or pulsating tail duration. Reasonable candidate mechanisms for fast dissipation within neutron stars are mutual friction of superfluid components with flux tubes or normal components, or a deep mechanical failure of the NS crust \citep{2015MNRAS.449.2047L,2017ApJ...841...54T,2023ApJ...947L..16L}.

\subsection{High Energy Neutrinos}

High energy neutrino emission requires strong undamped particle acceleration which generally cannot be realized in optically thick conditions. As such, high energy neutrino emission is generally only permitted prior to the bulk thermalization of the confined fireball of the pulsating tail, when photon and plasma densities are low. As magnetars are not rotationally driven, the relevant potential (and maximum) realizable voltage drop is that induced by quakes and global oscillations in low-twist conditions. This was computed in the context of FRBs in in the appendix of \cite{Wadiasingh2020}. Strong particle acceleration, and pair cascades in these circumstances could also produce nearly-simultaneous FRB-like emission from magnetized neutron stars. Since some ion acceleration likely also occurs in such transient gaps, this will produce high energy neutrinos via either $p+\gamma \rightarrow \Delta$ \citep{2003ApJ...595..346Z} or proton curvature radiation of pions  \citep{2008JCAP...08..025H}.  Purely leptonic processes may also produce neutrinos, but with much reduced efficiency.

The MGF case differs in several ways from that expected in FRBs. Firstly, the magnetosphere prior to bulk thermalization of leptons ought to be more monopolar owning to prodigious plasma sourced from the crust. This drives up field curvature radii and thus reduces the pair production opacity \citep{2022ApJ...940...91H}. These outflows are transient as baryon-rich plasma departs without replacement, leaving a charge-starved state where gap discharges may develop as the magnetospheric solution relaxes back from a quasi-monopolar mass-loaded configuration to a force-free dipolar one.

The crustal disturbance requires a charge density $ \rho_{\rm burst} \sim  (\xi/\lambda) (\Omega_{\rm osc}/c) B/2$ where $\xi$ is the dislocation amplitude ($\xi/\lambda$ is the characteristic crust strain), $\Omega_{\rm osc}$ is an oscillation frequency\footnote{One does not need to excite well-defined normal modes here, and one can adopt an identification of a characteristic disturbance timescale $\Delta t \leftrightarrow 1/\Omega_{\rm osc}$. These disturbances or spectrum of oscillations are impulsive and necessarily damp quickly, but on timescales longer than the bulk Comptonization one.}, and $\lambda$ is a characteristic wavelength \citep{Wadiasingh2019}. The gap lengthscale for at-threshold curvature pair cascades without radiation reaction is \citep{Wadiasingh2020}
\begin{equation}
h_{\rm gap} \sim \frac{7}{3} \left(\frac{2}{\pi}\right)^{3/7} \left(\frac{\lambda}{\xi} \right)^{3/7} \left( \frac{B_{\rm cr}}{B} \right)^{3/7} \left( \frac{\lambda_C^2 \rho_{\rm c}^2 c^3}{\Omega_{\rm osc}^3} \right)^{1/7}.
 \end{equation}
where $\lambda_C$ is the reduced Compton wavelength, $\rho_c$ is the field curvature, $B_{\rm cr}$ is the quantum critical field. The gap electric field is $E_{\rm gap} \sim 4 \pi \rho_{\rm burst} h_{\rm gap}$ resulting in a characteristic potential drop of $10$ TeV to $1$ PeV:
\begin{equation}
e \Phi_{\rm gap} \sim  2 \pi e \rho_{\rm burst} h_{\rm gap}^2 \sim 1  \, \left( \frac{B_{15} \xi_4 \Omega_{\rm osc, 3} \rho_{c,8}^4}{\lambda_5}  \right)^{1/7} \, \, {\rm PeV} \sim  10 \,  \left( \frac{B_{14} \xi_1 \Omega_{\rm osc, 2} \rho_{c,6}^4}{\lambda_5}  \right)^{1/7} \, \, {\rm TeV} 
 \end{equation}
that has moderate sensitivity to field curvature radius $\rho_c$, which may be large for a MGF case as noted above. The luminosity of charges, $L_p \sim  2 \pi \rho_{\rm burst}^2 h_{\rm gap}^2 A_{\rm act} c$ is a bound on the high energy neutrino luminosity (and also beaming-corrected FRB luminosity sourced from pair cascades in low-twist conditions, see e.g., \citealt{2023MNRAS.519.3923C}). Here $A_{\rm act}$ is the active surface area of the magnetar permitting such acceleration. This luminosity bound is,
\begin{equation}
L_\nu \lesssim L_p \sim 10^{41} \, A_{\rm act,12} \, \rho_{c,8}^{4/7}\, \left(\frac{B_{15} \Omega_{\rm osc,3} \xi_4}{\lambda_5} \right)^{8/7} \, \rm erg \, s^{-1}
\label{lumbound}
\end{equation}
which will be realized until bulk Comptonization and ambient high photon number density terminates the gap discharges. This relaxation occurs on the Spitzer timescale \citep{1962pfig.book.....S,1983MNRAS.202..467S},
\begin{equation}
    \tau_{\rm Compt} \sim \frac{1}{n_e \sigma c \, \log \Lambda}    
\end{equation}
where $ \log \Lambda \sim 3-10$ the Coulomb logarithm, $\sigma \sim \sigma_T$ the Thomson (electron) cross section, and $n_e \sim \kappa \rho_{\rm burst}/e$ the pair density, $\kappa \sim 10^{2}-10^5$ the pair multiplicity. $\tau_{\rm Compt}$, for self-consistency, ought to be commensurate with $t_{\rm rise}$ of MGFs but many multiples of the light crossing timescale of a gap. In strong fields the cross section is also altered and in a Rosseland mean formulation, modifications scale as $\sigma_{\rm res} \sim \sigma_{\rm T} (\Theta_e/{\cal B})^2$ where ${\cal B} = B/B_{\rm cr}$ and $\Theta_e \equiv k_b T_e/(m_e c^2) \sim 0.01-1$ is the dimensionless pair plasma temperature \citep{2013MNRAS.434.1398V,2016MNRAS.461..877V,2023MNRAS.518..810D}. Yet this cross section is also dependent on the polarization state and angle of photons scattering, and Monte Carlo calculations suggest in high optical depth regimes mixing of polarization states reduces the impact of such cross section modifications \citep{2021MNRAS.500.5369B}. Using $\rho_{\rm burst}$ above, we obtain, $\tau_{\rm Compt} \gtrsim 100 \, \lambda_5 \kappa_2^{-1} B_{15}^{-1} \nu_3^{-1} \xi_4^{-1}$ $\mu$s. This Spitzer timescale, with less extreme parameters, could also be associated to the lag of a $2-3$ ms of the observed Comptonized hard X-rays compared to FRB sub-pulses in the April 2020 activity of SGR 1935+2154 \citep{2020ApJ...898L..29M,2023ApJ...953...67G,2023arXiv231016932G}.

The high energy neutrinos bounded by Eq.~\ref{lumbound} should be collimated along  magnetic polar axis similar to magnetospheric models of FRBs, albeit possibly in a different direction to where the radio emission decouples from the magnetosphere. Assuming radius-to-frequency mapping, the directionality of neutrinos and FRB emission ought to be more collinear at higher radio frequencies, and nearly simultaneous. Due to this beaming, isotropic-equivalent neutrino fluxes may significantly exceed the pair luminosity bound Eq.~\ref{lumbound} depending on the collimation of discharges prior to pair-photon thermalization. The magnetar oscillation neutrino luminosity considered here with $\Omega_{\rm osc} \sim 1$~kHz is analogous to a millisecond magnetar considered in \cite{2008JCAP...08..025H}, with similar detectability conclusions. In the MGF case, the isotropic-equivalent fluence of high energy neutrinos is roughly $E_{\nu, \rm MGF, iso} \sim \tau_{\rm Compt} L_p/f_{b,\nu} \sim 10^{38}-10^{41}$~erg; the chief source of uncertainty here is the timescale and locales over which particle acceleration is sustained. Nearby MGFs within the local group ought to be promising for high energy neutrino searches with cubic kilometer scale detectors.

\section{Discussion and Prospects}
\label{sec:discuss}
\subsection{Observational constraints on magnetar energetics over their lifetimes}
\label{sec:ObsEnergy}
The magnetic energy stored in the toroidal and poloidal fields of conventional magnetars is released through several channels over their lifetime. This available magnetic free energy is mostly contained in the crust\footnote{Even in some configurations with core fields, e.g. \citet[][]{2024MNRAS.528.5178S}.}, and the activity of conventional magnetars generally requires a zero-field core boundary condition for interesting crust field evolution on short timescales \citep[e.g.,][]{2002MNRAS.337..216H,2007A&A...470..303P,2013MNRAS.434..123V,2014MNRAS.438.1618G,2016PNAS..113.3944G,2020ApJ...903...40D,2023MNRAS.518.1222D}.  Aside from the short bursts and MGFs, magnetars emit a sizable fraction of their total magnetic energy through their persistent emission across the electromagnetic spectrum, yet, strongly peaking in X-rays and gamma-rays \citep[see][for a review]{2019RPPh...82j6901E}. Moreover, magnetars randomly enter periods of enhanced X-ray emission, concurrent to bursting activity, during which their persistent emitted power increases by up to 1000 fold \citep{cotizelati18MNRAS}. These outbursts last anywhere from weeks to years during which the emission decays back to its baseline level. Neutrino emission, driven by the bursting activity and/or the transient surface heating, could also be triggered \citep[e.g.,][]{yakovlev04ARAA,pons12ApJ:mag,guepin17AA}. Additionally, strong increase in spin-down rate, or torque, has been observed in a large fraction of magnetars following outbursts \citep[e.g.,][]{dib2014}. This is likely due to an enhanced wind and/or a twisted field configuration, both of which can be the result of magnetic energy dissipation. The sum of these components will limit the available energy for the bursting components observed in magnetars.

The X-ray persistent emission consists of a hot, surface thermal emission dominating in the energy range 0.1-10 keV, and a hard non-thermal tail of magnetospheric origin extending to energies $\gtrsim200$~keV \citep[][]{kuiper06ApJ}. The peak of the latter component has never been measured, yet, should lie in the energy range of a few hundred keV to 1 MeV. The soft X-ray emission has been shown to correlate with the magnetars spin-down age, $t_{\rm sd}$ following $L\propto t_{\rm sd}^{-c}$, with $c$ in the range $0.6-1$ \citep[][see also \citealt{seo23JKAS, kaspi10ApJ, marsden01ApJ}]{enoto2019RPPh}. Moreover, the hard-to-soft X-ray flux ratio strongly correlates with the spin-down age $L_{\rm h}/L_{\rm s} \propto t_{\rm sd}^{-0.7}$ \citep{enoto17ApJS}. Utilizing these two observational constraints and integrating over a period of $10^2-10^6$ years, we estimate an average total energy of about  $\sim10^{47}$~erg emitted by a magnetar as surface thermal and magnetospheric X-ray emission during its lifetime. This order-of-magnitude estimate considers $t_{\rm sd}$ as the true age of the magnetar $\tau$, yet the former has already been shown to be a poor indicator (e.g. \citealt{Beniamini2019}). A substantial fraction of the magnetars $t_{\rm sd}$ is derived using temporal properties measured during outbursts, and, as discussed above, $\dot{\nu}$ is typically larger during these episodes, leading to younger ages. Other indicators, such as the presence of a supernova remnant, could provide better estimates, yet these instances are scarcer and suffer from other systematics, e.g., degree of association.

Outbursts are a ubiquitous property of magnetars having been observed from almost all of the known sources \citep{cotizelati18MNRAS}. The total energy emitted in each outburst through the enhanced persistent soft X-ray emission is on average $10^{42}$~erg, with relatively small variation with $t_{\rm sd}$. The most uncertain element about outburst epochs is their recurrence rate and its variation within the population, which is mainly due to the short timespan for which we have been sensitive to magnetar activity. For instance, SGR~0526$-$66 in the LMC has been quiet for almost 50 years since its 1979 MGF, while on the other extreme end, 1E 1048.1$-$5937 shows quasi-periodic outbursts every 5 years since its monitoring with RXTE began \citep{archibald20ApJ}. For consistency with our analysis in \S \ref{sec:analysis}, we consider the rate of bursts (at a given energy $E$ that is sufficiently small compared to that of MGFs) to be proportional to $A{\rm (\tau)}E_t$ (see Eq. \ref{eq:dndev2}). From Eq. \ref{eq:Atau}, we can re-write $A{\rm (\tau)}$ in terms of $E_{\rm B}{\rm (\tau)}$, $s$, and $\alpha$ as $A{\rm (\tau)}E_t\sim A_0E_t(1+\tau/\tau_{\rm d,0})^{x}$ where $x=(2-\alpha-2s)/\alpha$, and $A_0E_t=0.05$~yr$^{-1}$ is the initial outburst rate. Hence, for $\tau_{\rm d,0} =10^4$~yr, $s=1.7$, $\alpha = 0$, and integrating over the period $10^2-10^6$ years, we find that a magnetar will emit in their active lifetime approximately $4\times10^{44}$~erg through enhanced soft X-ray radiation. This estimate excludes the enhancement of the hard X-ray tail emission that has been observed in a few sources with {\sl NuSTAR}. In most of these cases, the emission at energies $10$-$80$~keV has been equivalent or smaller than that of the soft X-ray emission \citep[e.g.,][]{kaspi14ApJ}.

A portion of the outburst energy is released through a particle wind as evidenced by (1) the increased torque $|\dot\nu|$ on the star during these epochs, sometimes by as much as one order of magnitude \citep{Harding1999,2006MNRAS.368.1717B,Beniamini+20}, and, (2) in the case of the SGR~1806$-$20 MGF, an expanding radio-emitting wind nebula \citep{Granot2017}. For the latter case, the expansion rate and luminosity decay of the nebula imply a total particle wind energy of the order of $10^{44}$~erg, or about $1\%$ of the total energy emitted in the MGF spike \citep[e.g.][]{Granot+06}. For the former, we also consider the case of the SGR~1806$-$20 outburst for which an extended increased torque $\delta\nu/\nu\approx0.03$ was observed for a period of 10 years, the largest of any magnetar \citep{younes17ApJ:1806}. This is equivalent to a total energy release $\int \dot{E}_{\rm rot}\,dt \sim 3\times10^{43}$~erg.

Neutrinos during bursts and outbursts are definitely produced, yet, with uncertain neutrino energies and fluxes. During MGFs, which possess non-thermal spectra and are likely due to emission from a mildly relativistic outflow, neutrino flux might result from proton-proton or photohadronic interactions with thermal radiation, as well as prolific MeV energy scale neutrino production within the crust and core \citep[e.g.,][]{1996ApJ...473..322T}. The strongest constraint for such a neutrino energy release is derived at the time of the SGR~1806$-$20 MGF using the AMANDA-II detector, $E_{\rm neutrino}\lesssim10^{45}$~erg \citep{icecube06PhRvL:1806}, or $<10\%$ of the electromagnetic counterpart. We note that the same detector also provided an upper-limit on the TeV $\gamma$-ray emission during the 2004 GF, $E_{\rm TeV-\gamma}\lesssim10^{44}$~erg, an order of magnitude smaller than the neutrino one.  
As for outbursts, neutrino emission is thought to emanate from the inner crust though plasmon and pair annihilation \citep{pons12ApJ:mag}. This might be the cause of the maximum luminosity of $\sim10^{36}$~erg~s$^{-1}$ observed at the onset of magnetar outbursts. While there is no direct observational constraint on this neutrino flux, several attempts at modeling the X-ray outburst decay curve with neutron-star cooling curves required a total energy deposited in the crust to be a factor of $\lesssim10$ larger than the observed total outburst energy, where the excess is lost to neutrino cooling \citep{cotizelati18MNRAS,camero14,scholz14,an18}. For a theoretical discussion of neutrino production mechanisms in magnetars see \S \ref{sec:theoreticalneutrino}.

The above simple observationally-motivated estimates imply that the non-flare magnetic energy losses of a magnetar are mainly dissipated through their persistent soft and hard X-ray emission, which is, to within an order of magnitude, about $10^{47}$~erg. This energy output can be related to the minimum initial internal magnetic field strength required to power the magnetar's X-ray output. To provide a conservative estimate, even if we reduce this nominal estimate above of the persistent emission output by a factor of 3-4, the result is $B_0\gtrsim 4\times 10^{14}$\,G.
The persistent emission cannot be too energetically dominant over MGFs, considering that the energy of even one MGF can be $10\%$ of the persistent energy release, and that integrated over the lifetime the energy release through GFs can potentially be much larger still.  Furthermore, we see that if $B_0$ is $\gtrsim 10^{15}$\,G, then it would require most of the energy to be released through MGFs, i.e. $f_{\rm fl}\to 1$. Overall, the considerations based on observational limits on X-ray output support our choice for the range of $0.1<f_{\rm fl}<1$ considered in \S \ref{sec:analysis}.

\subsection{Prospects for future MGF electromagnetic detections}
\label{sec:MGFfuturedet}
We outline the detection capabilities of several instruments for Magnetar Giant Flares (MGFs) by examining their energy ranges, sensitivities, and detection distances (Table\,\ref{tab:mgf_instruments}). Konus-WIND, Fermi-GBM, and INTEGRAL SPI-ACS can detect MGFs up to 15-20 Mpc, while Swift/BAT has the furthest detection range at approximately 25 Mpc, and INTEGRAL IBIS is more limited with a range of around 10 Mpc. Although these distances provide useful context, a more detailed discussion on MGF detection distances can be found in Appendix A of \cite{2021ApJ...907L..28B}.
Each instrument's sensitivity to MGFs, indicated by a limiting fluence ($\Phi_{\rm lim}$), varies. Most have a sensitivity around 
$2\times10^{-6}$ erg/cm$^{2}$, except for Swift/BAT, which is more sensitive at approximately $9.8\times10^{-7}$ erg/cm$^{2}$. These sensitivities are tailored to the hard spectra observed in MGFs, differing from the sensitivities for GRB detection due to the distinct spectral profiles.
The approximate distance within which MGFs can be confidently associated to local galaxies ($r_{\rm cc}$) varies based on instrument type. For coded aperture mask instruments like Swift/BAT and INTEGRAL, which have better localization capabilities, this distance is around 30 Mpc, constrained mainly by the effective area of the instruments. In contrast, for instruments like Fermi-GBM with poorer localization, the distance is approximately 10 Mpc, limited by the localization capabilities of the Interplanetary Network (IPN) triangulation, which depends on concurrent detections by other instruments.
Taken at face value, current observational data present a slightly lower detection rate of MGFs than expected (see Fig. \ref{fig:MGFsGRBratiocontour}). In particular, considering the number of MGFs seen by Fermi-GBM, Swift (which is both more sensitive and has a larger $r_{\rm cc}$) should have observed at least a few MGFs. The fact that none have been recorded, suggests a strong spectral dependence on the energy involved in these events that makes many fainter MGFs harder to detect. Additionally, the low-energy trigger window is not optimized for hard spectra like those of MGFs.
It is also possible that some of the samples above include as of yet unidentified MGF candidates. We advise readers to use these observational samples with caution. The most reliable sample currently available is the one presented in \cite{2021ApJ...907L..28B} which combines information from different instruments (see \S \ref{sec:modelindconst}) and which is used for deriving the main results in this paper.

\begin{table}
    \centering
    \begin{tabular}{lccccc}
        \toprule
        & Konus-WIND & Fermi-GBM & INTEGRAL SPI-ACS & INTEGRAL IBIS & Swift/BAT \\
        \midrule
        Energy range (keV) & 20--20000 & 8--40000 & 80--18000 & 15--10000 & 15--150 \\
        Detection Distance (Mpc) & 13--16 & 15--20 & 15--20 & 10 & 25 \\
        $\Phi_{\text{lim}}$ (erg cm$^{-2}$) & $2.00 \times 10^{-6}$ & $2.00 \times 10^{-6}$ & $2.00 \times 10^{-6}$ & $6.13 \times 10^{-7}$ & $9.80 \times 10^{-7}$ \\
        $r_{\text{cc,GF}}$ (Mpc) & 10 (IPN) & 10 (IPN) & 10 (IPN) & 30 & 30 \\
        Instantaneous Sky Coverage (\%) & 100 & 70 & 100 & 2 & 17 \\
        Duty Cycle (\%) & 95 & 85 & 85 & 85 & 85 \\
        $4\pi$ yr equivalent coverage & 26.6 & 9.5 & 18.7 & 0.4 & 2.9 \\
        $N_{\rm eMGF}/N_{\rm sGRB}$\footnote{The values of $N_{\rm sGRB}$ are taken from the most recent available catalogs. We use the standard definition for sGRB \citep[$T_{90}<2\rm{s}$;][]{Kouveliotou1993}, as reported in the referenced catalogs. $N_{\rm eMGF}$ values here represent the number of extragalactic MGFs detected by each instrument within its mission lifetime up to the date of the referenced catalog. For Fermi-GBM and INTEGRAL SPI-ACS, the $N_{\rm eMGF}$ values are lower limits due to the absence of a defined $r_{\text{cc, GF}}$ without concurrent detections from other IPN missions.} & 4/494 & $\geq$3/650 & $\geq$2/194 & 1/6 & 0/138 \\
        \bottomrule
    \end{tabular}
    \caption{Instrument parameters for MGF detection. Taken from \cite{SwiftBAT2005SSRv..120..143B,2005A&A...438.1175R,2009ApJ...702..791M,2010A&A...523A..61K,2010AstBu..65..326M,2015MNRAS.447.1028S,2016ApJS..223...28N,2016ApJ...829....7L,IBAS_GRB,2020ApJ...893...46V,2021ApJ...907L..28B,2022ApJS..262...32L}.}
    \label{tab:mgf_instruments}
\end{table}
 
Looking ahead, future space missions and mission concepts, both pointed and monitor, hold great promise for advancing our understanding of extragalactic magnetars. New space missions such as the Lynx X-ray Observatory \citep{2019JATIS...5b1001G}, the Athena (Advanced Telescope for High-ENergy Astrophysics) mission \citep{2022cosp...44.2316B}, the Einstein Probe \citep{Yuan2024}, and the Advanced X-Ray Imaging Satellite (AXIS) \citep{AXISpaper, AXIStdamm_paper} will improve our ability to detect more of these events and study them in greater detail. AXIS and Lynx are improved successors of the Chandra X-ray Observatory, with superior sensitivity and spatial resolution. Designed to uncover the faintest X-ray signals from the deep universe, these focusing X-ray telescopes will have enough sensitivity to observe MGF tails out to several Mpc (contingent upon rapid repointing). Similar capabilities will be offered by the Athena mission offering better line sensitivity and high spectral resolution. The Einstein Probe, with its wide-field X-ray telescope, is specifically designed to monitor the sky for transient events and will be pivotal in detecting serendipitous occurrences. Future missions, such as COSI \citep[the COmpton Spectrometer and Imager,][]{COSIofficial} and AMEGO-X \citep[All-sky Medium Energy Gamma-ray Observatory eXplorer,][]{2022JATIS...8d4003C}, require detailed simulations to accurately estimate detection distances due to the complex interaction between their different energy ranges, sensitivities, and improved localization capabilities. Almost certainly, with wide-field capabilities and high sensitivity in the Compton regime, these instruments will be MGF factories. StarBurst is a basic scintillator mission with six times the effective area of GBM \citep{woolf2024development}. It will be a prolific detector of MGFs, but because of its limited localization capability will require improved searches to classify detected sGRBs as MGFs without detection by additional spacecraft.

We emphasize that the association to a host Galaxy is crucial to identify extragalactic MGFs. Future optical survey (e.g., the Vera Rubin Observatory) will identify more and more galaxies, but the limiting factor is the gamma-ray monitors' capability to constrain the events' arrival direction.

\section{Conclusions}
\label{sec:conclusion}
We have explored in this paper theoretical and observational aspects of the extragalactic population of MGFs. While they are intrinsically orders of magnitude more common than sGRBs, the current population of MGFs are `buried' as a few percent fraction of the observed sGRB population. At present, the most reliable way of distinguishing an MGF from a sGRB is localizing the former to a nearby galaxy (and thus revealing its much reduced energy and MGF nature). Interestingly, for a survey with a threshold fluence of $\Phi_{\rm lim}\gtrsim 5\times 10^{-9}\mbox{\,erg cm}^{-2}$, simply improving the sensitivity without also improving the localization precision and increasing the distance to within which host galaxies can confidently be associated ($r_{\rm cc,GF}$), will likely not lead to an increase in the detected ratio $N_{\rm MGF}/N_{\rm sGRB}$  and may even decrease it (Fig. \ref{fig:MGFsGRBratiocontour}). The energy band of observations is also a critical component. The lack of extragalactic MGF candidates in the Swift-BAT sample, despite its improved sensitivity and localization as compared with, e.g. GBM, suggests that if we wish to optimize MGF detectability, a priority ought to be better $\gtrsim 100$~keV gamma-ray sensitivity. Taking this into account, future large-area missions such as eASTROGAM, StarBurst and AMEGO-X are particularly promising as extragalactic MGF detectors. While detections by an all-sky surveys with well defined observation strategies are ideally suited for theoretical interpretation, a different approach is to target particularly promising galaxies, which are both nearby and have an enhanced star formation. Beyond the Milky Way, NGC 253, M82, M77, IC 342, the Circinus galaxy and NGC 6946 top this list of hosts with future potentially detectable MGFs.

The current sample of 3 Galactic and 6 extragalactic MGFs, is already contributing to our understanding of magnetar bursts. Using largely model independent parametrization of magnetar bursts and evolution, we have shown that the observed fraction of MGFs out of a sample of 250 sGRBs with $\Phi_{\rm lim}=2\times 10^{-6}\mbox{\,erg cm}^{-2}, r_{\rm cc,GF}=10$\,Mpc can be used to constrain typical magnetar properties. The observed blind-survey rate is mostly a function of three parameters: the typical (internal) birth field of magnetars ($B_0$), the product $f_{\rm fl}f_{\rm mag}$, and the combination $f_{E}f_{\rm dip}/f_b$ (see Table \ref{tab:paras} for definitions of the various $f$ factors). Those results are summarized in Fig. \ref{fig:AllowedParamSpace} and Eq. \ref{eq:Bconstraint}.
As an illustration, if the magnetar formation rate is $\approx 20\%$ of the CCSNe rate, the fraction of a magnetar's field energy channeled into MGFs is $>0.1$ and $f_E f_{\rm dip}/f_b\lesssim 0.3$ then a typical magnetar should be born with $B_0\approx 2\times 10^{14}-2\times 10^{15}$\,G. Including additional constraints from the energy budget required to power persistent emission (\S \ref{sec:ObsEnergy}), the allowed range narrows to $B_0\approx 4\times 10^{14}-2\times 10^{15}$\,G.
These estimates are consistent with independent arguments \citep{Beniamini2019,2024MNRAS.tmp.2386L}. 
In the future, using the methodology outlined in this work, one could use the distribution of MGFs (and sGRBs) as a function of energy and distance to further constrain the intrinsic properties, and in particular, to uncover a potential sub-population of magnetars born with substantially larger free energy of  magnetic energy than the rest. Such a population, while it may represent a minority of magnetars, and is only likely to affect the MGF distribution at the highest energies, is of much astrophysical importance, and might be related to transients such as FRBs \citep{1985Natur.313..202L,Popov2010,Kumar2017,CHIME2020,Bochenek+20,MBSM2020,LBK2022,BeniaminiKumar2024} (where the most magnetic NSs may account for a rare but highly prolific population of repeating sources) and in smaller numbers, potentially also to GRBs and superluminous SNe \citep{Usov1992,DaiLu1998,Wheeler2000,Thompson2004,Uzdensky2007,BGM2017,MBG2018}. On the other hand, a limited range of $B_0$ or a maximum inferred $B_0$ would constrain superconductor gap models in neutron stars for core-expelled Meissner states; According to some models, stars born with higher $B_0$ may not display prolific magnetar-like activity as the core field would not be entirely expelled \citep{2024MNRAS.tmp.2386L,2024Landersub}.

We also consider GWs from MGFs, and the SGWB contribution from magnetars. The two most promising magnetar GW channels are (A) impulsive acceleration (occurring on a timescale $\ll1$ ms) of baryon-loaded outflows and (B) global $f$-mode oscillations excited during MGFs. The $f$-mode case would result in local burst GW  events with GW frequencies of $\nu \sim 2$ kHz, in the LVK band and is of relevance to third generation facilities \citep{2017CQGra..34d4001A} such as Cosmic Explorer \citep{2021arXiv210909882E}, Einstein Telescope \citep{2010CQGra..27s4002P}, or facilities tuned toward slightly higher frequencies such as NEMO \citep{2020PASA...37...47A}. A Galactic or nearby extragalactic event would potentially even be detectable and set interesting neutron star mass and equation-of-state constraints \citep{2007PhRvD..76f2003A,2017CQGra..34p4002Q,2021ApJ...918...80M,2024ApJ...966..137A,2024MNRAS.tmp.1988B,2024arXiv241009151T} although the (poorly-understood) efficacy of $f$-mode excitation must be higher than standard expectations during the MGF trigger. Yet, perhaps this is not pessimistic, given that large timing anomalies and glitches with energetics $\sim 10^{41}$ erg have been observed in magnetars even in energetically much less violent events than MGFs \citep{dib2014,2020ApJ...896L..42Y,2024Natur.626..500H}. The SGWB $f$-mode magnetar contribution (peaks at approximately $\sim 1$ kHz) is also possibly significant, and potentially detectable by third generation facilities. It's shape and spectrum trace the cosmic history of magnetar formation, and the contributing distribution of magnetar masses and excitation spectrum modes beyond the fundamental $n=0,l=2$ mode \citep{2015PhRvD..91d4034C}. 

Acceleration of outflows in MGFs is another possible GW source, explored in this work for the first time to our knowledge. This is motivated by the confirmation of $t_{\rm rise} \lesssim 10-100\,\mu$s (unresolved) rise times in 200415A at burst onset \citep{Roberts2021Natur.589..207R} and the existence of collimated baryonic outflows in MGFs with kinetic energy comparable to that in the prompt electromagnetic emission. The emitted GW energy in this scenario is similar to or possibly even much higher than $f$-mode excitations, with the proviso that short $t_{\rm rise}$ which permits efficient GWs also pushes the characteristic frequency of GW emission beyond the frequency band of LIGO or third generation detectors. However, a $100 \,\mu$s is still relevant for LIGO-like experiments, and potentially detectable for Galactic events from the estimate Eq.~\ref{eq:outflow_hmax}. For shorter $t_{\rm rise} \lesssim 100 \,\mu$s the GW frequency is pushed beyond 3 kHz. Nevertheless, there is an established community of high-frequency GW experiments and concepts \citep{2021LRR....24....4A}, largely directed at exotic GW sources. For example, the Levitated Sensor Detector, currently under development \citep{2013PhRvL.110g1105A,PhysRevLett.128.111101,2022PhRvL.129e3604W}, is probing this relevant frequency range, with proposed upgrades potentially approaching a parameter space interesting for outflow GWs from MGFs, relevant to both Galactic burst sources as well as the SGWB. Generally, the issue of detectability of a SGWB (for either the $f$-mode or outflow scenario) is a nontrivial calculation \citep[e.g.,][]{2013PhRvD..88l4032T,2024arXiv241104029B} and depends on factors such as observational integration time, long-term stability of the GW experiment, background spectral shape, competing cosmic backgrounds and their characteristics, and potential frequency-dependent sky anisotropy associated large scale structure. This is deferred to a future work.

\appendix

\section{The Magnetar SGWB}
\label{sec:stoch}

The SGWB from a cosmological population of MGFs and bursts will be stationary, gaussian, unpolarized and nearly isotropic. The magnetar GW background in a particular frequency interval is the contribution of all magnetars within the past light cone (at different redshifts) at Earth which contribute to the observed spectrum in the present epoch.

 A weighting of ``$\nu F_\nu$" $\rightarrow \nu \, d E_{\rm GW}/d\nu$ spectral energy distribution by the number density of GW sources $d{\cal N}/(dV dz)$ in comoving volume element $dV$ and redshift interval $dz$ then yields the concomitant differential SGWB energy in element $dz$. Denote $\nu_{\rm rest}$ the comoving emitted frequency in the source frame and $\nu_{\rm obs} = \nu_{\rm rest}/(1+z)$ the received frequency by an observer at $z=0$. The energy density of GWs at the present epoch (which permits contributions from up to $z_{\rm max}$) is,
\begin{equation}
\Omega_{\rm GW}(\nu_{\rm obs}) = \int_0^{z_{\rm max}} \frac{dz}{1+z} \left(\frac{d{\cal N}}{dV dz} \right) \left[ \frac{d E_{\rm GW}^{\rm rest}}{d \log \nu_{\rm rest}} (\nu_{\rm rest}) \right]_{\nu_{\rm rest} = (1+z)\nu_{\rm obs}}
\label{eq:omega}
\end{equation}
where the $1+z$ factor in the denominator arises from the conversion of $E_{\rm GW}$ from the comoving to the observer frame \citep{2018gwv..book.....M}. Adopting standard $\Lambda$CDM cosmology and assuming stationarity, the differential source number density in a redshift element $dz$ may be expressed as a source production rate,
\begin{equation}
\frac{d{\cal N}}{dV dz} = \frac{d {\cal N}}{dV d t_{\rm rest}} \frac{1}{dz/dt_{\rm rest}} = \left(\frac{d{\cal N}}{dV d t_{\rm rest}} \right) \frac{1}{(1+z) H_0 {\cal E}(\Omega_{\rm M}, \Omega_{\rm \Lambda}, z)}
\label{eq:dNdVdz}
\end{equation}
where we set $\Omega_\Lambda = 0.69$, $\Omega_{\rm M} = 0.31$ and Hubble constant $H_0 = 68$ km s$^{-1}$ Mpc$^{-1}$ \citep{2016A&A...594A..13P} with ${\cal E}(z) = \sqrt{\Omega_{\rm M} (1+z)^3 + \Omega_\Lambda}$.

Since magnetars are young objects whose formation is approximately consistent with core collapse supernovae \citep{Beniamini2019,2021ApJ...907L..28B}, we assume here that their production rate will track the local SFR. Adopting Eq.~(60) of \cite{2022ApJS..259...20H} which matches recent James Webb Space Telescope inferences for the cosmic star formation  rate history \citep{2023ApJS..265....5H} up to $z \sim 15$, 
\begin{equation}
\frac{d{\cal N}}{dV d t_{\rm rest}} = \mathcal{R}_{\rm MGF}(z=0) \left(\frac{\Phi(z)}{\Phi_{0}} \right) =  63.4 \mathcal{R}_{\rm MGF}(z=0) \left[ 10^{0.22 (1+z)} + 2.4\times 10^{-0.3+0.5(1+z)} + 61.7(1+z)^{-3.13} \right]^{-1} \equiv \mathcal{R}_{\rm MGF}(z) \qquad  1 \leq z \lesssim 15.
\label{eq:Rmagz}
\end{equation}
%
%
The value of $\mathcal{R}_{\rm MGF}(z\!=\!0)$ may be connected to estimates in \S\ref{sec:MGFpop}, so that $\mathcal{R}_{\rm MGF}(z=0) = \mathcal{R}_{\rm mag}(z=0) f_{fl}/(f_E f_{\rm dip}) \sim 7 \times 10^4 (f_{\rm mag} f_{\rm fl}/0.06)(0.1/f_E f_{\rm dip})$ Gpc$^{-3}$ yr$^{-1}$ for burst energies $E_{c,t}\sim 4\times10^{45} [B_0/(5\times10^{14}{\rm\, G})]^{1/2}[(f_E f_{\rm dip})/0.1]$ erg.

Scaling Eq.~(\ref{eq:omega}) to the critical energy density $\Omega_{\rm c} = \rho_{\rm c} c^2 = 3 H_0^2c^2/(8 \pi G)$ \citep{1992PhRvD..46.5250C,1999PhRvD..59j2001A,2001astro.ph..8028P}, the SGWB is then given by

\begin{equation}
 \frac{\Omega_{\rm GW} (\nu_{\rm obs})}{\Omega_{\rm c}} = \frac{\nu_{\rm obs}}{\Omega_{\rm c} H_0} \int_0^{z_{\rm max}} dz \, \frac{\mathcal{R}_{\rm MGF}(z)}{(1+z) {\cal E}(\Omega_{\rm M}, \Omega_{\rm \Lambda}, z) }  \left[ \frac{d E_{\rm GW}}{d\nu_{\rm rest}} (\nu_{\rm rest}) \right]_{\nu_{\rm rest} = (1+z)\nu_{\rm obs}}
\end{equation}
identical to other derivations \citep[e.g.,][]{2018PhRvL.120i1101A}.
The spectrum of GWs in the rest frame may be related \citep[e.g. Eq.~(1.160) of][]{2007gwte.book.....M} to the Fourier transform $\Tilde{h}_{+,\times}$ of strain amplitudes of GWs via 
\begin{equation}
\frac{d E_{\rm GW}}{d \nu_{\rm rest}} (\nu_{\rm rest}) = \frac{\pi c^3}{2 G} \nu_{\rm rest}^2 d_{\rm c}^2 \int d \Omega \left( |\Tilde{h}_+ (\nu_{\rm rest})|^2 + |\Tilde{h}_\times (\nu_{\rm rest})|^2 \right)
\end{equation}
where the integral here is regarded as average over all solid angles.

Consider now asymptotic estimates of the SGWB. For the $f$-mode case \S\ref{sec:fmode}, a Dirac delta function approximation may be made to the spectrum (Eq.~\ref{eq:fmodedEgwdnu}), $dE_{\rm GW}/d\nu \rightarrow E_{\rm GW} \,\delta(\nu_{\rm rest}-\nu_0)$. This eliminates the integration over $z$ via a change of variables, viz. $\delta\left[\nu_{\rm obs} (1+z)-\nu_0\right] dz = (1/v_{\rm obs}) \delta(z_0-z) dz$ with $z_0 \equiv \nu_0/\nu_{\rm obs} -1$. As an order-of-magnitude estimate, note that the function Eq.~\ref{eq:dNdVdz} with Eq.~\ref{eq:Rmagz} peaks at $z\approx 0.91$ at value,
\begin{equation}
   \left. \frac{d{\cal N}}{dV dz} \right|_{\rm peak} \sim  \frac{1.8 \, \mathcal{R}_{\rm MGF}(z=0)}{H_0}.
\end{equation}
Thus, the magnitude of the SGWB at $\nu_{\rm obs} \sim 1$ kHz with $z_0 \approx 0.91$ is,
\begin{equation}
\left. \frac{\Omega_{\rm GW} (\nu_{\rm obs}= 1 \, \rm kHz)}{\Omega_{\rm c}} \right|_{f\rm-mode} \sim  \left. \frac{1.8 \, E_{\rm GW} \mathcal{R}_{\rm MGF}(z=0)}{H_0 \Omega_{\rm c} } \right|_{z=0.91} \approx 8 \times 10^{-22} \left( \frac{E_{\rm GW}}{10^{38} \,\, \rm erg} \right) \left( \frac{\mathcal{R}_{\rm MGF}(z=0)}{7\times10^4 \, \,\rm  Gpc^{-3}\, yr^{-1}} \right). 
\end{equation}
For the outflow case, \S\ref{sec:GWoutflow}, a similar but less accurate delta function approximation may be made owing to the steep and narrow frequency dependence of Eq.~\ref{eq:outflowdGWdnu}, $dE_{\rm GW}/d\nu \approx 2\pi E_{\rm GW} \,\delta(\omega_{\rm rest} -\omega_{\rm max})$ with $\omega_{\rm max}\approx 4.6/t_{\rm rise}$. This approximation results in,
\begin{eqnarray}
  &  \left. \frac{\Omega_{\rm GW}(\nu_{\rm obs}\!=\! 30 \rm kHz)}{\Omega_{\rm c}} \right|_{\rm outflow} \sim  10^{-20} \left(\frac{m_{\rm outflow}}{10^{-7} M_\odot} \right)^2 \left(\frac{v_{\rm final}}{0.7 c} \right)^4 \left(\frac{10 \, \mu \rm s}{t_{\rm rise}}\right) \left( \frac{\mathcal{R}_{\rm MGF}(z=0)}{7\times10^4 \, \,\rm  Gpc^{-3}\, yr^{-1}} \right)\nonumber \\ &
  \sim 10^{-22} \left(\frac{B_0}{5\times 10^{14} \mbox{ G}} \right)^4 \left(\frac{f_E f_{\rm dip} \eta_{\rm kin}^2}{0.1} \right)\left(\frac{f_{\rm fl} f_{\rm mag} }{0.06} \right)\left(\frac{10 \, \mu \rm s}{t_{\rm rise}}\right).
\end{eqnarray}

\bigskip\bigskip
\noindent {\bf ACKNOWLEDGEMENTS}

\medskip

We thank Sam Lander, Abhishek Desai, Jessie Thwaites, Marcos Santander, Dan Kocevski, and Cole Miller for helpful discussions.
P.B. acknowledges support from a NASA grant 80NSSC24K0770, a grant (no. 2020747) from the United States-Israel Binational Science Foundation (BSF), Jerusalem, Israel and by a grant (no. 1649/23) from the Israel Science Foundation.
C.C. and Z.W. acknowledge support by NASA under award numbers 80GSFC21M0002 and 80GSFC24M0006. This research has made use of the NASA Astrophysics Data System.

\end{document}